\documentclass[sn-mathphys,Numbered]{sn-jnl}

\usepackage{graphicx}%
\usepackage{multirow}%
\usepackage{amsmath,amssymb,amsfonts}%
\usepackage{amsthm}%
\usepackage{mathrsfs}%
\usepackage[title]{appendix}%
\usepackage{xcolor}%
\usepackage{textcomp}%
\usepackage{manyfoot}%
\usepackage{booktabs}%
\usepackage{algorithm}%
\usepackage{algorithmicx}%
\usepackage{algpseudocode}%
\usepackage{listings}%

\usepackage{lineno}
\usepackage{pdflscape} 
\usepackage{float}

\usepackage{subcaption}

\theoremstyle{thmstyleone}%
\theoremstyle{thmstyletwo}%

\theoremstyle{thmstylethree}%

\newcommand{\co}[0]{CO\textsubscript{2} }
\newcommand{\cd}[0]{CO\textsubscript{2}}

\begin{document}

\title[Article Title]{Design Optimization and Global Impact Assessment of Solar-Thermal Direct Air Carbon Capture}
\author*[1,2]{\fnm{Zhiyuan} \sur{Fan}}\email{zf2198@columbia.edu}


\author*[1]{\fnm{Bolun} \sur{Xu}}\email{bx2177@columbia.edu}

\affil*[1]{\orgdiv{Earth and Environmental Engineering}, \orgname{Columbia University}, \orgaddress{\street{500 W. 120th Street \#510}, \city{New York}, \postcode{10027}, \state{NY}, \country{US}}}

\affil*[2]{\orgdiv{Center on Global Energy Policy}, \orgname{Columbia University}, \orgaddress{\street{1255 Amsterdam Avenue}, \city{New York}, \postcode{10027}, \state{NY}, \country{US}}}



\abstract{
The dual challenge of decarbonizing the economy and meeting rising global energy demand underscores the need for scalable and cost-effective carbon dioxide removal technologies. Direct air capture (DAC) is among the most promising approaches, but its high energy intensity, particularly the thermal energy required for sorbent regeneration, remains a critical barrier to cost reduction and sustainable deployment. 
This study explores solar-thermal DAC systems that combine concentrated solar thermal technology with low-cost sand-based thermal energy storage to meet this demand. We analyze the techno-economic performance of such systems in both grid-connected and stand-alone configurations. Results show that solar-thermal DAC can achieve annual capacity factors exceeding 80\% and \co removal costs as low as \$160–\$200 per ton, making it competitive with leading DAC technologies. An optimal 6000 ton/yr modular system design takes \(<\)1 km\textsuperscript{2} land-use requirement. The proposed system operates most efficiently with short-cycle sorbents that align with solar availability. The stand-alone Solar-DAC systems, which rely solely on solar energy for both electricity and thermal energy, are particularly promising in regions with high solar capacity and sandy terrain, exhibiting minimal ambient sensitivity from temperature and humidity. In areas with sedimentary basins suitable for \co storage, solar-powered DAC offers a lower-cost alternative to geothermal heating, which often faces geological and economic constraints.
}

\keywords{Direct air capture (DAC), Solar, Concentrates solar thermal (CST), Sand thermal energy storage, Climate change, Power market, Optimization}



\maketitle

\section{Introduction}\label{sec1}

Our society faces the intertwined challenges of climate change and energy demand surge. The year 2024 marked the first year in which global average temperature surpassed the 1.5 \textdegree C threshold~\cite{copernicus_climate_change_service_copernicus_2024}, with rising atmospheric \co concentrations identified as a primary driver of global warming. Simultaneously, the rapid growth of energy-intensive sectors such as data centers, manufacturing, and electrified infrastructure has intensified the search for scalable, low-carbon energy sources. Synthetic fuels have emerged as a promising solution to meet this demand, and \co is a critical feedstock in their production~\cite{o2024co2}. Direct air capture (DAC) is one of the most scalable approaches to removing \co directly from the atmosphere~\cite{mcqueen_review_2021, izikowitz_carbon_2021}, with the captured \co either permanently sequestered via geological storage or mineralization~\cite{alcalde_estimating_2018}, or utilized to displace fossil-derived carbon in applications such as construction materials~\cite{liu_new_2021} and synthetic fuels~\cite{parigi_power--fuels_2019}.

Despite recent advancements in DAC system energy efficiency, its significant energy demand is still the major limitation to its scalable and sustainable deployments~\cite{sabatino_comparative_2021}. DAC operation requires a significant amount of electricity and heat, ranging from 167-305 kWh of electricity and 1.4-3.2 MWh of thermal energy per ton of \co capture, driven by various regeneration temperatures and heating process design~\cite{mcqueen_review_2021, fasihi_techno-economic_2019, realmonte_inter-model_2019}. Such extensive energy consumption makes it the largest cost component for DAC's operational expenditure (OPEX), also posing significant risks to its sustainability. While sourcing electricity from the power grid, the DAC system operation inevitably adds scope 2 emission due to fossil generators in the grid generation mixture, which sometimes even lead to net \co emissions instead of net-removal~\cite{terlouw_life_2021, deutz_life-cycle_2021}. 

The thermal energy requirement for solvent/sorbent materials' regeneration poses even higher risks than electricity consumption. Firstly, the regeneration process requires certain temperature targets, typically 100 \textdegree C target for solid sorbent DAC systems~\cite{leonzio_environmental_2022, azarabadi_sorbent-focused_2019, sinha_systems_2017} 
and 800 \textdegree C target for liquid solvent processes involving calcination, such as potassium hydroxide (KOH) liquid solvent or calcium hydroxide Ca(OH)\textsubscript{2} mineralization \cite{sabatino_comparative_2021, keith_process_2018, mcqueen_use_2024}. Secondly, the sources of low-carbon heat are often restrictive as thermal energy cannot be easily transferred over long distances~\cite{thiel_decarbonize_2021}. Many studies propose industrial waste heat as an alternative thermal energy source, but it faces challenges such as heat transfer logistics, consistency in availability, and temperature compatibility \cite{fasihi_techno-economic_2019}\cite{deutz_life-cycle_2021}. Current DAC system developers harvest thermal energy either through dedicated low-carbon heat sources such as geothermal energy \cite{faircloth_passive_2023}, through electrification process by purchasing verified renewable power \cite{paulsen_techno-economic_2024}, or unspecified.

Solar energy provides both clean electricity through photovoltaics (PV) and low-carbon heat via concentrated solar thermal (CST) collectors for DAC systems \cite{zhang_concentrated_2013, li_solar_2024, prats-salvado_solar-powered_2024}.  
This study explores the design of solar-thermal DAC systems that utilize solar thermal energy and electricity. We focus on a solid sorbent DAC system with a regeneration temperature of 100 \textdegree C, which is within the comfortable range of concentrated solar thermal collectors. The system is complemented with sand thermal storage to improve the operational capacity factor. We evaluate the operation of solar-thermal DAC systems in two scenarios: (1) grid-connection operation representative electricity markets in the U.S.; (2) stand-alone operation powered by PV and battery energy storage, applicable globally. Detailed thermodynamic simulation and operational analyses show that solar-thermal DAC is technically feasible and economically competitive compared to alternative DAC designs while offering reliable and verifiable low-emission carbon removal. 



\section{Solar-thermal DAC Design and Simulation}\label{sec2}

\begin{figure}[h!]
    \centering
        \begin{subfigure}[b]{0.75\textwidth}
            \centering
            \includegraphics[width=\textwidth]{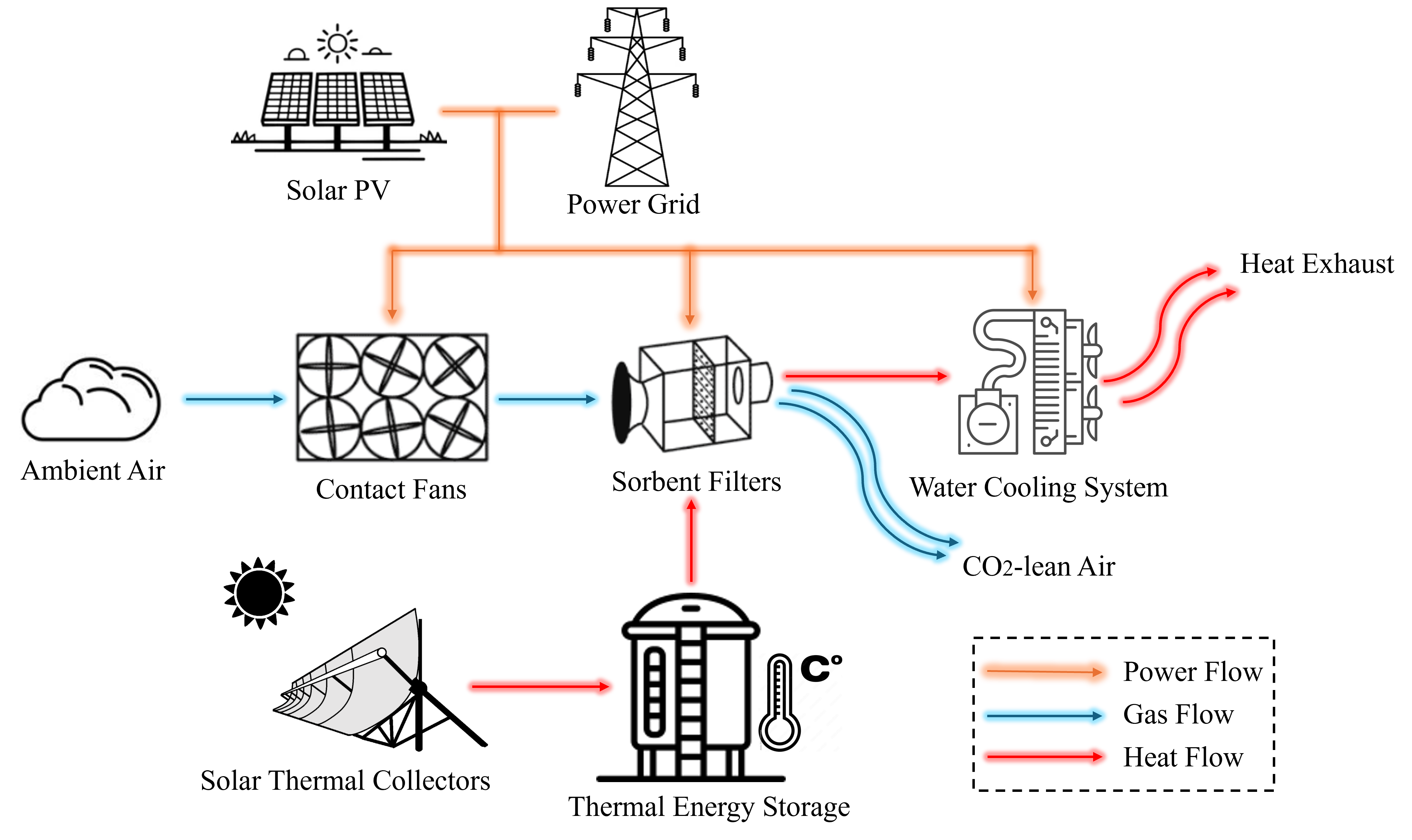}
        \end{subfigure}
        \hfill
        \begin{subfigure}[b]{0.95\textwidth}  
            \centering 
            \includegraphics[width=\textwidth]{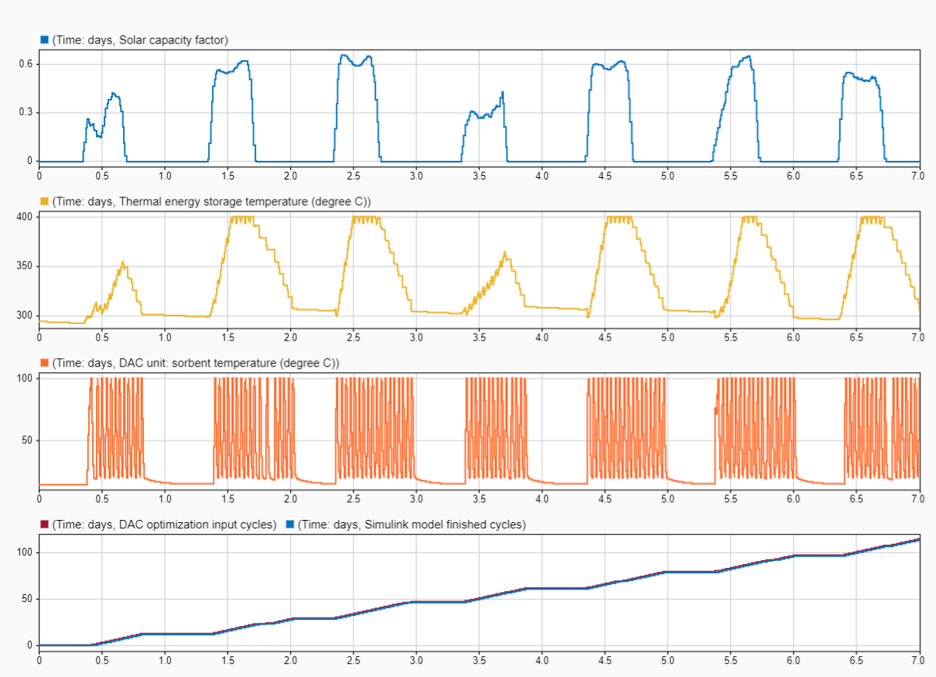} 
        \end{subfigure}
    \caption{\textbf{solar-thermal DAC schematic and sample week thermodynamic simulation.} The schematic diagram (upper) shows the single unit solid-sorbent DAC system design. A sample week thermodynamic simulation results using MATLAB Simscape (lower) shows the solar capacity factor input (row-1), the sand energy storage temperature profile (row-2), and resulted DAC sorbent temperature cycles (row-3). The optimization used in this study simplifies the simulation thermodynamic assumption, and the DAC regeneration cycle comparison (row-4) show that operation optimization can achieve \(>\)99\% simulation cycles. This study utilizes a type of MOF \cite{azarabadi_sorbent-focused_2019, sinha_systems_2017} sorbent enabling a 1-hour cycle time, minimizing solar curtailment by promptly utilizing thermal energy.}
    \label{tab:fig2}
\end{figure}

We propose a conceptual DAC design that supplies thermal energy through solar heating, addressing the challenges of thermal energy demand and sustainability. The design includes concentrated solar thermal (CST) and sand thermal energy storage to meet the thermal load (see \textcolor{blue}{Figure}~\ref{tab:fig2}). CST converts solar radiation into thermal energy at certain inlet temperatures ranging from 200–500 \textdegree C, storing it in sand storage. Excess thermal energy is curtailed when the storage temperature reaches the inlet temperature. During sorbent regeneration, heat is transferred from storage to DAC sorbents, maintained at 100 \textdegree C for the required desorption period, after which a water-cooling system cools the DAC system to ambient temperature and initiates a new adsorption cycle. Electricity, needed for both adsorption and desorption processes at varying power levels, can be sourced either from grid interconnections or on-site solar PV plus battery energy storage.

We select a specific MOF \cite{sinha_systems_2017} as the representative DAC sorbent primarily for its shorter cycle time, which is approximately 1 hour. This enables more flexible DAC operation in response to solar power volatilities. 
In contrast, typical amine-functionalized sorbents~\cite{leonzio_environmental_2022} exhibit much longer cycle times from 8 hours up to days, making them less suitable for utilizing the available solar thermal energy efficiently. Secondly, the MOF's rapid cycle time and high energy consumption compared to amine-functionalized sorbents enable rigorous stress testing of our thermodynamic simulation tool and optimization strategy, which are based on very high temporal resolution data. 

Thermal energy storage is essential for solar-thermal DAC systems, significantly boosting their capacity factor from below 25–30\% (typical of solar-only systems) to over 85\% in a cost-effective manner. This study examines sand-based thermal energy storage as an alternative to the more commonly discussed molten salt systems, offering several advantages: lower cost (\$4–\$10/kWh \cite{kraemer_nrel_2022, tetteh_improved_2024, poulose_power_2022} vs. \$20–\$30/kWh \cite{soto_performance_2024,gonzalez-roubaud_review_2017}), greater temperature flexibility due to the absence of phase changes\cite{bhatnagar_molten_2022}, and simplified operation without corrosive materials. Additionally, sand is abundant in desert regions—ideal locations for solar-thermal DAC, making deployment more practical despite its lower volumetric energy density.

A detailed thermodynamic simulation using MATLAB-Simscape is used as benchmark verification of several critical thermodynamic design properties: (1) minimum thermal source temperature 300 \textdegree C for efficient heat exchange to 100 \textdegree C regeneration temperature. (2) maximum DAC modular mass/capacity at 6000 ton-\cd/year, limited by heat transfer coefficient and surface area. This study optimizes the solar-thermal DAC design parameters and tests its performance under grid-connection and stand-alone scenarios with global impact analysis.


\section{Design Optimization and Grid-interactive Economics}\label{sec3}

We examine the design tradeoffs of grid-connected solar-thermal DAC systems. As the CST heating efficiency decreases with higher required temperature \cite{eck_modelling_2005, jiang_implementation_2014}, the trade-off between the inlet target temperature, storage capacity, and solar thermal heating capacity leads to an optimal design space (see \textcolor{blue}{Figure}~\ref{tab:fig2}). While increasing the solar concentration ratio can increase solar efficiency at a given temperature, it disproportionally increases the unit CAPEX compared to just adding CST capacity directly. Both CST and DAC systems exhibit scaling benefits, but the heat transfer coefficient imposes an upper limit on size.  For consistency, we define a modular DAC system with a 6000 ton-\cd/year capacity as the baseline, based on simulation results, and use it as the benchmark for further analysis. This modular design requires approximately 0.43 km\textsuperscript{2} of land for solar PV and CST equipment, based on typical U.S. land-use requirements\cite{ong_land-use_2013}, so the entire facility should occupy no more than 1 km\textsuperscript{2} of land.

\begin{figure}[h!]
    \centering
        \begin{subfigure}[b]{0.49\textwidth}
            \centering
            \includegraphics[width=\textwidth]{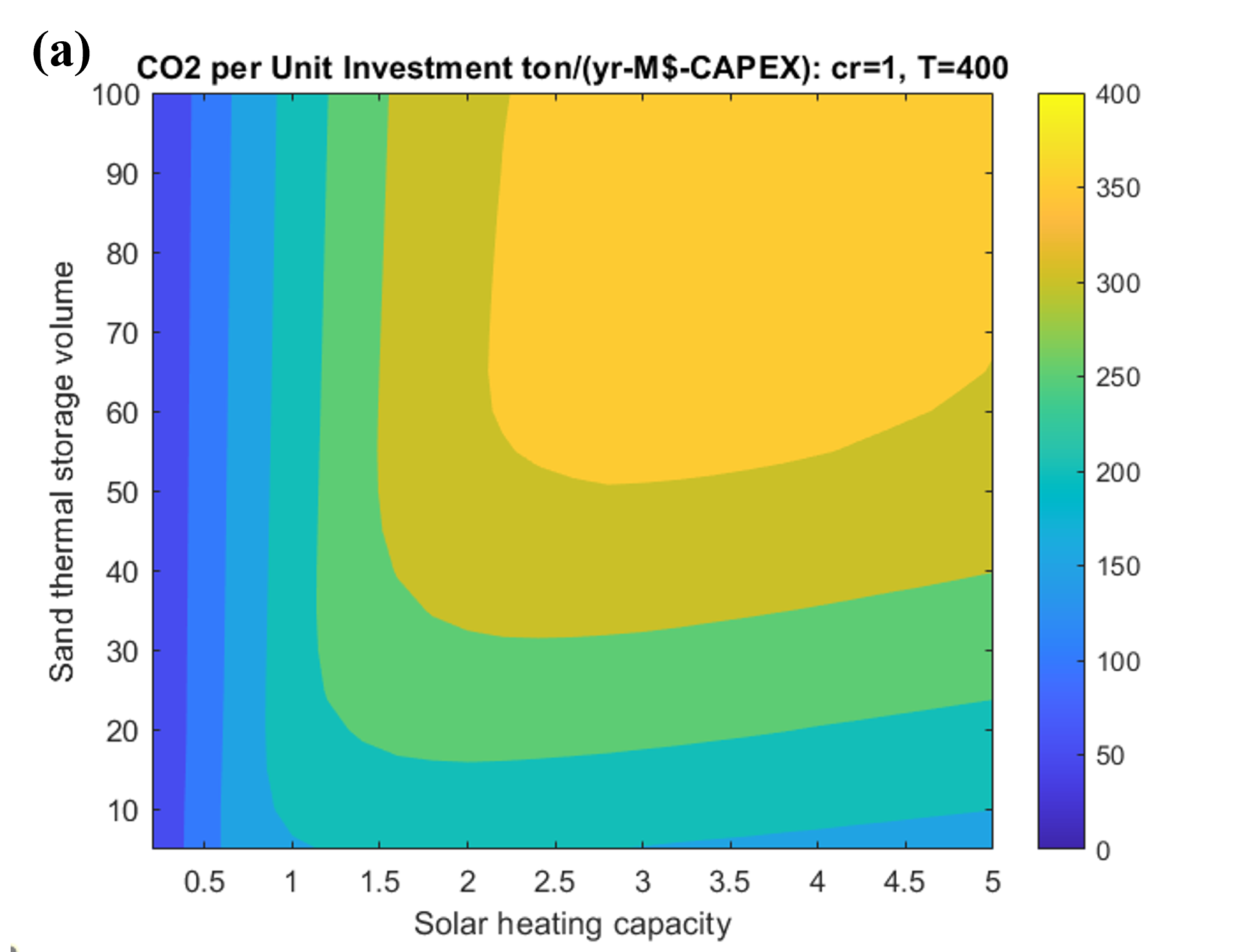}
        \end{subfigure}
        \hfill
        \begin{subfigure}[b]{0.49\textwidth}  
            \centering 
            \includegraphics[width=\textwidth]{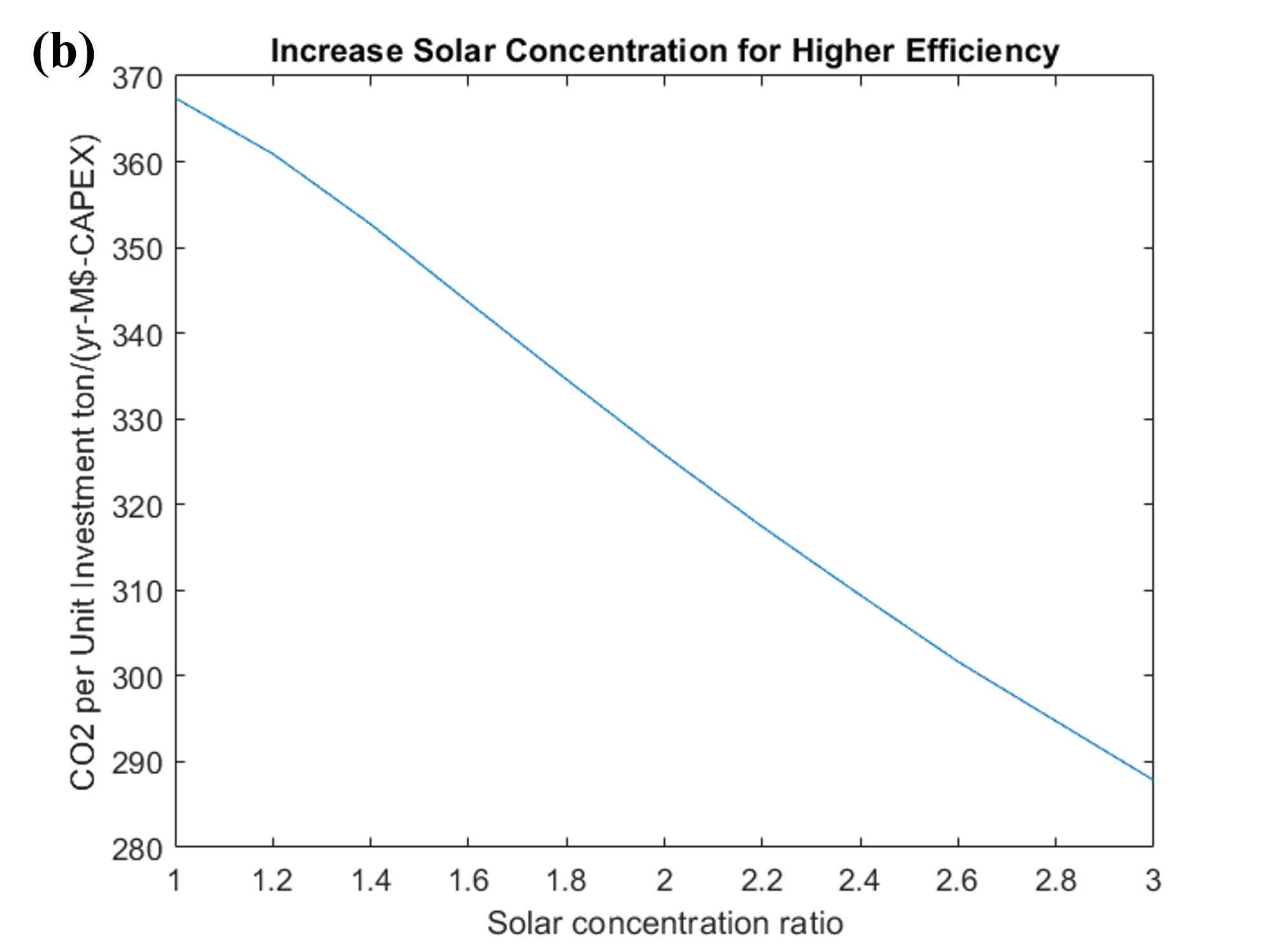} 
        \end{subfigure}
        \begin{subfigure}[b]{0.49\textwidth}  
            \centering 
            \includegraphics[width=\textwidth]{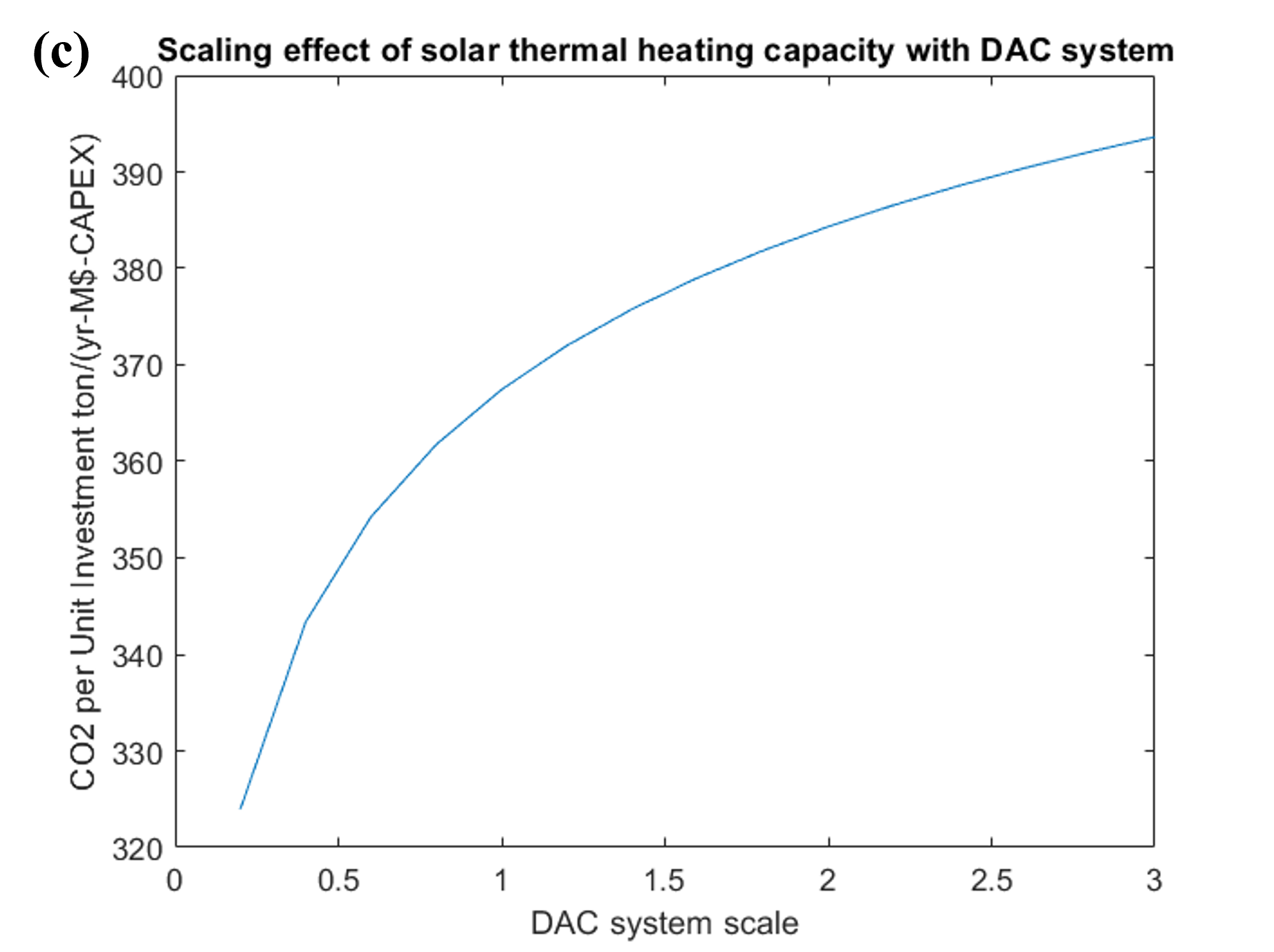} 
        \end{subfigure}
        \begin{subfigure}[b]{0.49\textwidth}  
            \centering 
            \includegraphics[width=\textwidth]{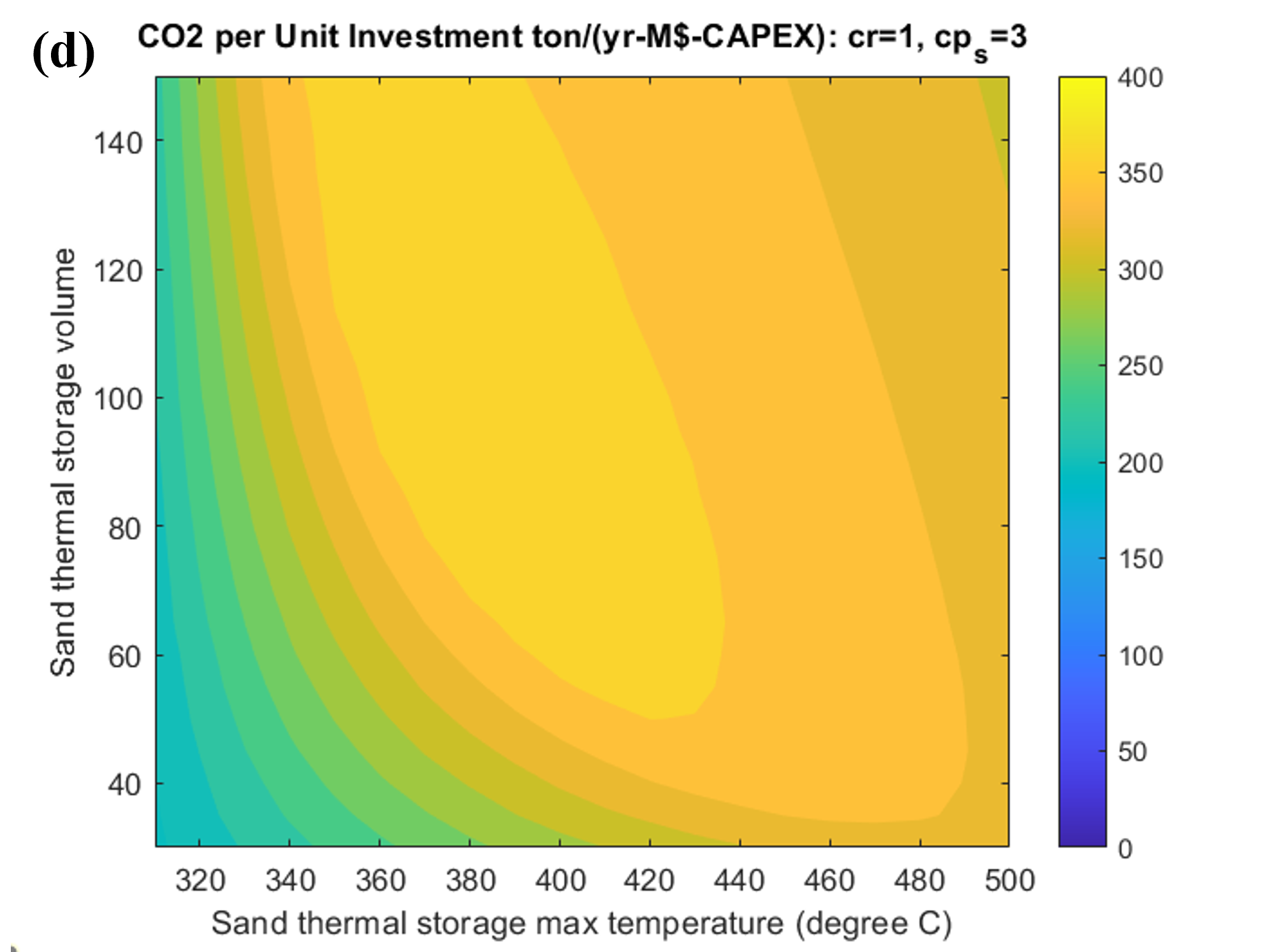} 
        \end{subfigure}
    \caption{\textbf{solar-thermal DAC design parameter optimization using TX power market connection as example.} (a) optimal CST heating capacity vs sand storage capacity; (b) impact of solar concentration ratio for net-\co abatement per unit CAPEX; (c) scaling effect of DAC and CST CAPEX, unit 1 = 6000 ton-\cd/year capacity; (d) optimal sand storage max temperature vs sand storage capacity, higher max temperature meaning higher usable heat for the same storage capacity, but lower solar heating efficiency for CST. Incentive selling price = \$200/ton-\cd, which is sufficiently high to support nearly 100\% operational capacity factor. Eventual optimal design parameters for 6000 ton-\cd/year modular solar-thermal DAC: CST heating capacity = 3 MWh/5min, sand storage capacity = 70 MWh/100\textdegree C operation range, sand storage max temperature = 400 \textdegree C, no additional solar concentration. When capacity factor of solar-thermal DAC is sufficiently high (\(>\)80\%), this optimal design settings are very robust across different power markets and stand-alone solar PV powered systems.}
    \label{tab:fig2}
\end{figure}


We consider western Texas as the representative deployment site for its abundant solar energy and the extensive sedimentary basin for \co storage, where the largest commercial DAC project is currently being developed~\cite{theguardian2023carbon}. The optimal levelized cost of \co (LC\cd) achieves \$220/ton with optimized system design and operation, considering the wholesale electricity price volatility of the Texas grid. The levelized cost decomposition is as follows: DAC system CAPEX 43\%, sorbent materials cost and O\&M 18\%, electricity 16\%, thermal energy from solar 23\%.  


However, grid-connection scenarios also present risks. Notably, although electricity accounts for just 19\% of total energy consumption (with thermal energy making up the remaining 81\%) in the representative DAC technology, its cost is nearly equivalent to that of thermal energy. This is especially concerning given the steady rise in average electricity prices across the U.S. over the past decade and increased volatility driven by higher renewable penetration~\cite{us_eia_electric_2025}. Moreover, consumers also incur additional fixed charges and peak demand charges that may take around 30\% of the electricity cost~\cite{us_doe_evaluating_2025}. In addition, the interconnection queueing process has emerged as a major bottleneck amid growing electricity demand from electrification and data center expansion. As of 2023, the median interconnection wait time reached five years ~\cite{rand_queued_2024}, with growing uncertainty as the grid undergoes a rapid transition.. 

\section{Stand-alone Solar-DAC System and Global Impact}\label{sec4}

\begin{figure}[h!]
    \centering
        \begin{subfigure}[b]{0.99\textwidth}  
            \centering 
            \includegraphics[width=\textwidth]{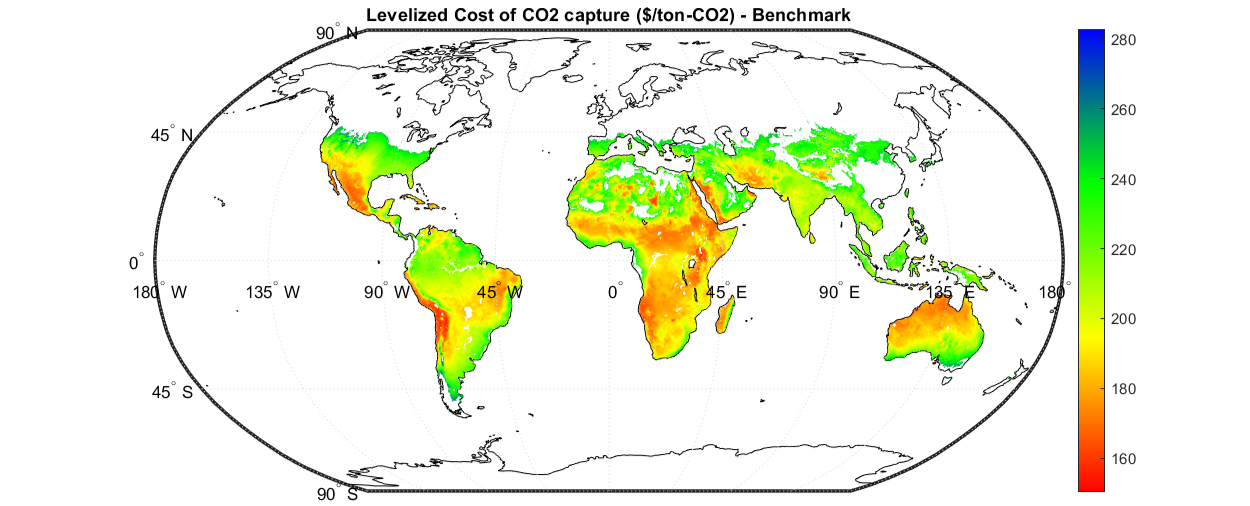} 
        \end{subfigure}
        \begin{subfigure}[b]{0.99\textwidth}  
            \centering 
            \includegraphics[width=\textwidth]{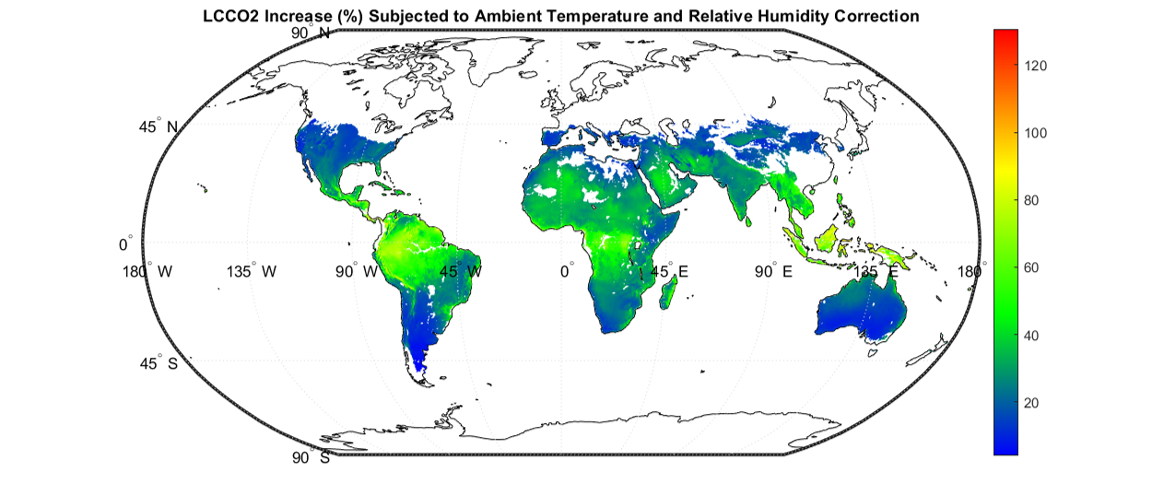} 
        \end{subfigure}
    \caption{\textbf{Stand-alone Solar-DAC System (power+thermal) global deployment cost analysis and potential ambient sensitivity using presentative sorbent technology.} (upper) global mapping of solar stand-alone LC\cd. The analysis focuses on regions with solar capacity factor \(>\)15\% with 107713 points globally, some places without calibrated data, such as the Sahara desert, are excluded. (lower) ambient sensitivity to local temperature and relative humidity of a representative sorbent material. The sample sorbent is particularly sensitive to relative humidity, which significantly increases costs in humid tropical regions. In contrast, sandy terrains with lower humidity levels offer more favorable operating conditions for this DAC technology.}
    \label{tab:fig5}
\end{figure}

We analyze a stand-alone solar-DAC system that relies solely on on-site photovoltaics (PV) and battery storage for electricity, eliminating the need for grid connections. This setup improves flexibility for rural deployment and avoids long interconnection delays.
\textcolor{blue}{Figure}~\ref{tab:fig5} shows the global LC\co results, where a similar design optimization framework to the grid-connected case study has been applied to achieve the minimal LC\cd.
A pre-optimized 6000 ton-\cd/year capacity modular stand-alone solar-DAC system design is used for deployment analysis, where the DAC power source is solar PV and battery energy storage. The levelized cost of \co (LC\cd) in ideal regions ranges from \$160-\$200/ton-\cd. In western Texas and California specifically, stand-alone solar-DRC systems are \$10-\$20 cheaper than grid-connection using wholesale electricity price in 2022, before factoring in grid interconnection charges.


Stand-alone solar-DAC systems can have out-sized impacts of \(<\)\$180/ton-\co cost in most regions with good solar radiation, especially (1) Gulf of California coastal to West Texas region; (2) Atacama Desert Solar Corridor in North Chile and Peru; (3) Red Sea coastal to Persian Gulf regions; (4) Sub-Saharan Africa; (5) North Australia.  Except for parts of Sub-Saharan Africa, these regions are largely covered in sandy terrain and are already recognized as emerging solar energy hubs. This aligns well with the proposed sand-based thermal energy storage, while the arid, low-density landscapes help reduce land and infrastructure costs.  Moreover, the solid sorbent technology does not require process water and potentially capture water vapors during operation \cite{drechsler_investigation_2020,rosa_water_2021}, making it suitable for deployment with recycled or alternative cooling methods and minimal water demand.

Based on 1 km\textsuperscript{2} land-use for the modular system, 1 Gt/year abatement capacity takes about 208,000 km\textsuperscript{2}. This is 30\% of total area of Texas or 15\% of the Great Australian Desert. The stand-alone solar-DAC systems deployed in sandy terrain alone could exceed global total potential capacity of 26.9 Gt/year, mostly in (1) Arabian Desert, 11.2 Gt/year; (2) Great Australian Desert, 6.6 Gt/year; (2) Kalahari Desert in Southern Africa, 4.3 Gt/year.

A research gap in solar-DAC system deployment globally is the atmospheric sensitivity to ambient temperature and relative humidity, which significantly affects energy consumption and \co capture capacity\cite{an_impact_2022, sendi_geospatial_2022}. For example, our representative MOF sorbent and other solid sorbents (including amine-functionalized sorbents) selectively bind \co but also capture water vapor at high humidity, reducing its efficiency. In contrast, liquid solvents using calcium hydroxide Ca(OH)\textsubscript{2} benefit from elevated humidity and temperature to enhance \co adsorption. Our global analysis of the potential increase in LC\co globally for the representative sorbent technology \textcolor{blue}{Figure~\ref{tab:fig5}(b)} show that the ambient sensitivity corrections is minimal in most sandy terrains identified as suitable for stand-alone solar-DAC system (\(<\)20\%). We demonstrate that robust DAC sorbent technology is already available for optimal solar-DAC deployment in ideal landscapes. However, in environments that deviate significantly from laboratory conditions, such as near-equatorial and rainforest regions, the worst-case scenario could double the costs, highlighting the need for further research into more resilient sorbents.

\section{Comparative Cost Analysis with Geothermal}

\begin{figure}[h!]
    \centering
    \includegraphics[width=0.99\linewidth]{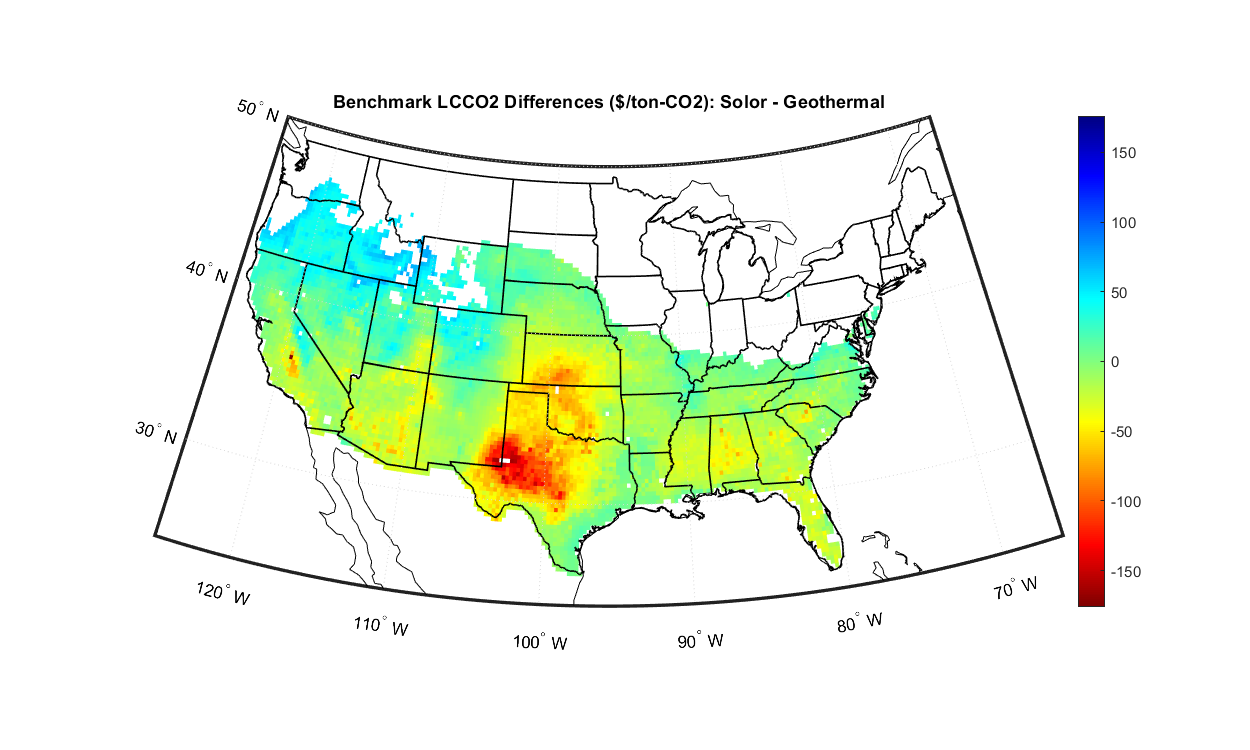}
    \caption{\textbf{Stand-alone Solar-DAC and Geothermal-DAC system comparison in the continental U.S. region.} Comparing the LC\co of stand-alone solar-DAC and stand-alone geothermal-DAC by mapping their differences (solar subtracting geothermal). Negative value shown in red color indicating in favor of solar deployment in the West Texas regions.}
    \label{tab:fig_geo}
\end{figure}

We conduct a comparative analysis between stand-alone DAC systems using solar and geothermal energy to highlight the cost-effectiveness of the solar approach. Geothermal energy has been the exclusive choice for low-carbon heat in existing commercial DAC projects. Using geothermal LCOE data \cite{aljubran_power_2024} from the United States, we calculate the LC\co for a stand-alone DAC system employing the same sorbent technology, under the assumption that all energy consumption is met by geothermal. \textcolor{blue}{Figure~\ref{tab:fig_geo}} illustrates the differences in LC\co between the two systems, where more negative values indicate a cost advantage for the solar-DAC system.

DAC systems using geothermal energy show comparable LC\co with solar energy in most regions, typically around \$160-\$210/ton-\cd. For instance, in the southwestern U.S., geothermal reservoirs exceeding  300\textdegree C at 5-7 km depth can achieve LCOE ranges between \$30-\$70/MWh~\cite{aljubran_power_2024}. However, regional differences are significant: solar energy is more cost-effective in West Texas due to high solar potential and costly geothermal extraction, while the Pacific Northwest favors geothermal energy due to lower solar availability. The cause of high geothermal LCOE in West Texas is sedimentary Permian basin geology with low geothermal temperature gradients and limited permeability, making high-temperature extraction difficult and expensive~\cite{ogland-hand_nationwide_2024}. Sedimentary basins, suitable for geological \co storage, can only harvest geothermal energy in relatively low temperatures (100 to 200\textdegree C) \cite{bielicki_promise_2023}. Even in regions with both geothermal potential and basaltic \co storage options, drilling must be spaced carefully to avoid conflicts \cite{oelkers_moving_2023}.

Our comparative case study in the United States reveals a potential conflict between geothermal energy costs and geological \co storage potential. For cost-effective DAC systems, proximity to verified geological storage is essential. Yet, many identified solar-rich regions, such as the Atacama Desert Solar Corridor, Red Sea coasts, and Sub-Saharan Africa, lack positively confirmed \co storage capacity. In contrast, regions like West Texas, South Africa, and the Xinjiang Tarim Basin offer both verified sequestration \cite{wei_proposed_2021,mcqueen_analysis_2021} and competitive costs (around \$200/ton-\cd) for solar-DAC system, presenting near-term opportunities. In these regions, the requirements for efficient geothermal reservoirs, including high-temperature gradient and permeable formations that facilitate water circulation,  cannot be met as they conflict with trapping \co in the supercritical phase under the cap-rock sealing layers. This potential conflict positions stand-alone solar-DAC systems as a promising alternative in sedimentary basins, with sites like the Permian Basin (USA), the Great Artesian Basin (Australia), and the Neuquén Basin (Argentina) combining abundant solar potential, ample geological storage, and favorable terrain without competing with costly geothermal drilling.

\section{Discussion}\label{sec6}

This paper simulated and optimized concentrated solar thermal (CST) coupled with sand thermal energy storage design to supply low-carbon heat for DAC systems. The tested DAC system uses solid-sorbent materials with a regeneration temperature of 100\textdegree C and approximately 1 hour per cycle of adsorption-desorption operation. Both grid-connection scenarios and stand-alone scenarios show a levelized cost of \co (LC\cd) around \$200/ton-\co with solar capacity factor \(>\)20\%, competitive with existing techno-economic analysis reported in the literature. It demonstrates that an optimized solar-thermal DAC system can operate at \(>\)80\% capacity factor annually and almost constantly through solar-abundant seasons, making solar thermal energy a feasible and ideal source of low-carbon heat for global large-scale DAC deployment.


Our analysis shows that the stand-alone solar-DAC system demonstrates cost-effectiveness comparable to grid-connected systems while offering several key advantages, including deployment flexibility without interconnection requirements, and insulation from electricity price risks amid anticipated demand surges driven by global electrification and data center expansion.
The stand-alone solar-DAC system has the potential to make a significant global impact, with numerous suitable deployment areas. These regions are often characterized by sandy terrains in arid zones, where land and sand-based energy storage costs are minimal. The representative sorbent's sensitivity to ambient temperature and relative humidity is well-suited for such environments. However, deploying the system in lower latitudes, such as near the equator or rainforest areas, would require different types of sorbents with enhanced tolerance to higher temperatures and humidity, which can be an important future research topic. 

The solar-DAC system avoids potential geothermal compatibility issues and drilling activity conflicts while addressing local \co geological storage limitations. They are particularly well suited for sedimentary basins with sandy terrains, as exemplified by West Texas in the U.S., which is currently being developed as a carbon capture and storage hub. In these areas, solar energy not only provides a competitive option for powering DAC systems but is likely the optimal choice. Moreover, the use of mature solar generation technologies like CST and PV reduces overall project risk, particularly given the emerging state of carbon sequestration markets, as these assets can be repurposed for other business applications if necessary.

The solar-thermal DAC system shifts the thermal energy and electricity cost from continuous operational expenditure (OPEX) to capital expenditure (CAPEX). The majority of the remaining OPEX is associated with sorbent material costs. This transition makes the DAC system more capital-intensive and, consequently, more sensitive to financial assumptions and policy changes, particularly the incentive per ton of \co removal. To ensure the success of solar-thermal DAC systems, low-cost green financing, and long-term stable policy support will be more critical than achieving purely engineering optimality.


\section{Methods}\label{sec7}

\subsection{Overview}

Our analysis is generally categorized into three parts. First, system conceptualization and simulation using MATLAB Simulink + Simscape. Second, DAC + solar thermal system design parameter and grid-interactive operation optimization. Third, using the pre-determined system design for global stand-alone solar-DAC system performance simulation, including sensitivity to ambient temperature and relative humidity. All simulation incorporate one year data (2022 sample year data with 5-min resolution for grid-connection study, 1-hour resolution for stand-alone global study) to support analysis for both daily and season patterns. For sustainability comparison, ``\co capture efficiency" is defined as below. For example, 80\% \co capture efficiency means 0.2 ton of \co is emitted to capture 1 ton of \co through DAC system.

\begin{equation}
    \eta_{CO_2} = \frac{CO_2^{net}}{CO_2^{captured}} = 1-\frac{CO_2^{emit}}{CO_2^{captured}}
\end{equation}

The levelized cost of \co (LC\cd) evaluated in this study incorporates both capital expenditures (CAPEX) and operational expenditures (OPEX). The CAPEX component encompasses equipment costs associated with initial investments, specifically including direct air capture (DAC) systems, solar-thermal concentrating systems (CST) paired with thermal energy storage, and photovoltaic (PV) solar installations integrated with battery energy storage. The OPEX component comprises energy expenditures and sorbent material costs (e.g., sorbent degradation), inclusive of operation and maintenance (O\&M) expenses. For stand-alone system analyses, energy-related OPEX is eliminated, with all energy infrastructure costs consolidated into the CAPEX framework. Costs associated with land acquisition, labor, and auxiliary infrastructure are excluded from the scope of this assessment.

It is acknowledged that this methodological approach aligns the LC\co boundary conditions with standard techno-economic analysis (TEA) conventions, facilitating direct comparison with prior literature. However, the exclusion of ancillary real-world expenditures—such as infrastructure, permitting, labor, and site preparation—may result in conservative cost estimates relative to full-scale commercial implementations. This simplification underscores the analytical focus on technology-centric cost drivers while recognizing potential deviations from comprehensive industrial cost structures.

\subsection{Simulation and Optimization}

To conceptualize the design of a solar-thermal energy system using Concentrated Solar Thermal (CST) technology, we first understand how solar radiation is converted into usable heat. Based on existing research, the efficiency of this conversion depends on three key factors \cite{eck_modelling_2005,jiang_implementation_2014}: (1) Sunlight intensity: measured by the solar radiation available at given time; (2) Target temperature: How hot the system needs to get that guarantee other subsystem performance, such as thermal energy storage and heat transfer efficiency; (3) Solar concentration ratio: How much solar radiation is focused onto the system (area of mirrors or lenses). These factors affect the eventual solar thermal efficiency: higher target temperatures reduce efficiency and more solar radiation or a higher concentration ratio improves efficiency. We combined existing research data with basic physics principles to create a semi-physics inspired function that links efficiency to these three factors, where solar radiation is a known input provided by solar time series data. Target temperature and concentration ratio are design parameters to be optimized during the system’s optimization to balance performance and cost. The solar-thermal collector efficiency is given by the function in form of:

\begin{equation}
\eta_{c_t}(DNI_t, cr, T) = 0.78 - \eta_{loss} = 0.78 - \alpha T^2 \cdot \frac{\beta}{DNI_t + m} \cdot \frac{\gamma}{cr}
\end{equation}

where the \(DNI_t\) is direct normal irradiance data for solar time series, \(cr\) is the solar concentration ratio and \(T\) is the target temperature given in \textdegree C. The efficiency \(\eta_{c_t}\) equals a baseline highest efficiency (0.78) minus losses \(\eta_{loss}\). Other parameters are obtained by fitting this equation to experimental data reported in literature.

Given the efficiency, the total amount of solar-thermal flux is given by:

\begin{equation}
s_t = DNI_t * cp * \eta_{c_t}
\end{equation}
where \(cp\) is the solar-thermal capacity, which measures the sizing of CST equipment.

A simulation model using MATLAB Simulink (signal package) + Simscape (thermdynamics and fluid mechanics package) was developed to simulate the thermodynamics behaviors that are sensitive to target temperature settings. The simulation model consists of several subsystems: (1) solar thermal source and storage subsystem; (2) DAC thermal subsystem; (3) water cooling subsystem; (4) sensor and control subsystem. The simulation model is parameterized and connected to the DAC operation optimization code. It receives signal as simulation input such as design parameters (e.g., target temperature) and control signals (e.g., when to initiate regeneration cycle which transfer heat from thermal storage to DAC thermal system). The DAC operation optimization framework not only streamlines high-fidelity thermodynamic simulations but also incorporates strategic operational scheduling to dynamically respond to volatile electricity prices, thereby aligning DAC energy consumption with cost-minimizing grid interactions. The optimization objective function maximizes the total profit of the DAC operation.
\begin{equation}
    max \sum_t \; \pi d_t - \lambda_t C_{t} (P^a u_t +  P^d v_t)-Sz_t
\end{equation}
where  \(\pi d_t\) is the total revenue by captured/desorbed \co \(d_t\) multiplied by incentive \(\pi\), minus the cost of power consumption \(\lambda_t(P^a u_t + P^d v_t)\), real-time price \(\lambda_t\) times power consumption for adsorption and desorption \(P^a \& P^d\) times binary activation status \(u_t \& v_t\), minus the cycle switching cost \(Sz_t\). CAPEX component is a fixed cost which is a constant for optimization for any given horizon, therefore does not affect the operational variables. The final LC\co will be corrected by adding the CAPEX. Similar to power system optimization, this operational optimization does NOT guarantee recovery of all costs for DAC system.

While the simulation model employs sub-second temporal resolution for thermodynamic precision, the optimization framework adopts a 5-minute sampling interval—synchronized with wholesale electricity price fluctuations—to enable real-time decision-making. This dual approach ensures computational tractability (7–10 seconds for a full-year optimization, a 300–500× speed improvement over conventional 45–60 minute thermodynamic simulations) while preserving \(>\)99\% solution quality (\(<\)1\% operational cycles deemed infeasible in simulation validation). Crucially, the model advances beyond simplification by integrating grid connectivity constraints and price-driven operational strategies, ensuring DAC systems function as responsive, grid-aware assets rather than isolated thermodynamic processes. More details about the simulation settings and full optimization problem formulation set up see \textcolor{blue}{Supplementary information}.

\subsection{Stand-alone and Global Analysis}

System Configuration and Global Analysis Methodology.
For the stand-alone solar-DAC system evaluated in this global analysis, system parameters—including the sizing of solar photovoltaic (PV) arrays, battery energy storage systems, solar thermal collectors, and sand-based thermal storage—were first optimized using a representative location (Texas, USA). These parameters were then applied uniformly across all global locations. While this approach ensures methodological consistency, the authors recognize that it may yield sub-optimal configurations, as site-specific solar profiles were not incorporated into individual optimizations. Consequently, the reported levelized cost of \co (LC\cd) represents an upper bound, with potential for reduction through location-specific parameter tuning.

Impact of Ambient Environmental Conditions.
The global analysis evaluates the influence of ambient temperature and relative humidity on two critical performance metrics of DAC systems: (1) energy efficiency and (2) \co capture rate. Hourly gridded global ground-level temperature and relative humidity data were used to calculate location-specific correction factors. These factors normalize energy efficiency and \co capture rates relative to standard laboratory conditions (20°C, 50\% RH). Deviations from these optimal conditions typically reduce both metrics, though regional variability exists.
The electricity consumption correction after numerical fitting is given by the following, using quadratic equation for temperature \(T\) fitting and exponential function for skewed symmetric behavior of relative humidity \(RH\):

\begin{equation}
    C_t = [1.9 + 0.01(T-20)](RH-0.4)^2 e^{RH-0.4}+[1.5+0.003(T-20)^2]
\end{equation}

The capture rate or abatement correction similarly:

\begin{equation}
    \eta_t = 65 - 0.01 T^2 - (T+20)(RH-0.4)^2
\end{equation}

Cost Implications and Methodological Constraints.
To quantify the LC\co impact of ambient conditions, operational optimization was re-executed using adjusted hourly energy consumption and \co capture rates derived from environmental data. However, this secondary optimization retained the original system sizing parameters, potentially compounding sub-optimality. As a result, the estimated cost penalty from ambient conditions may be conservatively high. Future work could mitigate this limitation through integrated design optimization that simultaneously accounts for local climate variability and component sizing.

\section{Data and Code Availability}

All data, code, and MATLAB Simscape Simulation models used in this study are available and can be directly accessed in the following \href{https://github.com/ZhiyuanF/Solar-DAC.git}{Github Repository}.

\section{Declaration of Interests}\label{sec8}

The authors declare no competing interests.

\section{Acknowledgments}\label{sec9}

To be finalized upon final decision.

\section{Declaration of generative AI and AI-assisted technologies in the writing process}

During the preparation of this work the author(s) used ChatGPT 4.0 only in order to improve readability. After using this tool/service, the author(s) reviewed and edited the content as needed and take(s) full responsibility for the content of the publication.


\bibliography{DAC_main}


\begin{thebibliography}{46}
\ifx \bisbn   \undefined \def \bisbn  #1{ISBN #1}\fi
\ifx \binits  \undefined \def \binits#1{#1}\fi
\ifx \bauthor  \undefined \def \bauthor#1{#1}\fi
\ifx \batitle  \undefined \def \batitle#1{#1}\fi
\ifx \bjtitle  \undefined \def \bjtitle#1{#1}\fi
\ifx \bvolume  \undefined \def \bvolume#1{\textbf{#1}}\fi
\ifx \byear  \undefined \def \byear#1{#1}\fi
\ifx \bissue  \undefined \def \bissue#1{#1}\fi
\ifx \bfpage  \undefined \def \bfpage#1{#1}\fi
\ifx \blpage  \undefined \def \blpage #1{#1}\fi
\ifx \burl  \undefined \def \burl#1{\textsf{#1}}\fi
\ifx \doiurl  \undefined \def \doiurl#1{\url{https://doi.org/#1}}\fi
\ifx \betal  \undefined \def \betal{\textit{et al.}}\fi
\ifx \binstitute  \undefined \def \binstitute#1{#1}\fi
\ifx \binstitutionaled  \undefined \def \binstitutionaled#1{#1}\fi
\ifx \bctitle  \undefined \def \bctitle#1{#1}\fi
\ifx \beditor  \undefined \def \beditor#1{#1}\fi
\ifx \bpublisher  \undefined \def \bpublisher#1{#1}\fi
\ifx \bbtitle  \undefined \def \bbtitle#1{#1}\fi
\ifx \bedition  \undefined \def \bedition#1{#1}\fi
\ifx \bseriesno  \undefined \def \bseriesno#1{#1}\fi
\ifx \blocation  \undefined \def \blocation#1{#1}\fi
\ifx \bsertitle  \undefined \def \bsertitle#1{#1}\fi
\ifx \bsnm \undefined \def \bsnm#1{#1}\fi
\ifx \bsuffix \undefined \def \bsuffix#1{#1}\fi
\ifx \bparticle \undefined \def \bparticle#1{#1}\fi
\ifx \barticle \undefined \def \barticle#1{#1}\fi
\bibcommenthead
\ifx \bconfdate \undefined \def \bconfdate #1{#1}\fi
\ifx \botherref \undefined \def \botherref #1{#1}\fi
\ifx \url \undefined \def \url#1{\textsf{#1}}\fi
\ifx \bchapter \undefined \def \bchapter#1{#1}\fi
\ifx \bbook \undefined \def \bbook#1{#1}\fi
\ifx \bcomment \undefined \def \bcomment#1{#1}\fi
\ifx \oauthor \undefined \def \oauthor#1{#1}\fi
\ifx \citeauthoryear \undefined \def \citeauthoryear#1{#1}\fi
\ifx \endbibitem  \undefined \def \endbibitem {}\fi
\ifx \bconflocation  \undefined \def \bconflocation#1{#1}\fi
\ifx \arxivurl  \undefined \def \arxivurl#1{\textsf{#1}}\fi
\csname PreBibitemsHook\endcsname

\bibitem[\protect\citeauthoryear{Service}{2024}]{copernicus_climate_change_service_copernicus_2024}
\begin{botherref}
\oauthor{\bsnm{Service}, \binits{C.C.C.}}:
Copernicus: 2024 is the first year to exceed 1.5°{C} above pre-industrial level
(2024)
\end{botherref}
\endbibitem

\bibitem[\protect\citeauthoryear{O’Brien et~al.}{2024}]{o2024co2}
\begin{barticle}
\bauthor{\bsnm{O’Brien}, \binits{C.P.}},
\bauthor{\bsnm{Miao}, \binits{R.K.}},
\bauthor{\bsnm{Shayesteh~Zeraati}, \binits{A.}},
\bauthor{\bsnm{Lee}, \binits{G.}},
\bauthor{\bsnm{Sargent}, \binits{E.H.}},
\bauthor{\bsnm{Sinton}, \binits{D.}}:
\batitle{Co2 electrolyzers}.
\bjtitle{Chemical Reviews}
\bvolume{124}(\bissue{7}),
\bfpage{3648}--\blpage{3693}
(\byear{2024})
\end{barticle}
\endbibitem

\bibitem[\protect\citeauthoryear{McQueen et~al.}{2021}]{mcqueen_review_2021}
\begin{barticle}
\bauthor{\bsnm{McQueen}, \binits{N.}},
\bauthor{\bsnm{Gomes}, \binits{K.V.}},
\bauthor{\bsnm{McCormick}, \binits{C.}},
\bauthor{\bsnm{Blumanthal}, \binits{K.}},
\bauthor{\bsnm{Pisciotta}, \binits{M.}},
\bauthor{\bsnm{Wilcox}, \binits{J.}}:
\batitle{A review of direct air capture ({DAC}): scaling up commercial technologies and innovating for the future}.
\bjtitle{Progress in Energy}
\bvolume{3}(\bissue{3}),
\bfpage{032001}
(\byear{2021})
\doiurl{10.1088/2516-1083/abf1ce} .
Accessed 2024-04-08
\end{barticle}
\endbibitem

\bibitem[\protect\citeauthoryear{Izikowitz}{2021}]{izikowitz_carbon_2021}
\begin{barticle}
\bauthor{\bsnm{Izikowitz}, \binits{D.}}:
\batitle{Carbon {Purchase} {Agreements}, {Dactories}, and {Supply}-{Chain} {Innovation}: {What} {Will} {It} {Take} to {Scale}-{Up} {Modular} {Direct} {Air} {Capture} {Technology} to a {Gigatonne} {Scale}}.
\bjtitle{Frontiers in Climate}
\bvolume{3},
\bfpage{636657}
(\byear{2021})
\doiurl{10.3389/fclim.2021.636657} .
Accessed 2025-02-17
\end{barticle}
\endbibitem

\bibitem[\protect\citeauthoryear{Alcalde et~al.}{2018}]{alcalde_estimating_2018}
\begin{barticle}
\bauthor{\bsnm{Alcalde}, \binits{J.}},
\bauthor{\bsnm{Flude}, \binits{S.}},
\bauthor{\bsnm{Wilkinson}, \binits{M.}},
\bauthor{\bsnm{Johnson}, \binits{G.}},
\bauthor{\bsnm{Edlmann}, \binits{K.}},
\bauthor{\bsnm{Bond}, \binits{C.E.}},
\bauthor{\bsnm{Scott}, \binits{V.}},
\bauthor{\bsnm{Gilfillan}, \binits{S.M.V.}},
\bauthor{\bsnm{Ogaya}, \binits{X.}},
\bauthor{\bsnm{Haszeldine}, \binits{R.S.}}:
\batitle{Estimating geological {CO2} storage security to deliver on climate mitigation}.
\bjtitle{Nature Communications}
\bvolume{9}(\bissue{1}),
\bfpage{2201}
(\byear{2018})
\doiurl{10.1038/s41467-018-04423-1} .
Accessed 2023-11-19
\end{barticle}
\endbibitem

\bibitem[\protect\citeauthoryear{Liu et~al.}{2021}]{liu_new_2021}
\begin{barticle}
\bauthor{\bsnm{Liu}, \binits{B.}},
\bauthor{\bsnm{Qin}, \binits{J.}},
\bauthor{\bsnm{Shi}, \binits{J.}},
\bauthor{\bsnm{Jiang}, \binits{J.}},
\bauthor{\bsnm{Wu}, \binits{X.}},
\bauthor{\bsnm{He}, \binits{Z.}}:
\batitle{New perspectives on utilization of {CO} 2 sequestration technologies in cement-based materials}.
\bjtitle{Construction and Building Materials}
\bvolume{272},
\bfpage{121660}
(\byear{2021})
\doiurl{10.1016/j.conbuildmat.2020.121660} .
Accessed 2023-11-19
\end{barticle}
\endbibitem

\bibitem[\protect\citeauthoryear{Parigi et~al.}{2019}]{parigi_power--fuels_2019}
\begin{barticle}
\bauthor{\bsnm{Parigi}, \binits{D.}},
\bauthor{\bsnm{Giglio}, \binits{E.}},
\bauthor{\bsnm{Soto}, \binits{A.}},
\bauthor{\bsnm{Santarelli}, \binits{M.}}:
\batitle{Power-to-fuels through carbon dioxide {Re}-{Utilization} and high-temperature electrolysis: {A} technical and economical comparison between synthetic methanol and methane}.
\bjtitle{Journal of Cleaner Production}
\bvolume{226},
\bfpage{679}--\blpage{691}
(\byear{2019})
\doiurl{10.1016/j.jclepro.2019.04.087} .
Accessed 2023-11-19
\end{barticle}
\endbibitem

\bibitem[\protect\citeauthoryear{Sabatino et~al.}{2021}]{sabatino_comparative_2021}
\begin{barticle}
\bauthor{\bsnm{Sabatino}, \binits{F.}},
\bauthor{\bsnm{Grimm}, \binits{A.}},
\bauthor{\bsnm{Gallucci}, \binits{F.}},
\bauthor{\bsnm{Van Sint~Annaland}, \binits{M.}},
\bauthor{\bsnm{Kramer}, \binits{G.J.}},
\bauthor{\bsnm{Gazzani}, \binits{M.}}:
\batitle{A comparative energy and costs assessment and optimization for direct air capture technologies}.
\bjtitle{Joule}
\bvolume{5}(\bissue{8}),
\bfpage{2047}--\blpage{2076}
(\byear{2021})
\doiurl{10.1016/j.joule.2021.05.023} .
Accessed 2023-11-19
\end{barticle}
\endbibitem

\bibitem[\protect\citeauthoryear{Fasihi et~al.}{2019}]{fasihi_techno-economic_2019}
\begin{barticle}
\bauthor{\bsnm{Fasihi}, \binits{M.}},
\bauthor{\bsnm{Efimova}, \binits{O.}},
\bauthor{\bsnm{Breyer}, \binits{C.}}:
\batitle{Techno-economic assessment of {CO2} direct air capture plants}.
\bjtitle{Journal of Cleaner Production}
\bvolume{224},
\bfpage{957}--\blpage{980}
(\byear{2019})
\doiurl{10.1016/j.jclepro.2019.03.086} .
Accessed 2023-10-18
\end{barticle}
\endbibitem

\bibitem[\protect\citeauthoryear{Realmonte et~al.}{2019}]{realmonte_inter-model_2019}
\begin{barticle}
\bauthor{\bsnm{Realmonte}, \binits{G.}},
\bauthor{\bsnm{Drouet}, \binits{L.}},
\bauthor{\bsnm{Gambhir}, \binits{A.}},
\bauthor{\bsnm{Glynn}, \binits{J.}},
\bauthor{\bsnm{Hawkes}, \binits{A.}},
\bauthor{\bsnm{Köberle}, \binits{A.C.}},
\bauthor{\bsnm{Tavoni}, \binits{M.}}:
\batitle{An inter-model assessment of the role of direct air capture in deep mitigation pathways}.
\bjtitle{Nature Communications}
\bvolume{10}(\bissue{1}),
\bfpage{3277}
(\byear{2019})
\doiurl{10.1038/s41467-019-10842-5} .
Accessed 2023-11-19
\end{barticle}
\endbibitem

\bibitem[\protect\citeauthoryear{Terlouw et~al.}{2021}]{terlouw_life_2021}
\begin{barticle}
\bauthor{\bsnm{Terlouw}, \binits{T.}},
\bauthor{\bsnm{Treyer}, \binits{K.}},
\bauthor{\bsnm{Bauer}, \binits{C.}},
\bauthor{\bsnm{Mazzotti}, \binits{M.}}:
\batitle{Life {Cycle} {Assessment} of {Direct} {Air} {Carbon} {Capture} and {Storage} with {Low}-{Carbon} {Energy} {Sources}}.
\bjtitle{Environmental Science \& Technology}
\bvolume{55}(\bissue{16}),
\bfpage{11397}--\blpage{11411}
(\byear{2021})
\doiurl{10.1021/acs.est.1c03263} .
Accessed 2023-11-21
\end{barticle}
\endbibitem

\bibitem[\protect\citeauthoryear{Deutz and Bardow}{2021}]{deutz_life-cycle_2021}
\begin{barticle}
\bauthor{\bsnm{Deutz}, \binits{S.}},
\bauthor{\bsnm{Bardow}, \binits{A.}}:
\batitle{Life-cycle assessment of an industrial direct air capture process based on temperature–vacuum swing adsorption}.
\bjtitle{Nature Energy}
\bvolume{6}(\bissue{2}),
\bfpage{203}--\blpage{213}
(\byear{2021})
\doiurl{10.1038/s41560-020-00771-9} .
Accessed 2023-07-20
\end{barticle}
\endbibitem

\bibitem[\protect\citeauthoryear{Leonzio et~al.}{2022}]{leonzio_environmental_2022}
\begin{barticle}
\bauthor{\bsnm{Leonzio}, \binits{G.}},
\bauthor{\bsnm{Mwabonje}, \binits{O.}},
\bauthor{\bsnm{Fennell}, \binits{P.S.}},
\bauthor{\bsnm{Shah}, \binits{N.}}:
\batitle{Environmental performance of different sorbents used for direct air capture}.
\bjtitle{Sustainable Production and Consumption}
\bvolume{32},
\bfpage{101}--\blpage{111}
(\byear{2022})
\doiurl{10.1016/j.spc.2022.04.004} .
Accessed 2023-10-31
\end{barticle}
\endbibitem

\bibitem[\protect\citeauthoryear{Azarabadi and Lackner}{2019}]{azarabadi_sorbent-focused_2019}
\begin{barticle}
\bauthor{\bsnm{Azarabadi}, \binits{H.}},
\bauthor{\bsnm{Lackner}, \binits{K.S.}}:
\batitle{A sorbent-focused techno-economic analysis of direct air capture}.
\bjtitle{Applied Energy}
\bvolume{250},
\bfpage{959}--\blpage{975}
(\byear{2019})
\doiurl{10.1016/j.apenergy.2019.04.012} .
Accessed 2023-10-31
\end{barticle}
\endbibitem

\bibitem[\protect\citeauthoryear{Sinha et~al.}{2017}]{sinha_systems_2017}
\begin{barticle}
\bauthor{\bsnm{Sinha}, \binits{A.}},
\bauthor{\bsnm{Darunte}, \binits{L.A.}},
\bauthor{\bsnm{Jones}, \binits{C.W.}},
\bauthor{\bsnm{Realff}, \binits{M.J.}},
\bauthor{\bsnm{Kawajiri}, \binits{Y.}}:
\batitle{Systems {Design} and {Economic} {Analysis} of {Direct} {Air} {Capture} of {CO} $_{\textrm{2}}$ through {Temperature} {Vacuum} {Swing} {Adsorption} {Using} {MIL}-101({Cr})-{PEI}-800 and mmen-{Mg} $_{\textrm{2}}$ (dobpdc) {MOF} {Adsorbents}}.
\bjtitle{Industrial \& Engineering Chemistry Research}
\bvolume{56}(\bissue{3}),
\bfpage{750}--\blpage{764}
(\byear{2017})
\doiurl{10.1021/acs.iecr.6b03887} .
Accessed 2023-05-26
\end{barticle}
\endbibitem

\bibitem[\protect\citeauthoryear{Keith et~al.}{2018}]{keith_process_2018}
\begin{barticle}
\bauthor{\bsnm{Keith}, \binits{D.W.}},
\bauthor{\bsnm{Holmes}, \binits{G.}},
\bauthor{\bsnm{St.~Angelo}, \binits{D.}},
\bauthor{\bsnm{Heidel}, \binits{K.}}:
\batitle{A {Process} for {Capturing} {CO2} from the {Atmosphere}}.
\bjtitle{Joule}
\bvolume{2}(\bissue{8}),
\bfpage{1573}--\blpage{1594}
(\byear{2018})
\doiurl{10.1016/j.joule.2018.05.006} .
Accessed 2024-10-29
\end{barticle}
\endbibitem

\bibitem[\protect\citeauthoryear{McQueen and Drennan}{2024}]{mcqueen_use_2024}
\begin{barticle}
\bauthor{\bsnm{McQueen}, \binits{N.}},
\bauthor{\bsnm{Drennan}, \binits{D.}}:
\batitle{The use of warehouse automation technology for scalable and low-cost direct air capture}.
\bjtitle{Frontiers in Climate}
\bvolume{6},
\bfpage{1415642}
(\byear{2024})
\doiurl{10.3389/fclim.2024.1415642} .
Accessed 2025-02-17
\end{barticle}
\endbibitem

\bibitem[\protect\citeauthoryear{Thiel and Stark}{2021}]{thiel_decarbonize_2021}
\begin{barticle}
\bauthor{\bsnm{Thiel}, \binits{G.P.}},
\bauthor{\bsnm{Stark}, \binits{A.K.}}:
\batitle{To decarbonize industry, we must decarbonize heat}.
\bjtitle{Joule}
\bvolume{5}(\bissue{3}),
\bfpage{531}--\blpage{550}
(\byear{2021})
\doiurl{10.1016/j.joule.2020.12.007} .
Accessed 2025-02-17
\end{barticle}
\endbibitem

\bibitem[\protect\citeauthoryear{Bailey and Johnston}{2023}]{faircloth_passive_2023}
\begin{bchapter}
\bauthor{\bsnm{Bailey}, \binits{K.}},
\bauthor{\bsnm{Johnston}, \binits{J.}}:
\bctitle{Passive {Direct} {Air} {Capture}: {Breathing} {Cities}}.
In: \beditor{\bsnm{Faircloth}, \binits{B.}},
\beditor{\bsnm{Pedersen~Zari}, \binits{M.}},
\beditor{\bsnm{Thomsen}, \binits{M.R.}},
\beditor{\bsnm{Tamke}, \binits{M.}} (eds.)
\bbtitle{Design for {Climate} {Adaptation}},
pp. \bfpage{603}--\blpage{614}.
\bpublisher{Springer},
\blocation{Cham}
(\byear{2023}).
\doiurl{10.1007/978-3-031-36320-7_38} .
\bcomment{Series Title: Sustainable Development Goals Series}.
\burl{https://link.springer.com/10.1007/978-3-031-36320-7_38}
Accessed 2025-02-17
\end{bchapter}
\endbibitem

\bibitem[\protect\citeauthoryear{Paulsen et~al.}{2024}]{paulsen_techno-economic_2024}
\begin{barticle}
\bauthor{\bsnm{Paulsen}, \binits{M.M.}},
\bauthor{\bsnm{Petersen}, \binits{S.B.}},
\bauthor{\bsnm{Lozano}, \binits{E.M.}},
\bauthor{\bsnm{Pedersen}, \binits{T.H.}}:
\batitle{Techno-economic study of integrated high-temperature direct air capture with hydrogen-based calcination and {Fischer}–{Tropsch} synthesis for jet fuel production}.
\bjtitle{Applied Energy}
\bvolume{369},
\bfpage{123524}
(\byear{2024})
\doiurl{10.1016/j.apenergy.2024.123524} .
Accessed 2025-02-17
\end{barticle}
\endbibitem

\bibitem[\protect\citeauthoryear{Zhang et~al.}{2013}]{zhang_concentrated_2013}
\begin{barticle}
\bauthor{\bsnm{Zhang}, \binits{H.L.}},
\bauthor{\bsnm{Baeyens}, \binits{J.}},
\bauthor{\bsnm{Degrève}, \binits{J.}},
\bauthor{\bsnm{Cacères}, \binits{G.}}:
\batitle{Concentrated solar power plants: {Review} and design methodology}.
\bjtitle{Renewable and Sustainable Energy Reviews}
\bvolume{22},
\bfpage{466}--\blpage{481}
(\byear{2013})
\doiurl{10.1016/j.rser.2013.01.032} .
Accessed 2025-02-17
\end{barticle}
\endbibitem

\bibitem[\protect\citeauthoryear{Li et~al.}{2024}]{li_solar_2024}
\begin{barticle}
\bauthor{\bsnm{Li}, \binits{S.}},
\bauthor{\bsnm{Chen}, \binits{R.}},
\bauthor{\bsnm{Wang}, \binits{J.}},
\bauthor{\bsnm{Deng}, \binits{S.}},
\bauthor{\bsnm{Zhou}, \binits{H.}},
\bauthor{\bsnm{Fang}, \binits{M.}},
\bauthor{\bsnm{Zhang}, \binits{H.}},
\bauthor{\bsnm{Yuan}, \binits{X.}}:
\batitle{Solar thermal energy-assisted direct capture of {CO2} from ambient air for methanol synthesis}.
\bjtitle{npj Materials Sustainability}
\bvolume{2}(\bissue{1}),
\bfpage{11}
(\byear{2024})
\doiurl{10.1038/s44296-024-00014-y} .
Accessed 2025-03-18
\end{barticle}
\endbibitem

\bibitem[\protect\citeauthoryear{Prats-Salvado et~al.}{2024}]{prats-salvado_solar-powered_2024}
\begin{barticle}
\bauthor{\bsnm{Prats-Salvado}, \binits{E.}},
\bauthor{\bsnm{Jagtap}, \binits{N.}},
\bauthor{\bsnm{Monnerie}, \binits{N.}},
\bauthor{\bsnm{Sattler}, \binits{C.}}:
\batitle{Solar-{Powered} {Direct} {Air} {Capture}: {Techno}-{Economic} and {Environmental} {Assessment}}.
\bjtitle{Environmental Science \& Technology}
\bvolume{58}(\bissue{5}),
\bfpage{2282}--\blpage{2292}
(\byear{2024})
\doiurl{10.1021/acs.est.3c08269} .
Accessed 2025-03-18
\end{barticle}
\endbibitem

\bibitem[\protect\citeauthoryear{Kraemer}{2022}]{kraemer_nrel_2022}
\begin{botherref}
\oauthor{\bsnm{Kraemer}, \binits{S.}}:
{NREL} {Results} {Support} {Cheap} {Long} {Duration} {Energy} {Storage} in {Hot} {Sand}
(2022)
\end{botherref}
\endbibitem

\bibitem[\protect\citeauthoryear{Tetteh et~al.}{2024}]{tetteh_improved_2024}
\begin{barticle}
\bauthor{\bsnm{Tetteh}, \binits{S.}},
\bauthor{\bsnm{Juul}, \binits{G.}},
\bauthor{\bsnm{Järvinen}, \binits{M.}},
\bauthor{\bsnm{Santasalo-Aarnio}, \binits{A.}}:
\batitle{Improved effective thermal conductivity of sand bed in thermal energy storage systems}.
\bjtitle{Journal of Energy Storage}
\bvolume{86},
\bfpage{111350}
(\byear{2024})
\doiurl{10.1016/j.est.2024.111350} .
Accessed 2024-03-26
\end{barticle}
\endbibitem

\bibitem[\protect\citeauthoryear{Poulose et~al.}{2022}]{poulose_power_2022}
\begin{barticle}
\bauthor{\bsnm{Poulose}, \binits{T.}},
\bauthor{\bsnm{Kumar}, \binits{S.}},
\bauthor{\bsnm{Torell}, \binits{G.}}:
\batitle{Power storage using sand and engineered materials as an alternative for existing energy storage technologies}.
\bjtitle{Journal of Energy Storage}
\bvolume{51},
\bfpage{104381}
(\byear{2022})
\doiurl{10.1016/j.est.2022.104381} .
Accessed 2024-04-04
\end{barticle}
\endbibitem

\bibitem[\protect\citeauthoryear{Soto et~al.}{2024}]{soto_performance_2024}
\begin{barticle}
\bauthor{\bsnm{Soto}, \binits{G.J.}},
\bauthor{\bsnm{Wagner}, \binits{M.J.}},
\bauthor{\bsnm{Neises}, \binits{T.W.}},
\bauthor{\bsnm{Lindley}, \binits{B.A.}}:
\batitle{Performance analysis of integrated {Nuclear}-{Solar} {Energy} system sharing same molten salt thermal energy storage}.
\bjtitle{Progress in Nuclear Energy}
\bvolume{171},
\bfpage{105166}
(\byear{2024})
\doiurl{10.1016/j.pnucene.2024.105166} .
Accessed 2024-04-01
\end{barticle}
\endbibitem

\bibitem[\protect\citeauthoryear{González-Roubaud et~al.}{2017}]{gonzalez-roubaud_review_2017}
\begin{barticle}
\bauthor{\bsnm{González-Roubaud}, \binits{E.}},
\bauthor{\bsnm{Pérez-Osorio}, \binits{D.}},
\bauthor{\bsnm{Prieto}, \binits{C.}}:
\batitle{Review of commercial thermal energy storage in concentrated solar power plants: {Steam} vs. molten salts}.
\bjtitle{Renewable and Sustainable Energy Reviews}
\bvolume{80},
\bfpage{133}--\blpage{148}
(\byear{2017})
\doiurl{10.1016/j.rser.2017.05.084} .
Accessed 2025-02-17
\end{barticle}
\endbibitem

\bibitem[\protect\citeauthoryear{Bhatnagar et~al.}{2022}]{bhatnagar_molten_2022}
\begin{barticle}
\bauthor{\bsnm{Bhatnagar}, \binits{P.}},
\bauthor{\bsnm{Siddiqui}, \binits{S.}},
\bauthor{\bsnm{Sreedhar}, \binits{I.}},
\bauthor{\bsnm{Parameshwaran}, \binits{R.}}:
\batitle{Molten salts: {Potential} candidates for thermal energy storage applications}.
\bjtitle{International Journal of Energy Research}
\bvolume{46}(\bissue{13}),
\bfpage{17755}--\blpage{17785}
(\byear{2022})
\doiurl{10.1002/er.8441} .
Accessed 2025-02-17
\end{barticle}
\endbibitem

\bibitem[\protect\citeauthoryear{Eck and Steinmann}{2005}]{eck_modelling_2005}
\begin{barticle}
\bauthor{\bsnm{Eck}, \binits{M.}},
\bauthor{\bsnm{Steinmann}, \binits{W.-D.}}:
\batitle{Modelling and {Design} of {Direct} {Solar} {Steam} {Generating} {Collector} {Fields}}.
\bjtitle{Journal of Solar Energy Engineering}
\bvolume{127}(\bissue{3}),
\bfpage{371}--\blpage{380}
(\byear{2005})
\doiurl{10.1115/1.1849225} .
Accessed 2024-03-17
\end{barticle}
\endbibitem

\bibitem[\protect\citeauthoryear{Jiang}{2014}]{jiang_implementation_2014}
\begin{botherref}
\oauthor{\bsnm{Jiang}, \binits{L.}}:
Implementation of thermodynamically efficient collectors for medium high temperature applications.
Unpublished
(2014).
\doiurl{10.13140/RG.2.2.12519.04004} .
\url{http://rgdoi.net/10.13140/RG.2.2.12519.04004}
Accessed 2025-02-17
\end{botherref}
\endbibitem

\bibitem[\protect\citeauthoryear{Ong et~al.}{2013}]{ong_land-use_2013}
\begin{botherref}
\oauthor{\bsnm{Ong}, \binits{S.}},
\oauthor{\bsnm{Campbell}, \binits{C.}},
\oauthor{\bsnm{Denholm}, \binits{P.}},
\oauthor{\bsnm{Margolis}, \binits{R.}},
\oauthor{\bsnm{Heath}, \binits{G.}}:
Land-{Use} {Requirements} for {Solar} {Power} {Plants} in the {United} {States}
(2013)
\end{botherref}
\endbibitem

\bibitem[\protect\citeauthoryear{{The Guardian}}{2023}]{theguardian2023carbon}
\begin{botherref}
\oauthor{\bsnm{{The Guardian}}}:
Carbon Capture in Texas: the World's Biggest – Will It Work?
Accessed: 12 March 2025.
\url{https://www.theguardian.com/environment/2023/sep/12/carbon-capture-texas-worlds-biggest-will-it-work}
\end{botherref}
\endbibitem

\bibitem[\protect\citeauthoryear{EIA}{2025}]{us_eia_electric_2025}
\begin{botherref}
\oauthor{\bsnm{EIA}, \binits{U.}}:
Electric {Power} {Monthly}
(2025)
\end{botherref}
\endbibitem

\bibitem[\protect\citeauthoryear{DOE}{2025}]{us_doe_evaluating_2025}
\begin{botherref}
\oauthor{\bsnm{DOE}, \binits{U.}}:
Evaluating {Your} {Utility} {Rate} {Options}
(2025)
\end{botherref}
\endbibitem

\bibitem[\protect\citeauthoryear{Rand et~al.}{2024}]{rand_queued_2024}
\begin{botherref}
\oauthor{\bsnm{Rand}, \binits{J.}},
\oauthor{\bsnm{Manderlink}, \binits{N.}},
\oauthor{\bsnm{Gorman}, \binits{W.}},
\oauthor{\bsnm{Wiser}, \binits{R.}},
\oauthor{\bsnm{Seel}, \binits{J.}},
\oauthor{\bsnm{Kemp}, \binits{J.M.}},
\oauthor{\bsnm{Jeong}, \binits{S.}},
\oauthor{\bsnm{Kahrl}, \binits{F.}}:
Queued {Up}: {Characteristics} of {Power} {Plants} {Seeking} {Transmission} {Interconnection}
(2024)
\end{botherref}
\endbibitem

\bibitem[\protect\citeauthoryear{Drechsler and Agar}{2020}]{drechsler_investigation_2020}
\begin{barticle}
\bauthor{\bsnm{Drechsler}, \binits{C.}},
\bauthor{\bsnm{Agar}, \binits{D.W.}}:
\batitle{Investigation of water co-adsorption on the energy balance of solid sorbent based direct air capture processes}.
\bjtitle{Energy}
\bvolume{192},
\bfpage{116587}
(\byear{2020})
\doiurl{10.1016/j.energy.2019.116587} .
Accessed 2025-02-17
\end{barticle}
\endbibitem

\bibitem[\protect\citeauthoryear{Rosa et~al.}{2021}]{rosa_water_2021}
\begin{barticle}
\bauthor{\bsnm{Rosa}, \binits{L.}},
\bauthor{\bsnm{Sanchez}, \binits{D.L.}},
\bauthor{\bsnm{Realmonte}, \binits{G.}},
\bauthor{\bsnm{Baldocchi}, \binits{D.}},
\bauthor{\bsnm{D'Odorico}, \binits{P.}}:
\batitle{The water footprint of carbon capture and storage technologies}.
\bjtitle{Renewable and Sustainable Energy Reviews}
\bvolume{138},
\bfpage{110511}
(\byear{2021})
\doiurl{10.1016/j.rser.2020.110511} .
Accessed 2024-10-30
\end{barticle}
\endbibitem

\bibitem[\protect\citeauthoryear{An et~al.}{2022}]{an_impact_2022}
\begin{barticle}
\bauthor{\bsnm{An}, \binits{K.}},
\bauthor{\bsnm{Farooqui}, \binits{A.}},
\bauthor{\bsnm{McCoy}, \binits{S.T.}}:
\batitle{The impact of climate on solvent-based direct air capture systems}.
\bjtitle{Applied Energy}
\bvolume{325},
\bfpage{119895}
(\byear{2022})
\doiurl{10.1016/j.apenergy.2022.119895} .
Accessed 2024-04-08
\end{barticle}
\endbibitem

\bibitem[\protect\citeauthoryear{Sendi et~al.}{2022}]{sendi_geospatial_2022}
\begin{barticle}
\bauthor{\bsnm{Sendi}, \binits{M.}},
\bauthor{\bsnm{Bui}, \binits{M.}},
\bauthor{\bsnm{Mac~Dowell}, \binits{N.}},
\bauthor{\bsnm{Fennell}, \binits{P.}}:
\batitle{Geospatial analysis of regional climate impacts to accelerate cost-efficient direct air capture deployment}.
\bjtitle{One Earth}
\bvolume{5}(\bissue{10}),
\bfpage{1153}--\blpage{1164}
(\byear{2022})
\doiurl{10.1016/j.oneear.2022.09.003} .
Accessed 2024-04-08
\end{barticle}
\endbibitem

\bibitem[\protect\citeauthoryear{Aljubran and Horne}{2024}]{aljubran_power_2024}
\begin{barticle}
\bauthor{\bsnm{Aljubran}, \binits{M.J.}},
\bauthor{\bsnm{Horne}, \binits{R.N.}}:
\batitle{Power supply characterization of baseload and flexible enhanced geothermal systems}.
\bjtitle{Scientific Reports}
\bvolume{14}(\bissue{1}),
\bfpage{17619}
(\byear{2024})
\doiurl{10.1038/s41598-024-68580-8} .
Accessed 2025-02-17
\end{barticle}
\endbibitem

\bibitem[\protect\citeauthoryear{Ogland-Hand et~al.}{2024}]{ogland-hand_nationwide_2024}
\begin{botherref}
\oauthor{\bsnm{Ogland-Hand}, \binits{J.}},
\oauthor{\bsnm{Cairncross}, \binits{E.}},
\oauthor{\bsnm{Adams}, \binits{B.M.}},
\oauthor{\bsnm{Middleton}, \binits{R.S.}}:
Nationwide {Assessment} of {Sedimentary} {Basin} {Geothermal} {Power}
(2024).
\doiurl{10.31224/3685} .
\url{https://engrxiv.org/preprint/view/3685/version/5099}
Accessed 2025-03-18
\end{botherref}
\endbibitem

\bibitem[\protect\citeauthoryear{Bielicki et~al.}{2023}]{bielicki_promise_2023}
\begin{barticle}
\bauthor{\bsnm{Bielicki}, \binits{J.M.}},
\bauthor{\bsnm{Leveni}, \binits{M.}},
\bauthor{\bsnm{Johnson}, \binits{J.X.}},
\bauthor{\bsnm{Ellis}, \binits{B.R.}}:
\batitle{The promise of coupling geologic {CO2} storage with sedimentary basin geothermal power generation}.
\bjtitle{iScience}
\bvolume{26}(\bissue{2}),
\bfpage{105618}
(\byear{2023})
\doiurl{10.1016/j.isci.2022.105618} .
Accessed 2025-03-18
\end{barticle}
\endbibitem

\bibitem[\protect\citeauthoryear{Oelkers et~al.}{2023}]{oelkers_moving_2023}
\begin{barticle}
\bauthor{\bsnm{Oelkers}, \binits{E.H.}},
\bauthor{\bsnm{Gislason}, \binits{S.R.}},
\bauthor{\bsnm{Kelemen}, \binits{P.B.}}:
\batitle{Moving subsurface carbon mineral storage forward}.
\bjtitle{Carbon Capture Science \& Technology}
\bvolume{6},
\bfpage{100098}
(\byear{2023})
\doiurl{10.1016/j.ccst.2023.100098} .
Accessed 2025-03-28
\end{barticle}
\endbibitem

\bibitem[\protect\citeauthoryear{Wei et~al.}{2021}]{wei_proposed_2021}
\begin{barticle}
\bauthor{\bsnm{Wei}, \binits{Y.-M.}},
\bauthor{\bsnm{Kang}, \binits{J.-N.}},
\bauthor{\bsnm{Liu}, \binits{L.-C.}},
\bauthor{\bsnm{Li}, \binits{Q.}},
\bauthor{\bsnm{Wang}, \binits{P.-T.}},
\bauthor{\bsnm{Hou}, \binits{J.-J.}},
\bauthor{\bsnm{Liang}, \binits{Q.-M.}},
\bauthor{\bsnm{Liao}, \binits{H.}},
\bauthor{\bsnm{Huang}, \binits{S.-F.}},
\bauthor{\bsnm{Yu}, \binits{B.}}:
\batitle{A proposed global layout of carbon capture and storage in line with a 2 °{C} climate target}.
\bjtitle{Nature Climate Change}
\bvolume{11}(\bissue{2}),
\bfpage{112}--\blpage{118}
(\byear{2021})
\doiurl{10.1038/s41558-020-00960-0} .
Accessed 2025-02-17
\end{barticle}
\endbibitem

\bibitem[\protect\citeauthoryear{McQueen et~al.}{2021}]{mcqueen_analysis_2021}
\begin{bchapter}
\bauthor{\bsnm{McQueen}, \binits{N.}},
\bauthor{\bsnm{Kolosz}, \binits{B.}},
\bauthor{\bsnm{McCormick}, \binits{C.}}:
\bctitle{Analysis and {Quantification} of {Negative} {Emissions}}.
In: \bbtitle{{CDR} {Primer}},
(\byear{2021}).
\burl{https://cdrprimer.org/about}
\end{bchapter}
\endbibitem

\end{thebibliography}


\end{document}


\title[Article Title]{Design Optimization and Global Impact Assessment of Solar-Thermal Direct Air Carbon Capture -- Supplementary Information}

\author*[1,2]{\fnm{Zhiyuan} \sur{Fan}}\email{zf2198@columbia.edu}


\author*[1]{\fnm{Bolun} \sur{Xu}}\email{bx2177@columbia.edu}

\affil*[1]{\orgdiv{Earth and Environmental Engineering}, \orgname{Columbia University}, \orgaddress{\street{500 W. 120th Street \#510}, \city{New York}, \postcode{10027}, \state{NY}, \country{US}}}

\affil*[2]{\orgdiv{Center on Global Energy Policy}, \orgname{Columbia University}, \orgaddress{\street{1255 Amsterdam Avenue}, \city{New York}, \postcode{10027}, \state{NY}, \country{US}}}



\maketitle


\section{DAC Thermodynamic Simulation: MATLAB Simscape}\label{sec1}

\begin{landscape}
    \begin{figure}
        \centering
        \includegraphics[width=0.99\linewidth]{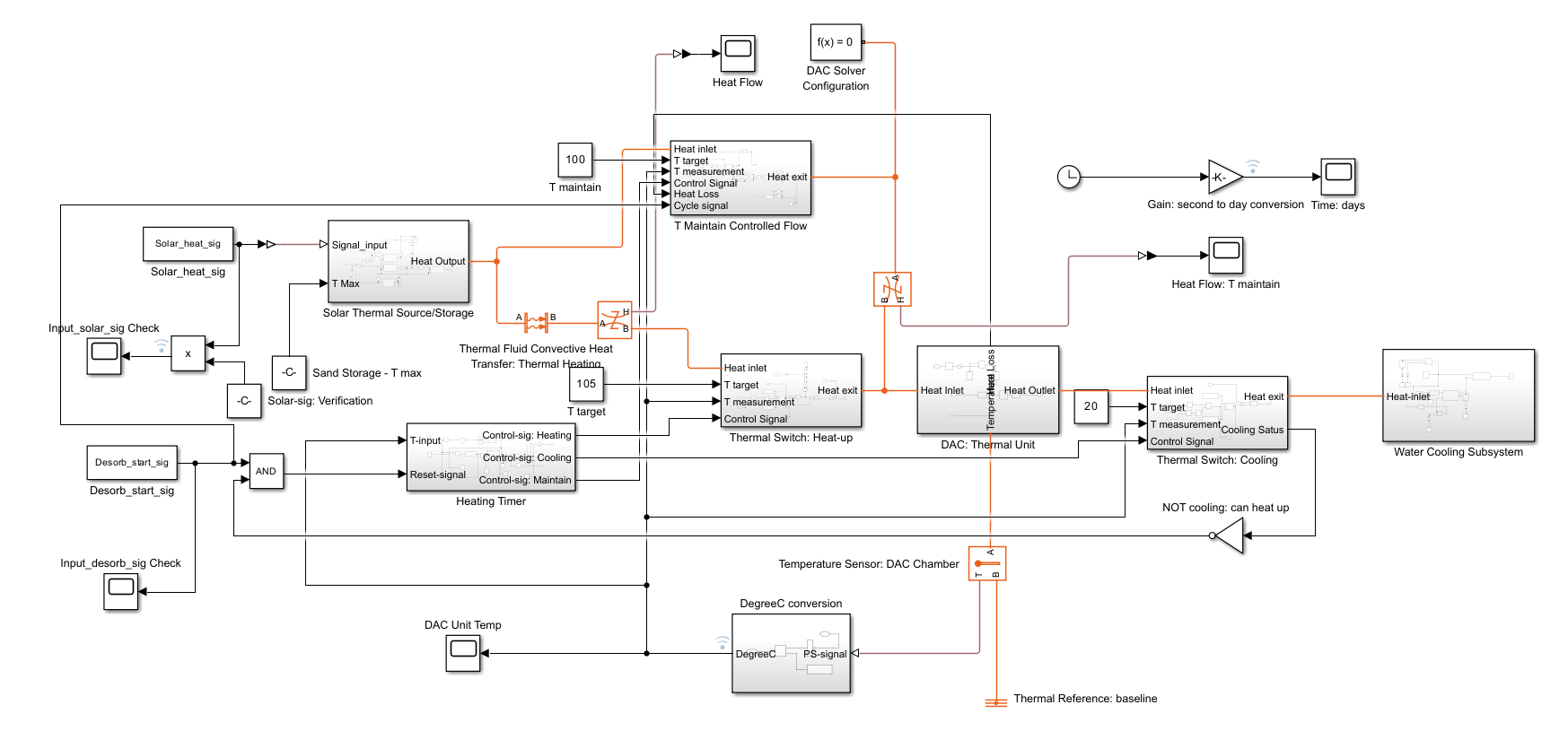}
        \caption{\textbf{MATLAB Simscape DAC system thermodynamic simulation structure.} The model structure showing here is parameterized and taking control signals from strategic operation optimization instead of simulation only. Zoom in details for each individual component and data inspector window can be found at the end of the Supplementary Information.}
        \label{tab:figS1}
    \end{figure}
\end{landscape}

\textcolor{blue}{Figure~\ref{tab:figS1}} illustrates a detailed thermal-fluid management model constructed in MATLAB/Simulink using the Simscape Fluids and Simscape Thermal libraries. The system is centered on a solar-thermal source, represented here by a “Solar Heat-Up Deck” that delivers heat to a dedicated thermal storage reservoir. In the parameterized simulation model connected to optimization model, the solar thermal flux data is automatically generated and fed to the simulation as input. From this reservoir, heat flow can be directed through a thermal switch that toggles between heating and cooling pathways, depending on DAC system temperature demands. The heating path circulates hot fluid (using forced convection steam heat transfer coefficient) to supply energy for processe: direct air capture (DAC) regeneration, while the cooling path leverages a thermal cooling subsystem to dissipate heat when required (cooling to ambient temperature for initiating adsorption again). A series of pumps, and flow/thermal controllers (using variable thermal resistor) regulate fluid/heat circulation, allowing for dynamic control of heat transfer. Temperature sensors and feedback control loops are integrated throughout the model to monitor the chamber environment (i.e., the DAC chamber which contains DAC thermal unit as a thermal mass), adjust heat transfer rates, and maintain user-defined setpoints. This multi-domain approach—combining thermodynamics, thermal energy storage, and control algorithms—enables simulation of both steady-state and transient behavior under variable conrtol signal or data input conditions. By capturing the interactions among heat sources, storage elements, and demand-side processes, the model facilitates comprehensive performance evaluation and design optimization of solar-based thermal systems and detailed DAC unit thermodynamic behavior, ultimately guiding system design parameters and verification of the simplified optimization model.

\section{Optimization Methodology: Extended}\label{sec2}

In this section, a more detailed version of the methodology for DAC operation optimization is presented, including all constraints for the mixed-integer linear programming (MILP) optimization model written in Julia, and the custom \textit{strategic bidding algorithm} algorithm written in MATLAB for cyclic DAC strategic operation which greatly improves the computational speed.

\subsection{MILP for Cyclic DACs}\label{subsec1}

The mixed-integer linear programming (MILP) is a rigorous optimization formulation for linearizing and solving inherently nonlinear and nonconvex process models. While MILP provides accurate profit-maximization results with optimality-bound guarantees, it is computationally expensive and practically unsolvable for technologies with long cycling periods or long optimization horizons.

Indexes:

\begin{itemize}
    \item \(t\): Time period index \(t \in \{1,2,...,T\}\), a total of T time periods in optimization horizon.
\end{itemize}

Binary decision variables:

\begin{itemize}
    \item \(u_t\): 1 if DAC is during the absorption phase at time period \( t\), otherwise zero. 
    \item \(v_t\): 1 if DAC is during the desorption phase at time period \( t\), otherwise zero. 
    \item \(z_t\): 1 if DAC is switched to a new cycle at time period \(t\), otherwise zero
    \item \(k_t\): introduce new sign-variable to determine the change of status at time period \(t\) to facilitate calculation of \(z_t\)
\end{itemize}

Continuous decision variables:

\begin{itemize}
    \item \(X_t\): state-of-saturation capacity of DAC system at time period \(t\) 
    \item \(a_t\): absorption amount of DAC system at time period \(t\)
    \item \(d_t\): desorption amount of DAC system at time period \(t\)
    \item \(h_t\): available thermal energy at time period \(t\)
\end{itemize}

System input parameters:

\begin{itemize}
    \item \(\lambda_t\): electricity price at time period \(t\), can subjected to modification of \(CO_2\) (see \(\lambda e_t\))
    \item \(s_t\): solar thermal charging at time period \(t\), given be data
    \item \(e_t\): electricity \(CO_2\)-intensity at time period \(t\)
    \item \(\rho_e\): \(CO_2\) value of carbon for electricity, 0 if for wholesale electricity price case, \(\pi\) if for carbon-tax adjusted electricity price case.  
    \item \(\lambda e_t\): electricity price without \(CO_2\)-intensity correction (\(\lambda_t = \lambda e_t + \rho_e e_t\))
    \item \(\eta_{t}\): adsorption rate correction factor at time period \(t\), which is a function of relative humidity and temperature.
    \item  \(C_{t}\): energy consumption correction factor at time period \(t\), which is a function of relative humidity and temperature.
    \item \(\pi\): \(CO_2\) value constant including selling price, subsidies, carbon-tax, etc.
\end{itemize}

DAC input parameters:

\begin{itemize}
    \item \(P^a\): electricity consumption for absorption phase per unit time period
    \item \(P^d\): electricity consumption for desorption phase per unit time period
    \item \(\bar{X}\): max available DAC capacity
    \item \(\bar{h}\): max available thermal energy storage capacity
    \item \(\eta\): thermal energy loss efficiency, \(0 \leq \eta \leq 1\)
    \item \(S\): switching cycle cost for consumption of sorbent
    \item \(H\): heating thermal energy load during desorption phasematerials
\end{itemize}

Piecewise linear approximation for absorption/desorption rate using quadratic coefficients:

\begin{itemize}
    \item \(\beta^{a}_{1}\): first-order coefficient for absorption
    \item \(\beta^{a}_{2}\): second-order coefficient for absorption
    \item \(\beta^{d}_{1}\): first-order coefficient for desorption
    \item \(\beta^{d}_{2}\): second-order coefficient for desorption
\end{itemize}

The objective function maximizes the total profit of the DAC operation.

\begin{equation}
    max \sum_t \; \pi d_t - \lambda_t C_{t} (P^a u_t +  P^d v_t)-Sz_t
\end{equation}
where  \(\pi d_t\) is the total revenue by captured (desorbed) \(CO_2\), minus the cost of power consumption corrected by ambient conditions at each time step \( \lambda_t C_{t} (P^a u_t +  P^d v_t)\), minus the cost of material consumption for switching cycles \(Sz_t\). 
\\
\bigbreak
Binary constraints for absorption/desorption status:

\begin{equation}
    u_t + v_t \leq 1
\end{equation}
where the DAC system can only be adsorption/desorption at one time period, it can be neither absorbing nor desorbing. 
\\
\bigbreak
Absorption and desorption rate (using inequality here to avoid contradiction with binary constraints, which is relaxation of the constraints), be mindful that only adsorption (capture) process is corrected by ambient condition of each time step, desorption is a controlled process which is not affected by ambient condition:
\begin{equation}
    a_t \leq \eta^{a}_{t} (\beta^{a}_{1} + \beta^{a}_{2} X_t)
\end{equation}
\begin{equation}
    d_t \leq \beta^{d}_{1} + \beta^{d}_{2} X_t
\end{equation}
additionally, the absorption rate and desorption rate shall be bounded by the binary variable for each as well.
\begin{equation}
    a_t \leq M u_t
\end{equation}
\begin{equation}
    d_t \leq M v_t
\end{equation}
where \(M\) is a sufficiently large number which does not bind the absorption and desorption rate if \(u_t\) and \(v_t\) are 1.

Be careful that although the below constraints are not required to be added explicitly, they shall be automatically satisfied with the above absorption and desorption rate constraints:
\begin{equation}
    0 \leq a_t \leq a_{max}
\end{equation}
\begin{equation}
    0 \leq d_t \leq d_{max}
\end{equation}
where absorption and desorption rates are always bounded by 0 and its maximum designed capacity
\\
\bigbreak
State-of-saturation capacity of DAC system updates with absorption/desorption rates:
\begin{equation}
    X_t-X_{t-1} = a_t - d_t
\end{equation}
where the change of state-of-saturation between time periods equals: adding absorption; subtracting desorption. Both absorption and desorption are corrected by the binary decision variables. 
\\
\begin{equation}
    0 \leq X_t \leq \bar{X}
\end{equation}
where the state-of-saturation is always lower bounded by 0, and upper bounded by its maximum capacity \(\bar{X}\)
\\
\bigbreak
Switching cycle constraints:
\begin{equation}
    k_0 = 0
\end{equation}
where the initial state of the sign-variable equals zero.
\begin{equation}
    -M(1-k_t) \leq k_{t-1} + (u_t - v_t) - 0.5
\end{equation}
\begin{equation}
    M k_t \geq k_{t-1} + (u_t - v_t) - 0.5
\end{equation}
where the sign function is defined here with a sufficiently large \(M\). The comparison is subtracted by 0.5 to avoid the possible case that \(k_{t-1} + (u_t - v_t) = 0\). The solution matrix for \(k_t\) for all possible combinations of sign-function is given in below table:

\begin{center}
\begin{tabular}{||c c c||} 
 \hline
 Variables & \(k_{t-1} = 0\) & \(k_{t-1} = 1\)\\ [0.5ex] 
 \hline\hline
 \(u_t-v_t\) = -1 & 0 & 0 \\
 \hline
 \(u_t-v_t\) = 0 & 0 & 1 \\
 \hline
 \(u_t-v_t\) = 1 & 1 & 1 \\ [1ex] 
 \hline
\end{tabular}
\end{center}

The above two equations together define the sign-variable \(k_t\) which is used to determine the binary cycle counting variable \(z_t\):
\begin{equation}
    z_t \geq k_t - k_{t-1}
\end{equation}
where \(z_t\) will be minimized in the objective function that only when \(k_t = 1\) and \(k_{t-1}=0\), \(z_t = 1\).

Eventually, the thermal storage and energy consumption behaviors. Thermal energy bounding constraint:
\begin{equation}
    0 \leq h_t \leq \bar{h}
\end{equation}

Thermal energy charging/discharging constraint:
\begin{equation}
    h_t \leq \eta h_{t-1} + s_t - v_t H
\end{equation}

\begin{figure}
    \centering
    \includegraphics[width=0.8\linewidth]{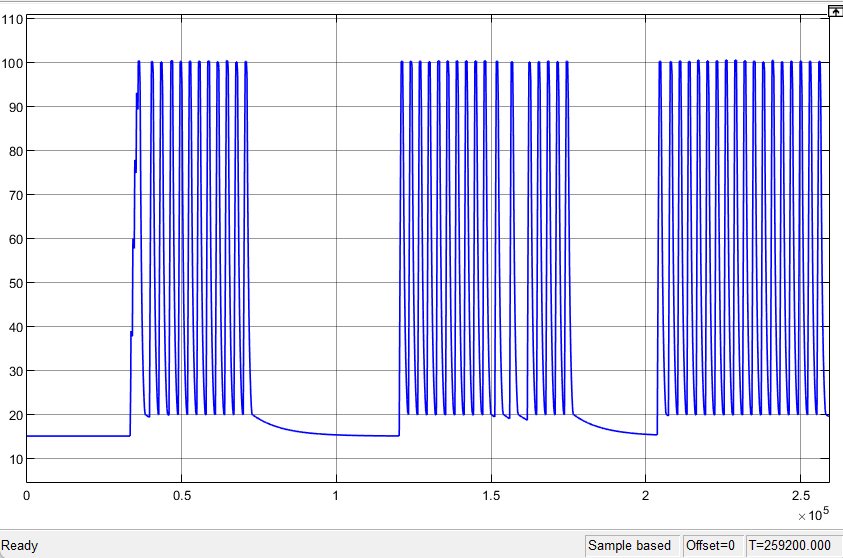}
    \caption{\textbf{3-days DAC thermal unit temperature profile.} It can seen clearly that the DAC system temperature is cycling between cooling target temperature (20 \textdegree C) and regeneration temperature (100 \textdegree C). In the Simscape simulation, the direct air capture (DAC) process is divided into four stages—adsorption, heating, desorption, and cooling. However, in the optimization model, these stages are consolidated into a two-step cycle that includes only adsorption and desorption, with the durations for heating and cooling effectively incorporated into these two steps. This streamlined approach emphasizes the importance of verifying whether the system can achieve sufficiently rapid heating and cooling to maintain practical adsorption and desorption times. Ultimately, the simulation results highlight a trade-off involving a minimum heating source temperature of 300 \textdegree C and a minimal flow rate for the water cooling system.}
    \label{tab:figS2}
\end{figure}

These two sets of thermal behavior constraints represents a significant simplification of the comlicated thermodynamic behavior that's been simulated in the previous mentioned MATLAB Simscape model. The key differences are: (1) completed ignoring the temperature profile of DAC unit itself (see \textcolor{blue}{Figure~\ref{tab:figS2}}), only tracking the thermal energy required for operating the system; (2) all thermal losses to ambient environment in various processes (e.g., storage, thermal switch, DAC unit) are combined into one single energy loss assign to the solar input or as correction of the thermal energy consumption; (3) the heat transfer during the regeneration process linearize the highly non-linear heat profile, which is very high during the heating up stage and moderate in maintaining the required temperature for enough time. These simplifications introduces only trivial differences in DAC operation that the optimization code and Simscape simulation achieve \(>\)99\% matching accuracy. 

The above formulation summarizes the optimization framework of the DAC system using the temperature-swing DAC technology (all 3 DAC technologies tested in this paper). In practice, the DAC input parameters including the piecewise linear approximation for absorption and desorption rate using quadratic coefficients will be determined by different DAC specifications. The actual optimization programs use a look-ahead framework (similar concept of MPC control) which looks longer optimization horizon but applies only the first few steps for action. 

\subsection{Custom Algorithm: Strategic Operation}\label{subsec1}

A threshold-based formulation is rooted in a custom \textit{strategic bidding} MATLAB algorithm, specifically designed to handle nonlinear processes. By participating in electricity markets through the bidding process, DAC operators can strategically operate during the low-price periods:  activate operation if prices are lower than the threshold and idle if higher, hence mimicking market clearing of flexible demand bids, potentially generating more profit at market price valleys. The optimization problem is therefore greatly simplified to optimizing only 1 variable, the price threshold for each optimization horizon, increasing the computational efficiency. 

The operation of the DAC follows exactly the same formulation as MILP, with additional key terms in this formulation including:
\begin{itemize}
\item \(\lambda_{\text{opt}}\): The price threshold, or the electricity price at which DAC operators are willing to operate.
\item \(c\): A data ``chunk" for processing. Each chunk is optimized individually.
\end{itemize}

A typical chunk size \(c\) is 24 hours, equivalent to 288 time steps (5-min temporal resolution), reflecting daily market price behaviors. However, the algorithm is designed with flexibility, allowing for adjustments in chunk size to cater to different market dynamics or operational needs for different DAC cycle times.

The DAC system's operation is determined by the threshold, \(\lambda_{\text{opt}}\). The decision to either accept or reject is defined as:
\[
\begin{cases} 
\text{Activate} & \text{if } \lambda_t(t)\leq \lambda_{\text{opt}} \\
\text{Idle} & \text{otherwise}
\end{cases}
\]

For each chunk \(c\), the algorithm employs an iterative process to find the optimal \(\lambda_{\text{opt}}\) that maximizes profit. It utilizes MATLAB's \texttt{fminsearch} function, which optimizes 1 variable to find the minimum of the objective function. Specifically, \texttt{fminsearch} is tasked to find the value of \(\lambda_{\text{opt}}\) that minimizes the negative profit objective function, effectively maximizing profit. 

To circumvent the issue of local minima and to ensure identification of the global minimum, the algorithm iterates over a series of initial guesses \(\lambda_{\text{series}}\) for \(\lambda_{\text{opt}}\) in the range from -10 to 500 USD. One may increase the granularity of the initial guesses to get a marginally better objective with a computation time trade-off. For each initial guess in the \(\lambda_{\text{series}}\), \texttt{fminsearch} conducts an optimization to locate a local maximum. Upon the completion of all such iterations, the algorithm chooses the highest profit from these local maxima as the global maximum. This approach reduces the risk of the optimization process settling for a suboptimal local maximum.

The final determination of \(\lambda_{\text{opt}}^*\) involves selecting the optimal value that corresponds to the maximum profit achieved across all the local optimizations:

\begin{equation}
\lambda_{\text{opt}}^* = \max_{\lambda_{\text{series}}} \text{fminsearch}(-\text{Objective}(c, \lambda_{\text{guess}})).
\end{equation}

The \(\max\) function in the equation is a representation of the selection process that identifies the highest value from all the local maxima obtained by \texttt{fminsearch}, starting from different initial guesses within the series. This comprehensive exploration across a diverse set of starting points helps in ensuring that the resulting \(\lambda_{\text{opt}}^*\) is the most favorable for maximizing profit.

\begin{figure}
    \centering
    \includegraphics[width=0.9\linewidth]{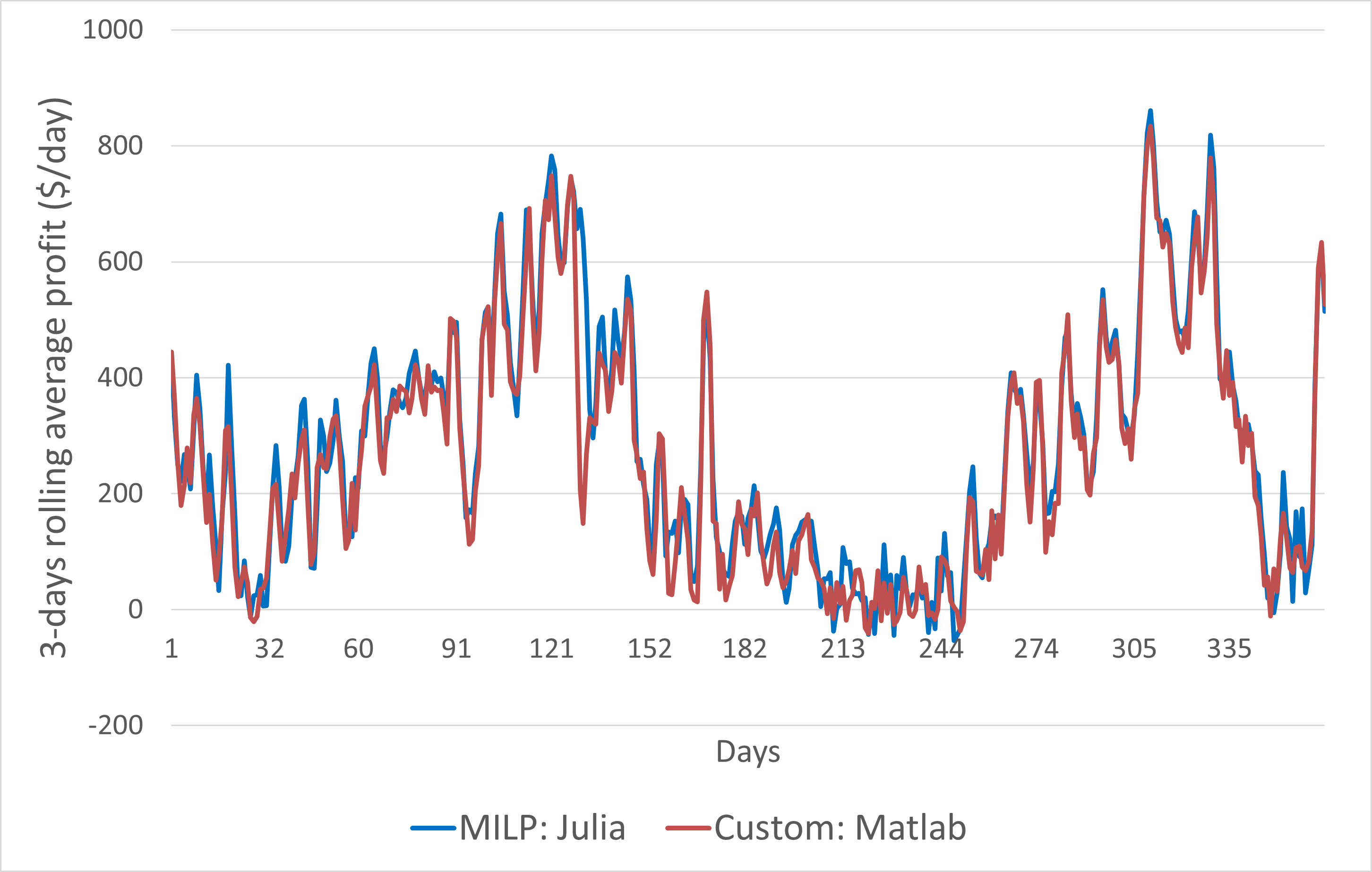}
    \caption{\textbf{3-days rolling average profit for MILP and Custom algorithm optimality comparison in NY.} based on MOF technology, incentive selling price = \$200/ton-\co. It shows that the customer algorithm pertains the optimality of rigurous MILP optimization method.}
    \label{tab:figS3}
\end{figure}

\subsubsection{Look-Ahead Mechanism:}

\begin{itemize}
    \item \(L\): Look-ahead parameter representing the number of future data points considered during optimization.
    \item \(t\): Starting time of the given chunk \(c\).
\end{itemize}

For a given chunk \(c\) starting at time \(t\), the algorithm factors in data from \(t\) to \(t+L\) to find the optimal \(\lambda_{\text{opt}}\). The optimization process employs the MATLAB function \texttt{fminsearch}:

\begin{equation}
\lambda_{\text{opt}}^* = \text{fminsearch}(-\text{Objective}(c, \lambda_{\text{opt}}, L))
\end{equation}

Where:
\[
\text{Objective}(c, \lambda_{\text{opt}} , L)
\]
is the profit obtained for chunk $c$ when using threshold \(\lambda_{\text{opt}}\)and considering $L$ future data points.

\subsubsection{Boost Mechanism:}

\begin{itemize}
    \item \textbf{X\_remain}: State-of-saturation at the end of the processed chunk. This represents the amount of CO\(_2\) credit still in the system that hasn't been cleared yet.
    \item \textbf{Profit\_chunk}: Total profit from the processed chunk.
    \item \textbf{CO2\_chunk}: Total amount of CO\(_2\) desorbed during the processed chunk
\end{itemize}

The boost mechanism is introduced to account for the potential value of the remaining CO\(_2\) in the system at the end of the DAC loop to avoid selling all CO2 at the end of each chunk to maximize profit, which will force the DAC operation to terminate at the end of each optimization horizon. The boost is calculated using the following formula:

\[
\text{boost} = \left( \frac{\text{Profit\_chunk}}{\text{CO2\_chunk}} \right) \times \text{X\_remain}
\]

Where:

\begin{itemize}
    \item \(\frac{\text{Profit\_chunk}}{\text{CO2\_chunk}}\) represents the average profit per unit of CO\(_2\) sold during the processed chunk. It indicates the expected value or profit obtained from selling one unit of CO\(_2\).
\end{itemize}

After calculating the boost, it's added to the profit for that chunk. This accounts for both the profit from the CO\(_2\) that was sold and the potential value of the CO\(_2\) that remains in the system:

\[
\text{Profit\_total} = \text{Profit\_chunk} + \text{boost}
\]

However, it's important to note that this inflated profit is not used when calculating the overall profits. It is solely used for the process of finding the optimal lambda. This ensures that the optimization process considers both actual and potential profits, valuing scenarios where there's a significant amount of CO\(_2\) left in the system at the end of the loop.

\subsection{Ambient correction calculations}\label{subsec1}

Assuming all DAC nominal capacity and power consumption is measured under standard lab environment with baseline temperature of 20 \(^{\circ}\)C and relative humidity of 50\%. We applied the ambient sensitivity from two studies: \cite{sendi_geospatial_2022} for cyclic DAC operation based on steam-assisted vacuum-pressure temperature swing adsorption with amine-functionalized solid sorbent; and \cite{an_impact_2022} for continuous liquid solvent KOH DAC. 

For cyclic DAC operation, energy consumption prefers mild relative humidity, where relative humidity either too high or low will increase energy consumption with very high sensitivity. Energy consumption is not very sensitive to ambient temperature, slightly in favor of lower temperature. From capture rate or abatement productivity perspective, it strongly in favor of low temperature and mild relative humidity.

The electricity consumption correction after numerical fitting is given by the following, using quadratic equation for temperature \(T\) fitting and exponential function for skewed symmetric behavior of relative humidity \(RH\):

\begin{equation}
    C_t = [1.9 + 0.01(T-20)](RH-0.4)^2 e^{RH-0.4}+[1.5+0.003(T-20)^2]
\end{equation}

The capture rate or abatement correction similarly:

\begin{equation}
    \eta_t = 65 - 0.01 T^2 - (T+20)(RH-0.4)^2
\end{equation}

For continuous KOH liquid solvent DAC, capture rate or abatement productivity prefers high temperature and high relative humidity. Little information about energy consumption's sensitivity on ambient conditions. It can be explained by the fact that majoroty of power consumption is used for driving moving equipments and thermal energy for regeneration of solvent at high temperature of 800 \(^{\circ}\)C, both insensitive of ambient conditions. 

The capture rate or abatement correction after numerical fitting is given by:

\begin{equation}
    \eta_t = 74 + 8(RH-0.5) + (T-20)
\end{equation}

\subsection{Solar Thermal Modeling}

The solar heating collector efficiency is a function of both input data (solar DNI) and design parameters.

\begin{equation}
\begin{split}
    \eta_{c_t} (DNI_t, cr, T) 
    &= 0.78 - \eta_{loss} \\
    &= 0.78 - \alpha T^2 * \frac{\beta}{DNI_t + m} * \frac{\gamma}{cr}\\    
\end{split}
\end{equation}
the function states that the collector efficiency is baseline (0.78) - loss. The loss is growing with \(T^2\) corrected by \(\alpha\). The loss will reduce when solar input (direct normal irradiation) \(DNI_t\) and solar concentration ratio \(cr\) is larger. For simplicity, the DNI is fitted using solar capacity factor instead of original \(W/m^2\). \(m\) is a small number to avoid calculation error when \(DNI_t\) is 0. After fitting the experimental data in literature, the coefficients are calculated to be.

\begin{itemize}
    \item \(\alpha = 8.8*10^{-7}\)
    \item \(\beta = 1.1\)
    \item \(m = 0.1\)
    \item \(\gamma = 1\)
\end{itemize}

So the solar heating collector efficiency is eventually given by:

\begin{equation}
    \eta_{c_t} (DNI_t, cr, T) = 0.78 - 8.8*10^{-7} *T^2 * \frac{1.1}{DNI_t + 0.1} * \frac{1}{cr}
\end{equation}

The available heat is given by the following:

Without thermal energy storage:
\begin{equation}
    s_t = DNI_t * cp * \eta_{c_t} 
\end{equation}

With thermal energy storage:
\begin{equation}
    s_t = DNI_t * cp * \eta_{c_t} * \eta_{th}(T)
\end{equation}
where \(\eta_{th}(T)\) is the charging efficiency of the thermal energy storage system, which is a function of target temperature T.

The results of above semi-physics fitting compared to experimental results in literature is shown:
\begin{figure}
    \centering
    \includegraphics[width=0.99\linewidth]{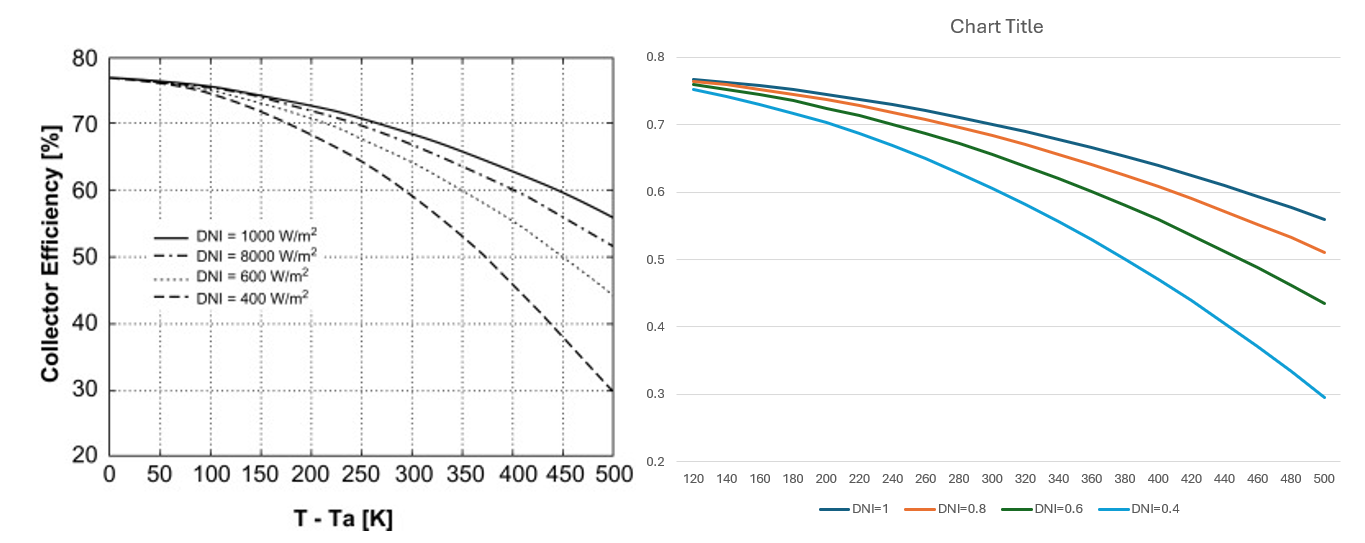}
    \caption{\textbf{Literature reported solar thermal efficiency and proposed semi-physics fitting results.} Additional solar concentration ratio \(cr\) = 1.}
    \label{tab:figS4}
\end{figure}

The solar Heating + Storage System CAPEX is given by function of the design parameters:

\begin{equation}
\begin{split}
    CAPEX_s (cp, cr, \bar{h}) 
    &= CAPEX_{sh} (cp, cr) + CAPEX_{th} (T, \bar{h}) \\
    &= CAPEX_{sh}^{unit}*cp*cr*(1-0.15)^{log_2(cp*cr)} + CAPEX_{th}^{unit}*\bar{h}\\    
\end{split}
\end{equation}

Sand thermal energy storage as a special type of storage. Sand is chemically very stable and the energy storage capacity is a function of both operation temperature and sand mass (size of the storage tank). Under the current formulation, the sand mass (size of the storage tank) is determined by the investment on the storage size, while the target temperature is determined by solar collector design.

\begin{equation}
    \bar{h}_{effect} = \bar{h} *\frac{T - T_{min}}{T_{unit}}
\end{equation}

Here the \(T_{unit}\) is the designed temperature increment for energy storage capacity \(\bar{h}\), and the effective energy storage capacity will be changing proportionally with the actual designed temperature increment (\(T - T_{min}\)).

\section{Data and Processing}\label{sec2}

NY, CA, and TX data collection and processing are presented, and the processed data input for the model can be accessed from \href{https://github.com/ZhiyuanF/Solar-DAC.git}{Github Repo}.

\subsection{NY}\label{subsec2}

5-minute resolution \href{https://www.nyiso.com/energy-market-operational-data}{NY pricing data} for the ``MHK VL'' zone was downloaded from NYISO's operational data website. This zone was chosen for its high number of industrial facilities, resulting in price fluctuations primarily based on demand.

Emission data was sourced from processing \href{https://www.nyiso.com/real-time-dashboard}{NYISO's Real-Time Fuel Mix} data archive, available at a 5-minute resolution for the year 2022. To calculate the Total Adjusted Emission Rate, the Total Demand (MW) was determined by summing up the energy generated from various fuel sources:

\begin{equation}
\text{Total Demand} = \sum_{i} \text{Demand}_{i}
\end{equation}

Where $i$ represents each fuel source.

Using standard emission rates for each fuel type, the Total Emission Rate (TER) for New York was determined by taking the dot product of the energy demand values and the respective emission values, resulting in a weighted sum of the emissions.

\begin{equation}
\text{Total Adjusted Emission Rate} = \frac{\text{TER}}{\text{Total Demand}}
\end{equation}

The Total Adjusted Emission Rate was then calculated using equation (18).

The NY case results are not shown since low solar capacity factor and very large gap to TX and CA cases. As shown in the solar map filter (see below), the NY region is not within the practical areas for solar-thermal DAC systems. 

\subsection{CA}\label{subsubsec2}

Pricing data was downloaded from \href{http://oasis.caiso.com/mrioasis/logon.do}{CAISO OASIS}, capturing 5-minute resolution Locational Marginal Price (LMP) data for the year 2022. The node selected for this study was \texttt{NEWHALL\_1\_N001}, located in Santa Clarita, California. This area was chosen due to the pronounced representation of the solar ``duck curve". 

5-min resolution \href{https://www.caiso.com/todaysoutlook/pages/emissions.html}{CA emission data}  (measured in ton-\cd/MWh) was obtained from CAISO's ``Today's Outlook''. The Total Emission Rate (TER) was then divided by the total demand in MW, sourced from the LMP data, to derive the Total Adjusted Emission Rate.

\subsection{TX}\label{subsubsec2}

Texas data from ERCOT using the \href{https://www.ercot.com/mp/data-products/data-product-details?id=NP6-785-ER}{real-time price (RTP)} data and \href{https://www.ercot.com/gridinfo/generation}{fuel mix} data. The TER calculation follows the same formula as NY and CA. This paper uses the price data from ``HB WEST" of the ERCOT settlement points in west Texas, this is where the largest ongoing DAC project, STRATOS, is located.

ERCOT has a 15-minute resolution for market clearing. To be consistent with the NY and CA 5-minute resolution comparison, this study reconstructed the 15-minute resolution data to 5-minute resolution by repeating each data point three times, where three 5-minute resolution time steps have the same data to create the 15-minute market clearing interval. Any missing data from sources are linearly extrapolated from the nearest available data.  

\subsection{Temperature and Relative Humidity Data}\label{subsubsec2}

\subsection{Grid-connection: Site-specific Data}

\begin{table}[h!]
    \centering
    \begin{tabular}{c|ccccccc}
         State & Location/Zone & Representative & Lat-1 & Lat-2 & Long-1 & Long-2 & Climate Remarks\\ \hline
         CA & Santa Clarita & Santa Clarita  & 34.5 & 34 & -119 & -118.5 & Hot-summer mediterranean\\
         TX & West Texas & Odessa & 32 & 31.5 & -102.5 & -102 & Cold/hot semi-arid \\
         NY & MHKVL & Watertown & 44 & 42 & -76 & -74.5 & Warm-summer humid continental\\
    \end{tabular}
    \caption{Location and description of ambient environmental data collection from Climate Data Store dataset}
    \label{tab:S2}
\end{table}

Ambient environmental data of hourly resolution temperature and relative humidity time series is downloaded from Climate Data Store \href{https://cds.climate.copernicus.eu/datasets/reanalysis-era5-pressure-levels?tab=overview}{Climate Data Store} ``ERA5 pressure levels'' based on 1000hPa atmospheric pressure in year 2022 \cite{copernicus_climate_change_service_era5_2018}. Details of data description about specific spot data retrieval is in the following table \textcolor{blue}{Table}~\ref{tab:S2}.

\subsection{Global Impact Data}

\begin{figure}[h!]
    \centering
        \begin{subfigure}[b]{0.9\textwidth}
        \centering
        \includegraphics[width=\textwidth]{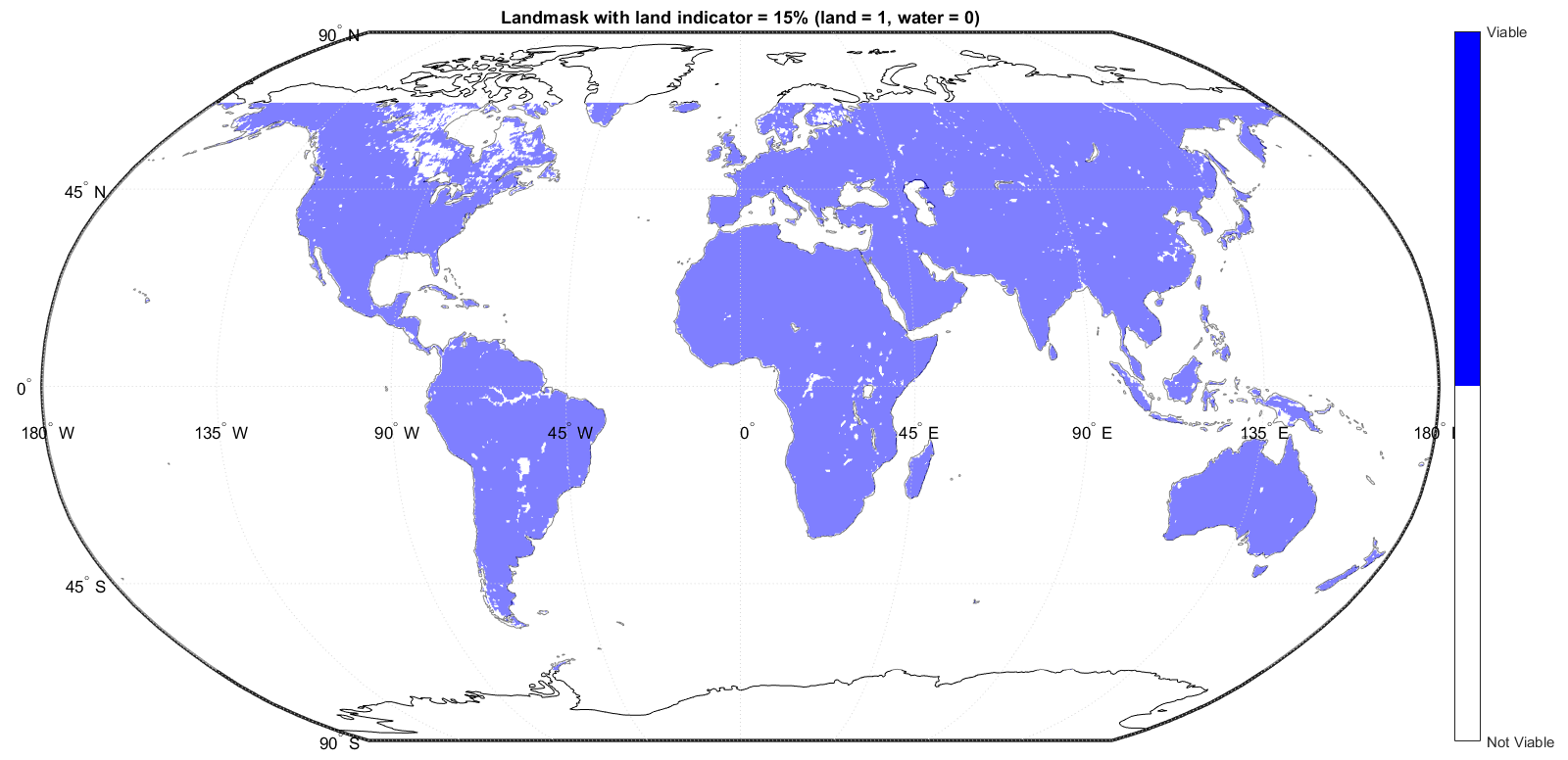}
    \end{subfigure}
    \hfill
    \begin{subfigure}[b]{0.9\textwidth}  
        \centering 
        \includegraphics[width=\textwidth]{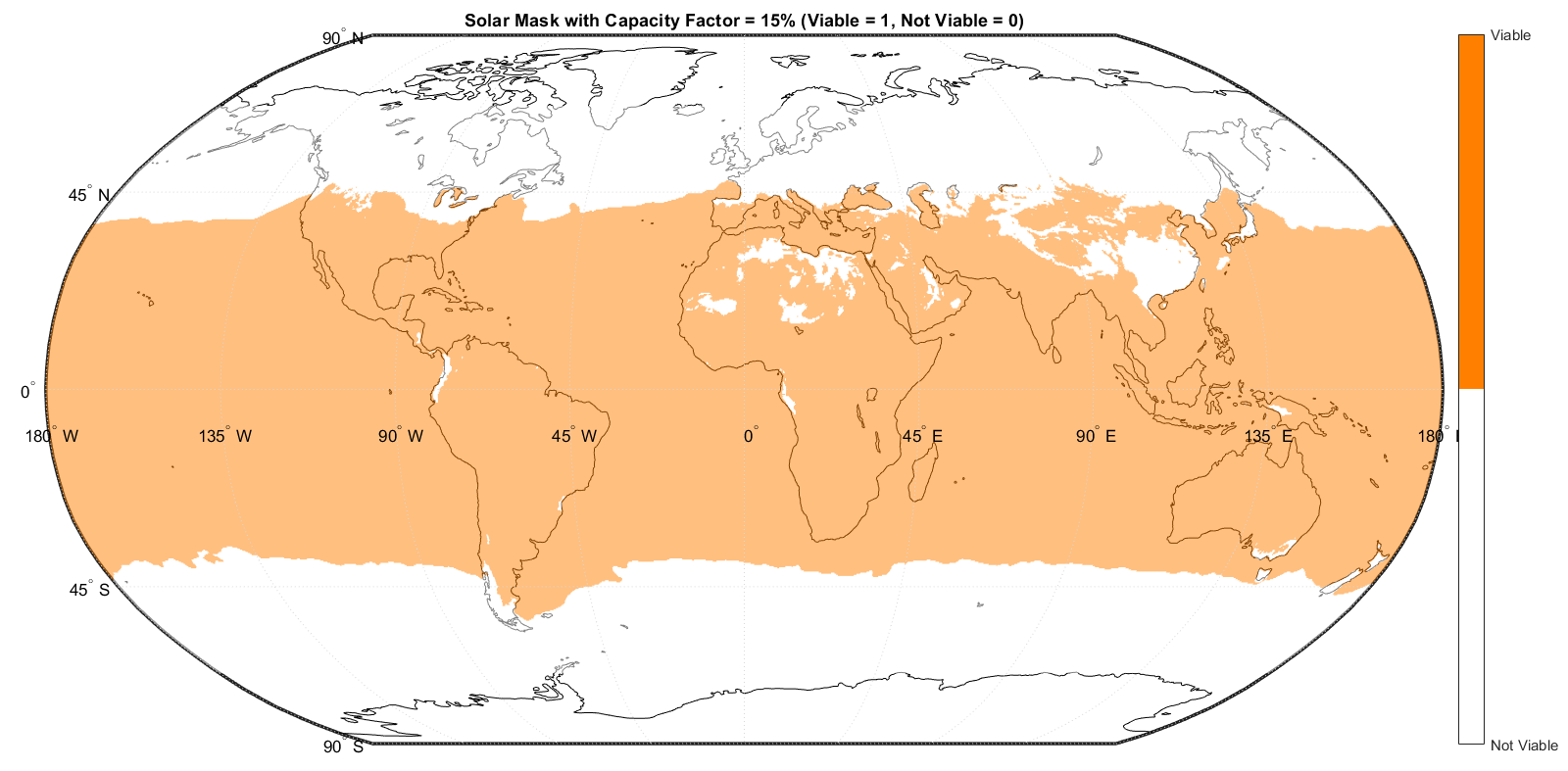} 
    \end{subfigure}
    \hfill
    \begin{subfigure}[b]{0.9\textwidth}  
        \centering 
        \includegraphics[width=\textwidth]{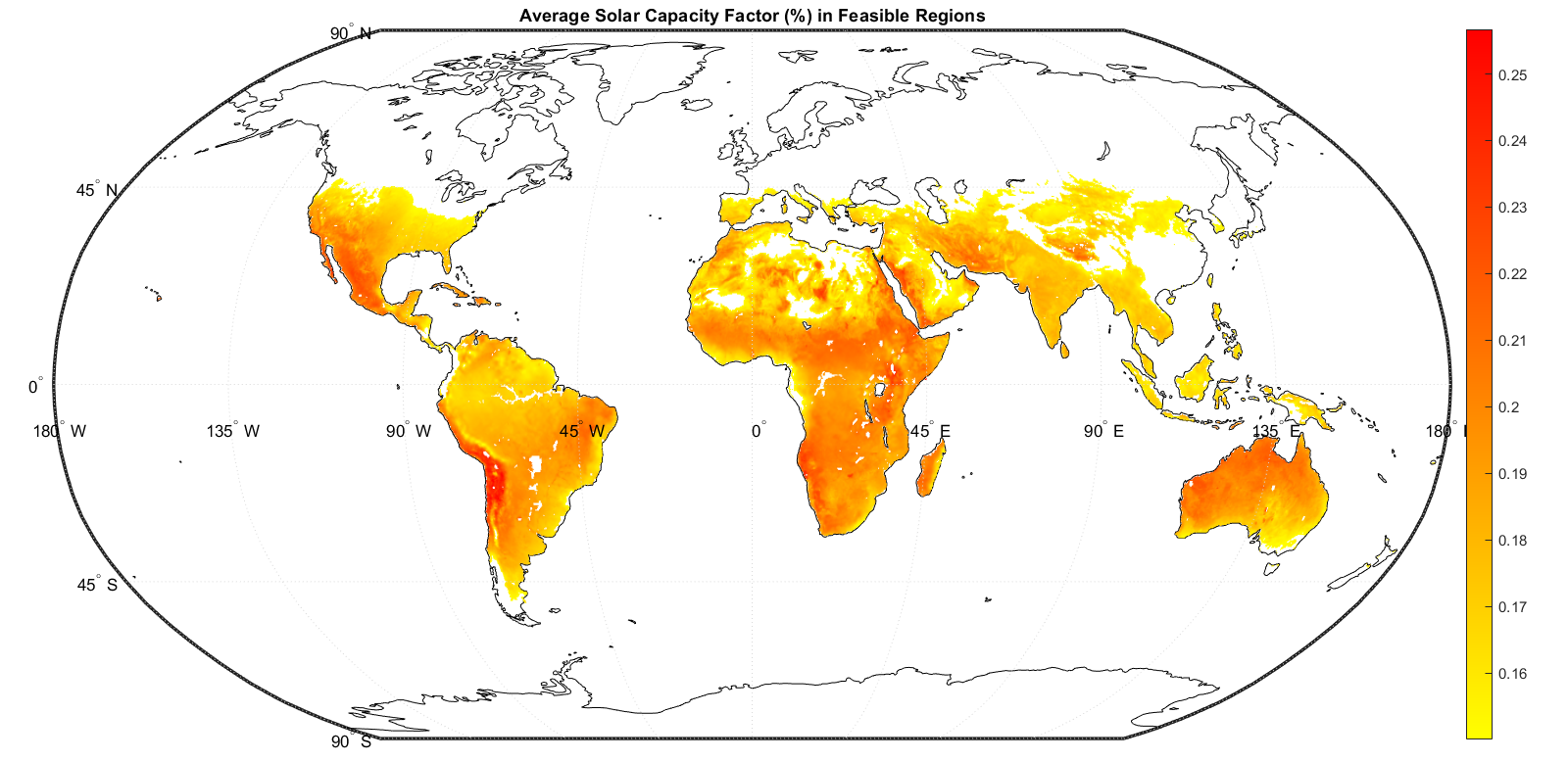} 
    \end{subfigure}
    \caption{\textbf{Land mask (upper) for removing all water surface data and solar capacity factor mask \(>\)15\% (middle) for removing non-solar feasible surface data and resulted solar-feasible study regions (lower) used for solar-DAC global impact study.} The combined masks will totally remove \(>\)75\% original global data. The total feasible data include 107713 points. Specifically, the Sahara Desert in north Africa showing lower solar capacity factor due to original ERA5 data issue, likely because lack of measurement and calibration.}
    \label{tab:figS5}
\end{figure}

The global study data is obtained from the exact same data source but expanded to global inclusion. To minimize the computation load, several masks on the original data is applied to eliminate point on water surface and low solar capacity regions, leading to the presented feasible study areas, see\textcolor{blue}{Figure~\ref{tab:figS5}}.

\section{DAC Power System Model Integration}\label{sec3}

To better understand the impact of DAC load to power system, both the power price and electricity carbon intensity, an integrated DAC + CAISO power system simulation is added here. The results show that the DAC's scale explored in this study has very limited impact on the overall power system. Even scaled up by \(\leq\) 100 times, the change of annual power consumption cost is \(\leq\) 5\%.

The power market model is using the open-sourced model from \cite{zheng_wecc-based_2023}, presenting a Western Electricity Coordinating Council (WECC) power market simulation tool with two-stage market clearing of both day-ahead unit commitment at hourly resolution and real-time price from economic dispatch with 5-min resolution, both consistent with CAISO power market.

To better understand the impact of unpredictable load change due to flexible DAC operation, the DAC load is added to real-time load forecast error instead of base-load that's within the prediction. The DAC load is added to Region 2, where solar duck-curve behavior is most prominent and average power price is similar to historical data. Meanwhile, considering the duck-curve driven DAC operation recovers the solar renewable profile, two different scenarios are generated: (1) DAC scaling without new PV capacity installation; (2) DAC scaling with new PV installation that matches DAC capacity.

The results of DAC power system integration model is presented in \textcolor{blue}{Table}~\ref{tab:S3}. Due to load added to real-time load forecast error, the day-ahead unit commitment would not see the DAC load and would dispatch the commited generators to satisfy the DAC load, which could be either curtailed solar PV generation or fossil generators. Therefore, adding PV with DAC load don't always results in lower costs or lower emissions. This problem can be resolved if the DAC load is added to day-ahead load, meaning predictable load. It can be seen that unless DAC is scaled up by 1000 times, meaning 0.8 GW power when turned on, it's impact on it's local locational marginal price and average power carbon intensity is very small, limited within 5\%. This proves that using price-taker and carbon-taker assumptions is well defended and very accurate, as the 5\% cost gap is smaller than optimization profitability gap from custom algorithm.

\begin{table}[h!]
    \centering
    \begin{tabular}{c|cccc} \hline
         Cases & DAC=1, SR=0 & DAC=10, SR=0 & DAC=100, SR=0 & DAC=1000, SR=0\\ \hline
         DAC power costs & +2.3\% & +2.5\% & +4.6\% & +43.7\%\\
         DAC \co emissions& +0.0016\% & +0.008\% & +0.023\% & +0.21\%\\ \hline
         Cases & DAC=1, SR=1 & DAC=10, SR=10 & DAC=100, SR=100 & DAC=1000, SR=1000\\ \hline
         DAC power costs & +2.94\% & +3.4\% & +4.2\% & +42.2\%\\
         DAC power costs & -0.0007\% & +0.009\% & +0.025\% & +0.36\%\\
    \end{tabular}
    \caption{\textbf{ Strategic operation results of DAC power system integration analysis compared to using price-taker and carbon-taker assumptions.} The cases are named after scaling of DAC. For example, "DAC=10" meaning the DAC is scale up by factor of 10 times as it's modeled in this study; "SR=10" meaning the solar renewable is added by scale of 10. DAC scale of 1 equals to 0.8 MW load on average, solar renewable scale of 1 equals to 2 MW of solar PV.}
    \label{tab:S3}
\end{table}

On the other hand, the integrated simulation with power market model is significantly more computationally expensive. A full year simulation using price-taker and carbon-taker assumptions with custom strategic algorithm takes seconds to solve, while the power market integrated model takes 6-8 hours, mostly repetitively solving unit commitment and economic dispatch optimization. Consider the wide ranges of scenarios and geography coverage in this study include thousands of simulations to run, the power market integrated model is practically impossible. Here we show the results that defend the price-taker and carbon-taker assumptions and all results presented in the manuscript is generated using the assumptions. 

\section{DAC Technology Models}\label{sec4}

Three different DAC technology samples are extracted from the literature and tested in this study. The details of the technology and parameters for model inputs are summarized in \textcolor{blue}{Table}~\ref{tab:table2}. As the short cycle MOF technology with about 1 hour cycle time most ideal for solar-thermal DAC considering reducing the solar curtailment, only MOF results is presented in the main text. 

\begin{table}[h!]
    \centering
    \begin{tabular}{cccc}
         & SI-AEATPMS & APDES-NFC-FD & MOF\\
        Cycle Switching Cost  & 213.36& 42.02& 115.60\\
        Adsorption Power Consumption & 0.357& 0.300& 0.642\\
        Desorption Power Consumption & 0.071& 0.060& 0.097\\
        \(\beta^{a}_{1}\) & 0.00099& 0.009434& 0.2 \\
        \(\beta^{a}_{2}\) & 0 & 0 & -0.2 \\
        \(\beta^{d}_{1}\) & 0 & 0 & 0 \\
        \(\beta^{d}_{2}\) & 0.088& 0.5& 0.4 \\
    \end{tabular}
    \caption{Key Parameters Input for Different DAC Technologies}
    \label{tab:table2}
\end{table}

In this study, two amine-functionalized sorbents, SI-AEATPMS and APDES-NFC-FD, are compared against the Nobel MOF sorbent. Leonzio et al.~\cite{leonzio_environmental_2022} juxtapose these amine-functionalized sorbents with three Metal-Organic Frameworks (MOFs) - MIL-101, MOF-177, and MOF-5, investigating their efficacy in capturing CO2 from dry air via a temperature swing adsorption system. To get more detailed non-linear adsorption/desorption rates behavior for the MOF sorbent technology, this paper also combined the results from \cite{azarabadi_sorbent-focused_2019}\cite{sinha_systems_2017} to calculate the \(\beta\)-values.

SI-AEATPMS, which stands for [N-(2-aminoethyl)-3-aminopropyl]trimethoxysilane grafted on silica gel, is created by drying silica gel beads and subsequently loading them with AEATPMS. This technology is characterized by a long, 89.6 hour cycle time, adsorbing for \( 3.06 \times 10^5 \) seconds, and desorbing for \( 1.67 \times 10^4 \) seconds.

Conversely, APDES-NFC-FD, which is 3-aminopropylmethyldiethoxysilane on nanofibrillated cellulose, is made by adding APDES to an NFC hydrogel and then undergoing a freeze-drying process.  This sorbent has a shorter, 9.6 hour cycle time, adsorbing for  \( 3.26 \times 10^4 \)seconds, and desorbing for \( 1.85 \times 10^4 \) seconds.

It's worth noting that Climeworks, a leading company in the Direct Air Capture (DAC) field, has endorsed these amine-based sorbents. In fact, APDES-NFC-FD is currently being used in their facility in Switzerland.


\subsection{Parameterization: Translating Source Data to Code Inputs}

\subsubsection{Calculation of Cycle Switching Cost}

\(S\) represents the total cost per cycle, which is the sum of various operational expenses. It is computed as:

\begin{equation}
S = S_{\text{sorbent material}} + S_{\text{thermal}}
\end{equation}

\subsubsection{Calculation of Power Consumption Rates}

The power consumption parameters are primarily focused on electrical energy. It's assumed that the regeneration energy is predominantly thermal, which can be sourced at a lower cost from waste heat. The most significant cost component is the electrical energy which is primarily used for fans.

\textbf{Adsorption Power Consumption:}
\begin{equation}
P_{\text{a}} = \frac{\text{Total Electricity Consumption (MWh/ton-CO}_2\text{)}}{\text{Number of 5-minute time steps for adsorption}}
\end{equation}

\textbf{Desorption Power Consumption:}
\begin{equation}
P_{\text{d}} = \frac{\text{Total Electricity Consumption (MWh/ton-CO}_2\text{)}}{\text{Number of 5-minute time steps for desorption}}
\end{equation}

\subsubsection{Calculation of Beta Coefficients}

The beta coefficients, which are used in the piecewise linear approximation for both absorption and desorption, were derived from the figures presented by Leonzio et al.\cite{leonzio_environmental_2022} Values were chosen to recover the number of time steps for the desired cycle length. Be minded that to keep consistent across different technologies, these \(\beta\)-values are calculated using unit capacity where \(\bar{X}\) = 1. 

\textbf{Absorption Coefficients:}

For both amine-functionalized sorbents, SI-AEATPMS and APDES-NFC-FD, the adsorption curve figure displays a long flat region before the ratio of outlet to inlet \(CO_2\) concentration reaches 1, indicating system saturation. This flat region preceding the breakthrough suggests a linear behavior, which allows for the extrapolation of a linear relationship. Because of this, in the case of SI-AEATPMS and APDES-NFC-FD, the secondary term for the beta values is set to zero:

\begin{itemize}
    \item \(\beta^{a}_{1}\): Represents the first-order coefficient for absorption.
    \item \(\beta^{a}_{2}\): Represents the second-order coefficient for absorption.
\end{itemize}

For the calculation of \( \beta^{a}_{1} \), a line is fit using two distinct points from the figure showing the loading for chemisorbents SI-AEATPMS and APDES-NFC-FD, for both the adsorption rates are effective linear and the second order coefficient is 0.

\textbf{Desorption Coefficients:}

For the desorption curve, the behavior is non-linear, so three distinct points from the figure (figure 8) showing CO2 concentration during the desorption step were selected and fit to the equation \( X \sim (1-\beta^{d}_{2})^{x} \), where \( X \) represents the state of saturation, ranging from 0 (completely unsaturated) to 1 (fully saturated). The derived relationship provides the value for \( \beta^{d}_{2} \).

\begin{itemize}
    \item \(\beta^{d}_{1}\): Represents the first-order coefficient for desorption, which is consistently set to zero unless explicitly specified otherwise.
    \item \(\beta^{d}_{2}\): Represents the second-order coefficient for desorption derived from the exponential decay fit of the desorption curve.
\end{itemize}

\subsubsection{Calculation of \( X_{\text{hat}} \)}

\( X_{\text{hat}} \) represents the maximum capture per one absorption cycle. It is derived from the ratio of the static plant capacity to the number of cycles per year. 

The number of cycles per year can be computed as:
\begin{equation}
\text{Number of cycles per year} = \frac{8760}{\text{cycle time (hours)}}
\end{equation}

Subsequently, \( X_{\text{hat}} \) is calculated with:
\begin{equation}
\bar{X} = \frac{\text{100\% Plant Capacity}}{\text{Number of cycles per year}}
\end{equation}


\section{Supplementary Results and Discussions}\label{sec6}

\subsection{Temporal behavior discussion with grid connection}

Power market pricing is highly volatile, driven by fluctuations in supply and demand that vary by time of day and season. Prices tend to be high when net demand is high, while low prices can result from either low overall demand (e.g., midnight) or an abundance of renewable energy generation. In regions like California, the ``duck curve" phenomenon highlights how solar power production peaks during midday, significantly reducing the price. Steep price peaks occur during sunset as the decline in solar power generation significantly increases net demand. Although the average daily power price typically ranges from \$20 to \$100, prices at specific hour or days can be far more volatile, dropping to zero or even negative during periods of renewable energy curtailment and soaring above \$1000 during price spikes.

\begin{figure}[h!]
    \centering
        \begin{subfigure}[b]{0.49\textwidth}
            \centering
            \includegraphics[width=\textwidth]{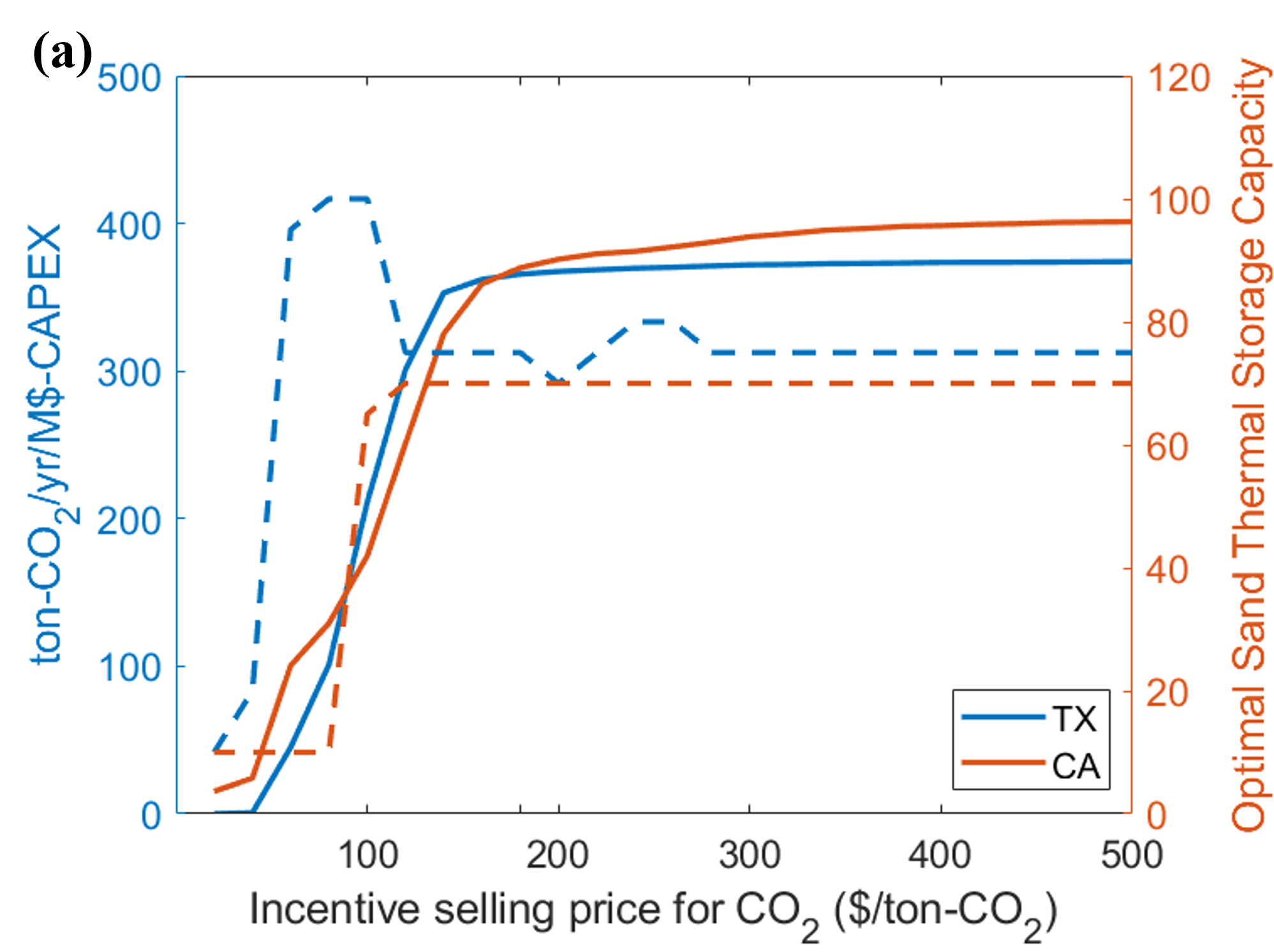}
        \end{subfigure}
        \hfill
        \begin{subfigure}[b]{0.49\textwidth}  
            \centering 
            \includegraphics[width=\textwidth]{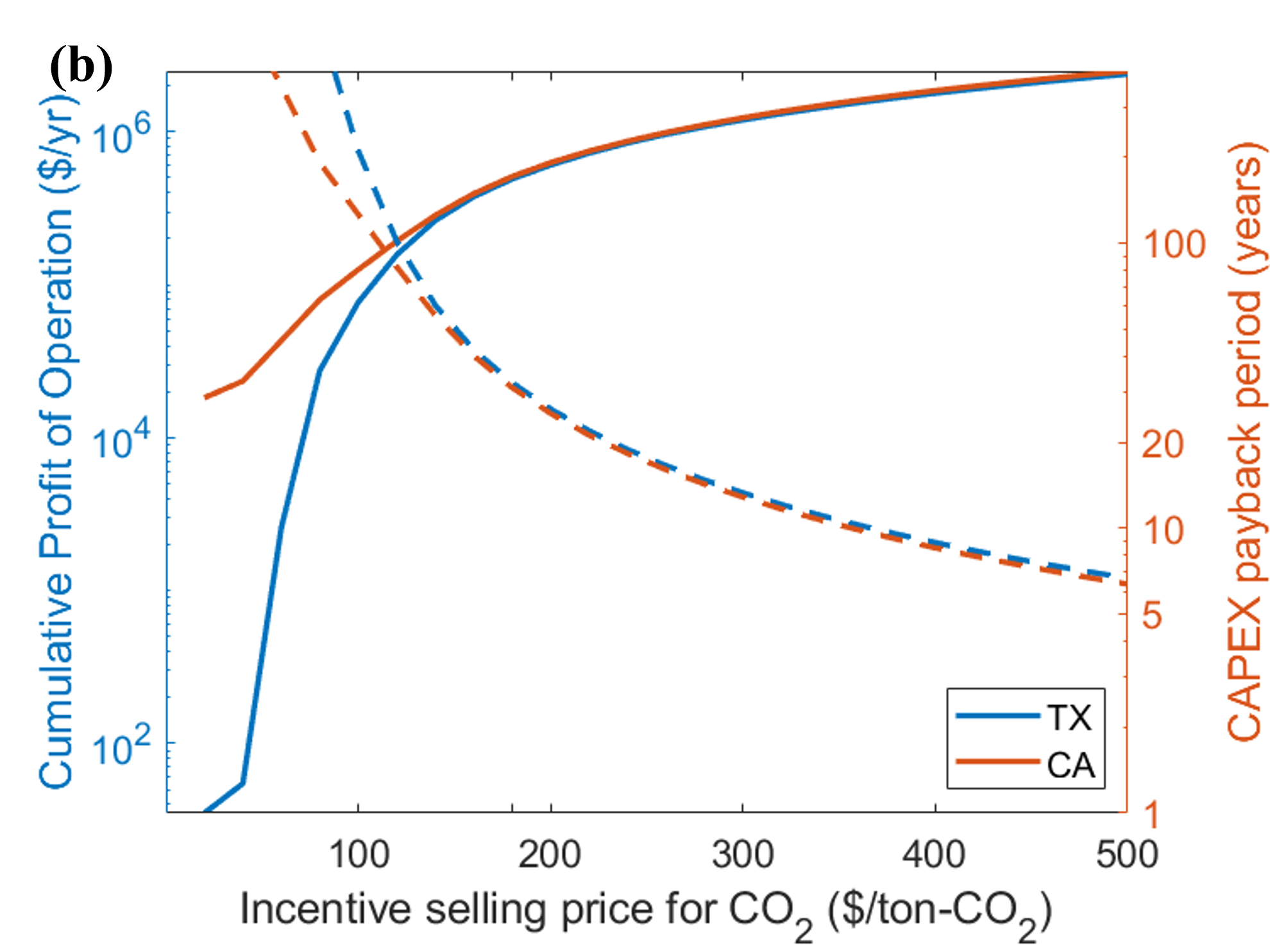} 
        \end{subfigure}
        \begin{subfigure}[b]{0.49\textwidth}  
            \centering 
            \includegraphics[width=\textwidth]{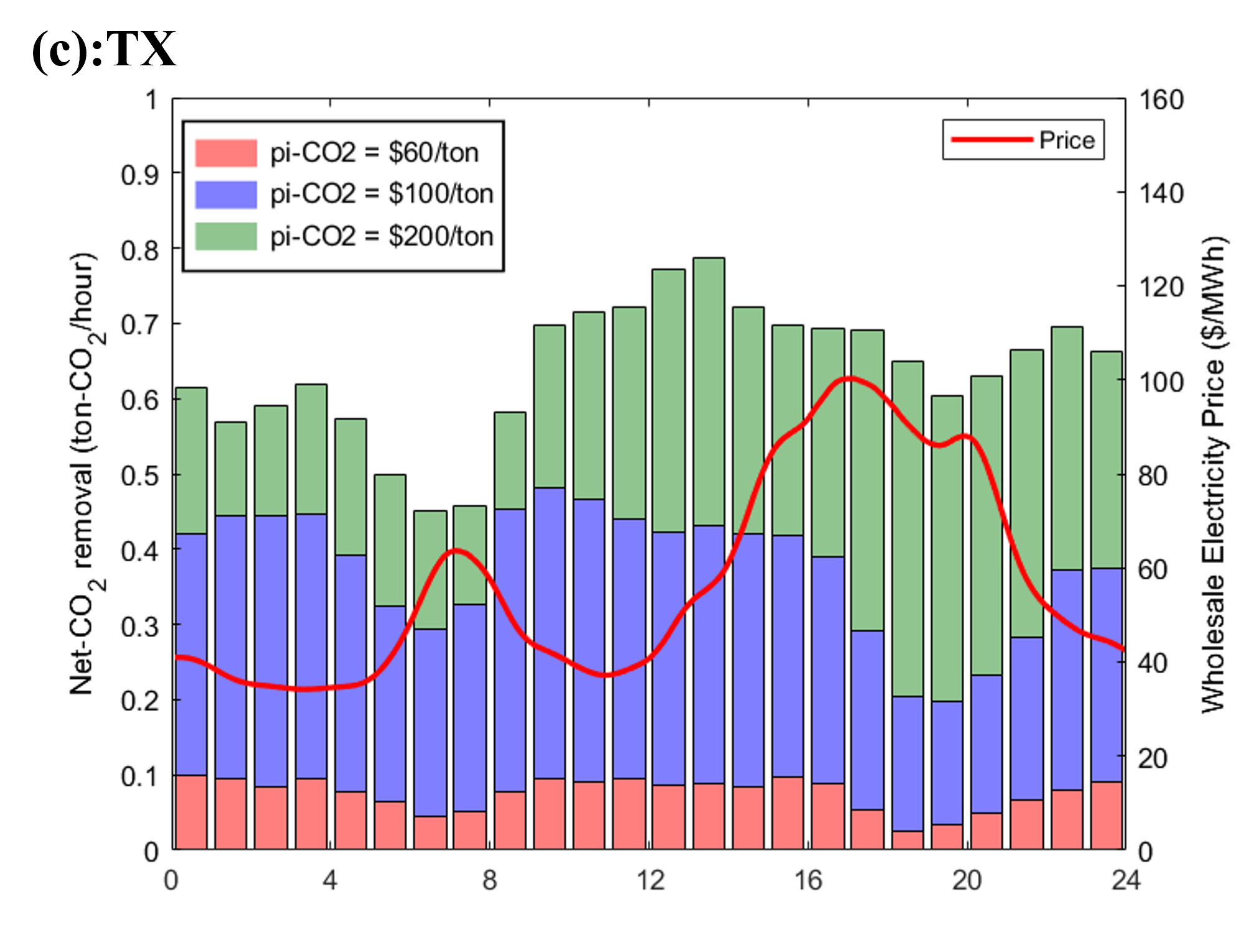} 
        \end{subfigure}
        \begin{subfigure}[b]{0.49\textwidth}  
            \centering 
            \includegraphics[width=\textwidth]{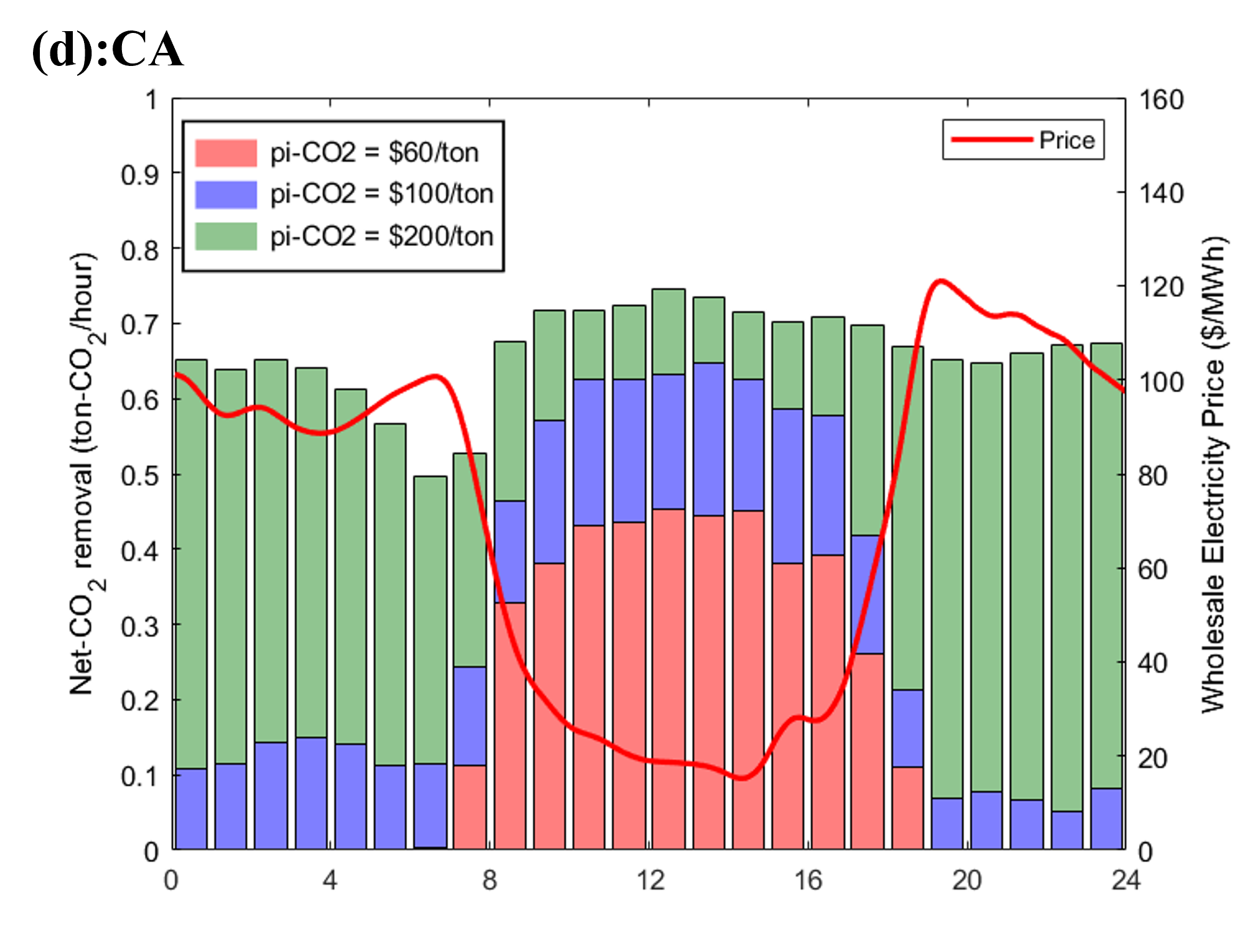} 
        \end{subfigure}
    \caption{\textbf{Solar-thermal DAC grid-interactive performance vs incentive selling prices and daily temporal behaviors in TX and CA power markets.} (a) \co removal per unit CAPEX (left, solid) and optimal thermal storage capacity (right, dashed) vs incentive selling prices; (b) operational profit (left, solid) and CAPEX payback period (right, dashed) vs incentive selling prices; (c)(d) DAC average hourly operation and average daily wholesale electricity price in TX and CA respectively with different incentive levels. Results averaged based on 2022 full-year optimization with 5-min data resolution. }
    \label{tab:fig4}
\end{figure}

Solar-thermal DAC system performs better in locations with solar-dominated electricity generation. As shown in \textcolor{blue}{Figure}~\ref{tab:fig4}, when connected to the solar-dominated California power market, solar-thermal DAC system demonstrate robust design optimization in thermal storage sizing and higher profits without sacrificing net-\co removal volume. When incentive selling prices of \co increase, California grid-connection firstly recovers the solar profile which have lower electricity prices without need of thermal storage. The ``solar synchronization'' also suggest DAC integration to solar-dominated power market can help absorb curtailed solar generation, leading to more cost efficient and sustainable operation (California has 7\% higher \co capture efficiency than Texas). California case performs slightly better in both \co abatement volume and profitability when incentive is sufficiently high despite the fact that the annual average power price in California (\$70/MWh) is higher than Texas (\$57/MWh). At low incentive levels (\(<\$150/ton\)), the solar-dominated California power market outperforms Texas by an order of magnitude in profitability and investment payback period, highlighting its long-term potential as the technology matures and incentives gradually decline. Additionally, deeper solar penetration is expected to intensify the "duck curve" phenomenon in the California grid.

While the main text and figures present results as annual averages or as representative daily averages over 365 days, it is important to note that significant seasonal and daily temporal variations exist. These variations necessitate high temporal resolution analyses. For instance, in the TX case, a noticeable discrepancy is observed: the electricity price peaks in July while \co removal reaches a valley in August. This counterintuitive result arises because aggregating daily data into monthly averages can mask the impact of outlier events. In July, although the average price is high due to several extremely expensive days, there are still eight days with average prices below \$60/MWh, which renders DAC operation marginally profitable and sustains a moderate capacity factor. Conversely, in August, despite a lower overall average price—attributable to the absence of extreme outliers—all 31 days exhibit prices exceeding \$60/MWh, rendering DAC operations unprofitable except during isolated, exceptional hours, and thus leading to a significantly reduced capacity factor. This mismatch underscores that aggregating higher-resolution data into lower-resolution averages can obscure critical dynamics that directly affect profitability. Consequently, employing high temporal resolution data—particularly the 5-minute resolution wholesale electricity prices—is essential for accurately capturing market volatility and optimizing operation scheduling for DAC systems. In contrast, relying on daily or annual averages can lead to substantial uncertainties in the analysis. Notably, while hourly-resolution solar data may suffice for modeling solar-driven processes such as thermal energy storage and DAC regeneration heat scheduling, the high variability of electricity prices with 5-min temporal resolution plays a non-trivial role in determining DAC profitability and therefore shall be carefully modeled.

\subsection{Global impact analysis}

\begin{figure}[h!]
    \centering
        \begin{subfigure}[b]{0.49\textwidth}
            \centering
            \includegraphics[width=\textwidth]{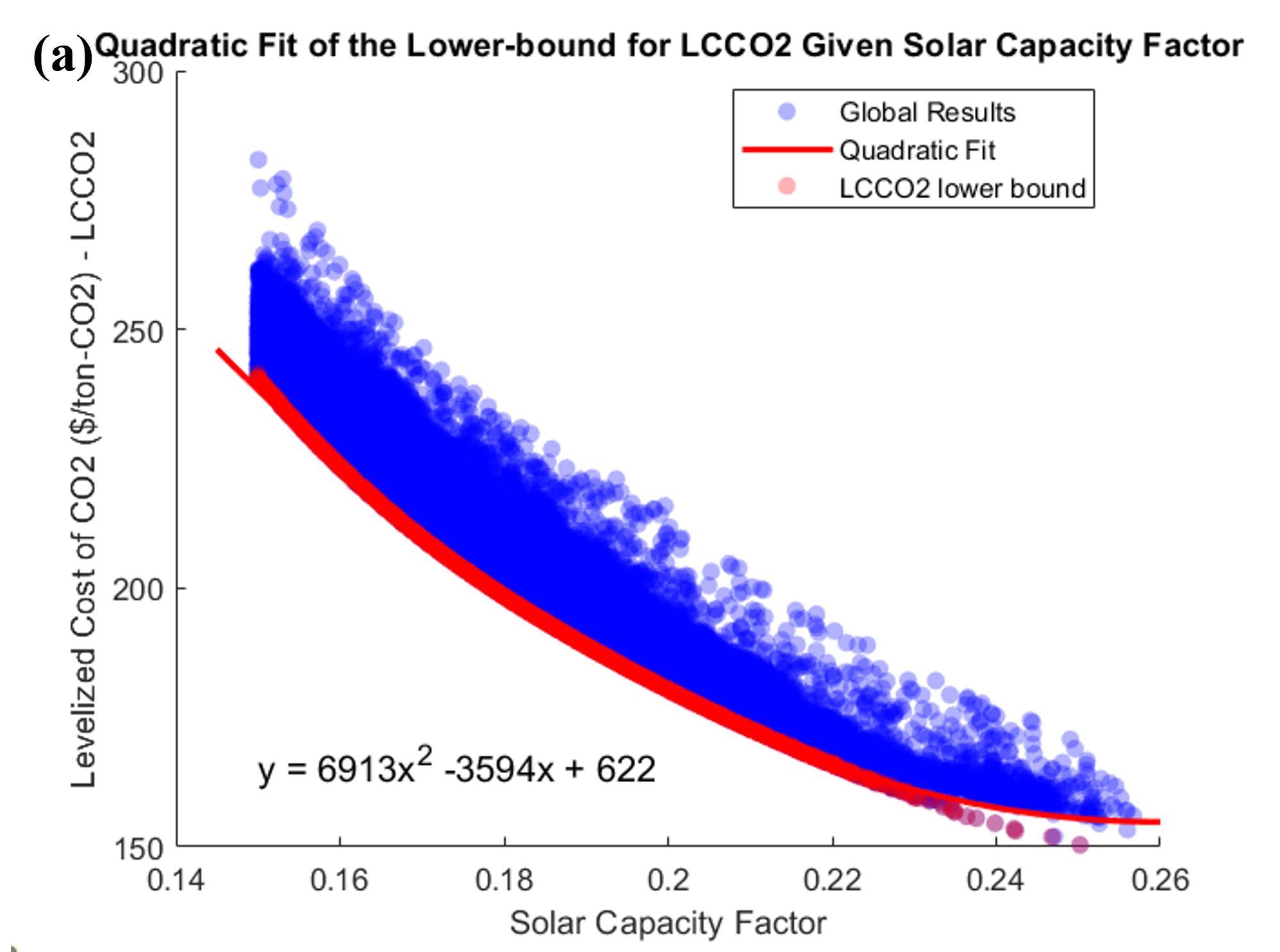}
        \end{subfigure}
        \hfill
        \begin{subfigure}[b]{0.49\textwidth}  
            \centering 
            \includegraphics[width=\textwidth]{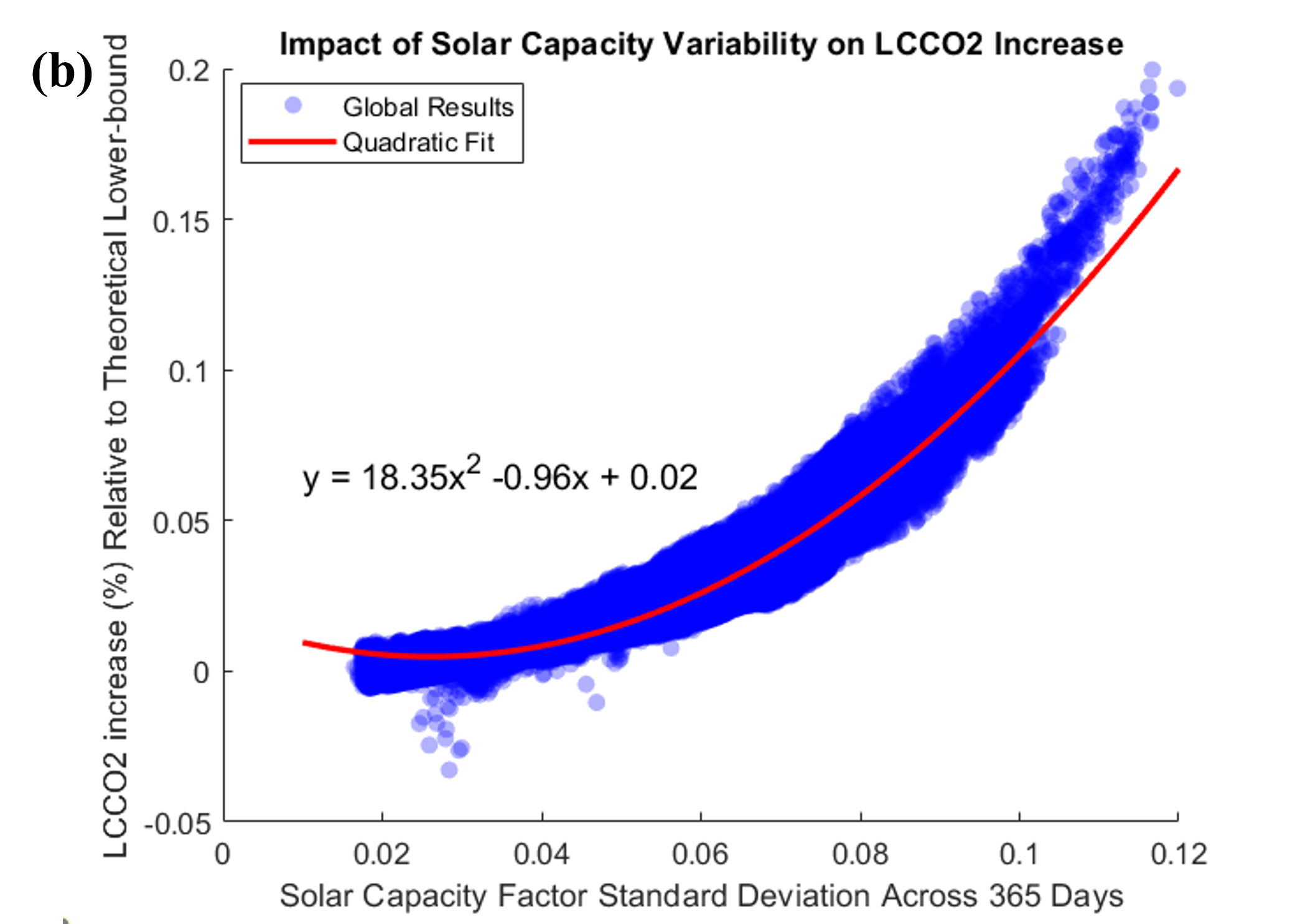} 
        \end{subfigure}
        \begin{subfigure}[b]{1\textwidth}  
            \centering 
            \includegraphics[width=\textwidth]{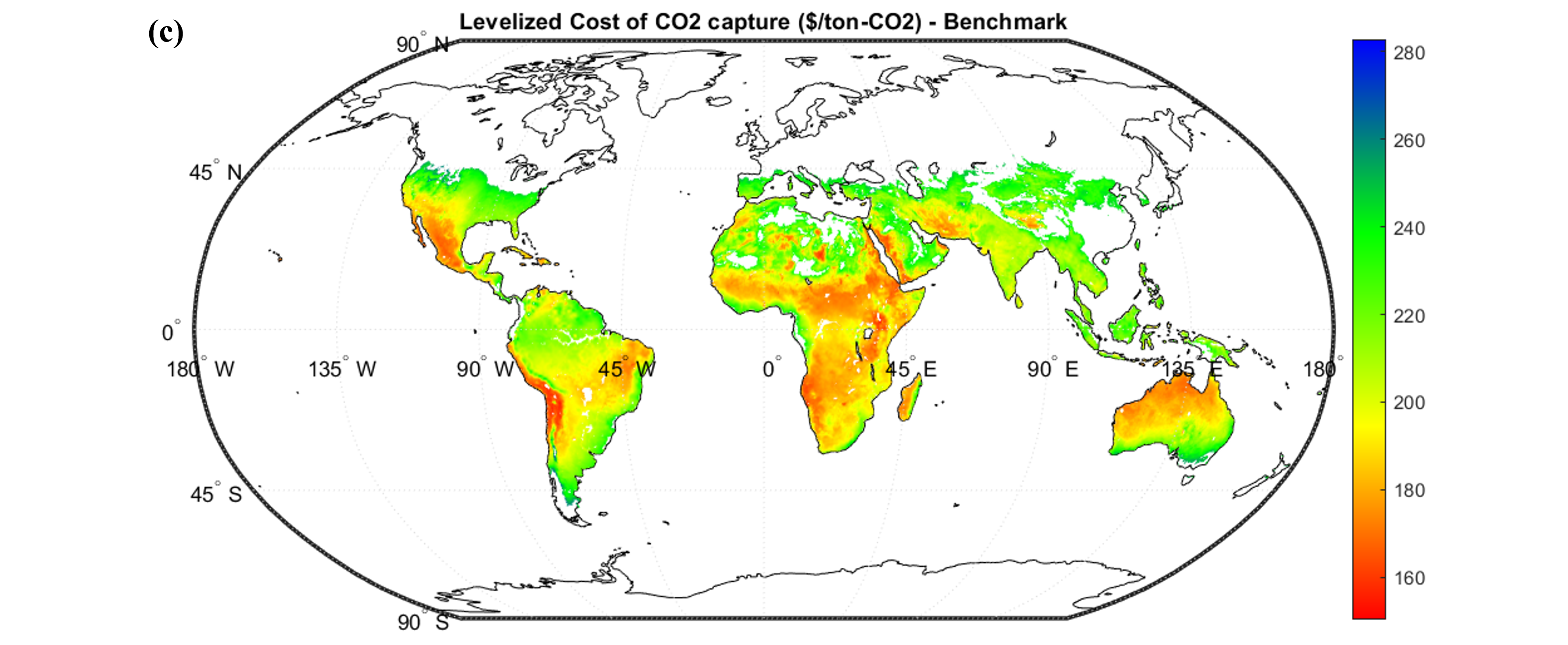} 
        \end{subfigure}
    \caption{\textbf{Stand-alone solar-driven DAC global deployment cost analysis.} (a) LCCO2 vs annual average solar capacity factor at each point of the global analysis, add quadratic fitting curve for the lower bound of LCCO2 given solar capacity factor; (b) daily average solar capacity factor standard deviation across the year best explains LCCO2 differences relative to lower bound; (c) global mapping of solar stand-alone LCCO2. Analysis focuses on regions with solar capacity factor >15\% with 107713 points globally, some place without calibrated data such as Sahara desert are excluded. }
    \label{tab:fig_S}
\end{figure}

It can be seen clearly that annual average solar capacity factor and capacity factor standard deviation across 365 daily average are most important predictors for solar-DAC performance. For each annual solar average capacity factor, there exist a lower bound for the LC\co of stand-alone solar-DAC systems, and this lower bound can be reached when the standard deviation across 365 daily average capacity factor is minimized.

\pagebreak

\subsection{MATLAB Simscape Model Details}

\begin{figure}[h!]
    \centering
    \includegraphics[width=1\linewidth]{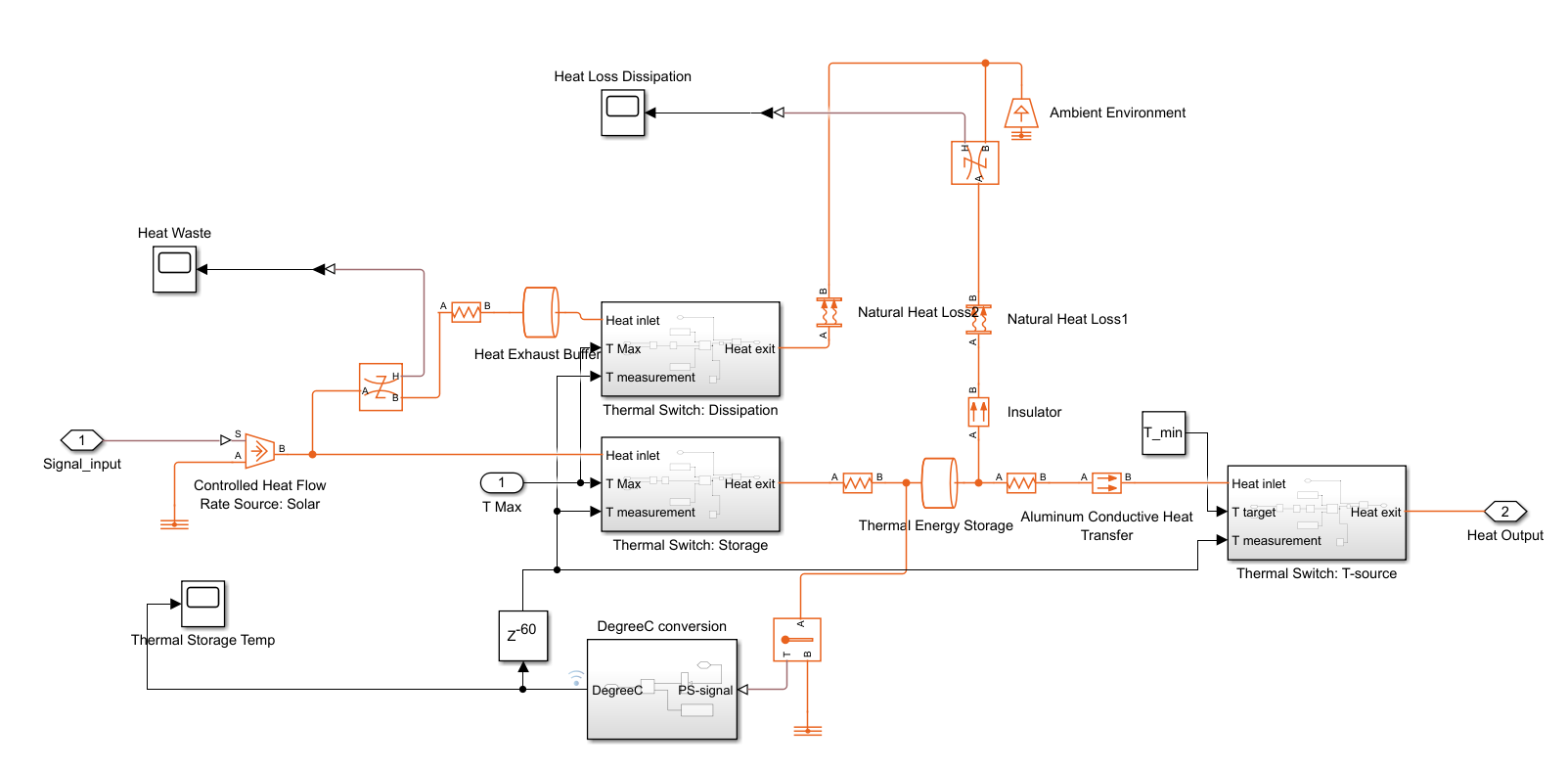}
    \caption{\textbf{MATLAB Simscape model: solar thermal and thermal storage subsystem details.}}
    \label{tab:figS6}
\end{figure}

\textcolor{blue}{Figures~\ref{tab:figS6}} to \textcolor{blue}{Figures~\ref{tab:figS12}} showing details about MATLAB Simscape model structure.

\begin{figure}[h!]
    \centering
    \includegraphics[width=1\linewidth]{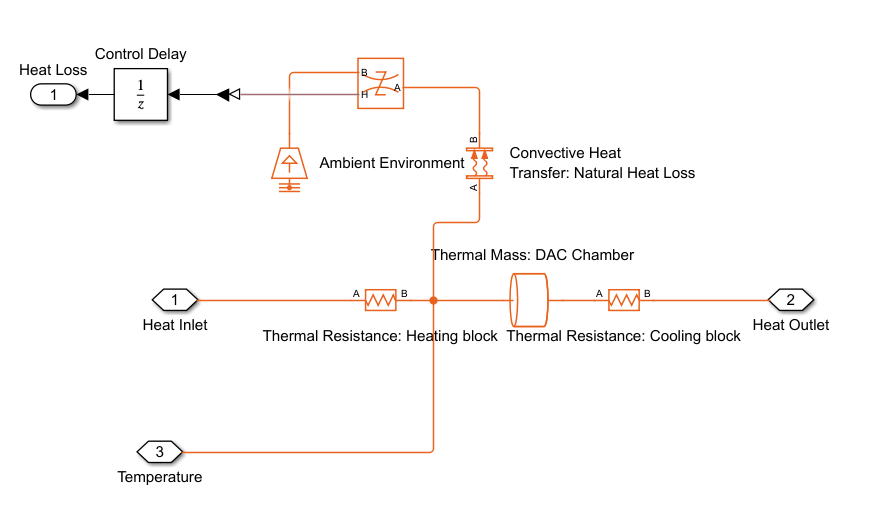}
    \caption{\textbf{MATLAB Simscape model: DAC unit subsystem by treating DAC chamber as a thermal mass.}}
    \label{tab:figS7}
\end{figure}

\begin{figure}[p!]
    \centering
    \includegraphics[width=1\linewidth]{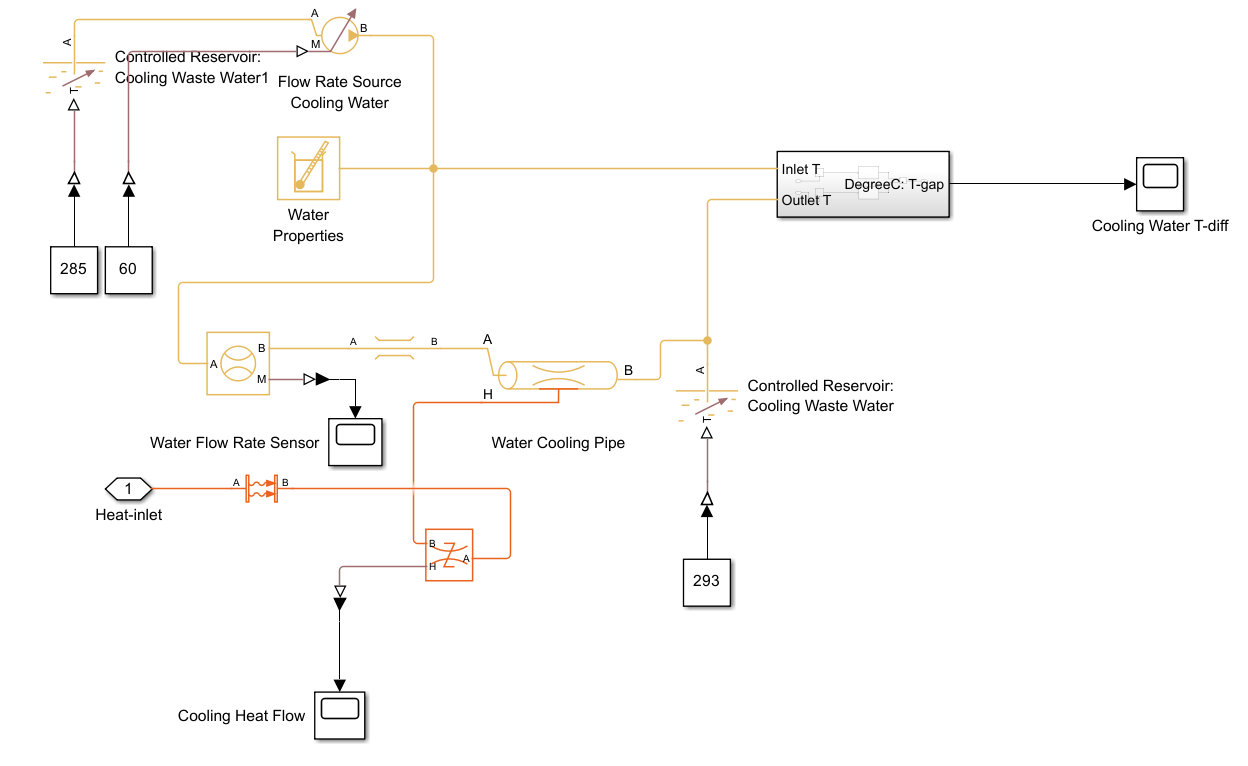}
    \caption{\textbf{MATLAB Simscape model: Water cooling subsystem (thermo-fluid with water properties).}}
    \label{tab:figS8}
\end{figure}

\begin{figure}[p!]
    \centering
    \includegraphics[width=1\linewidth]{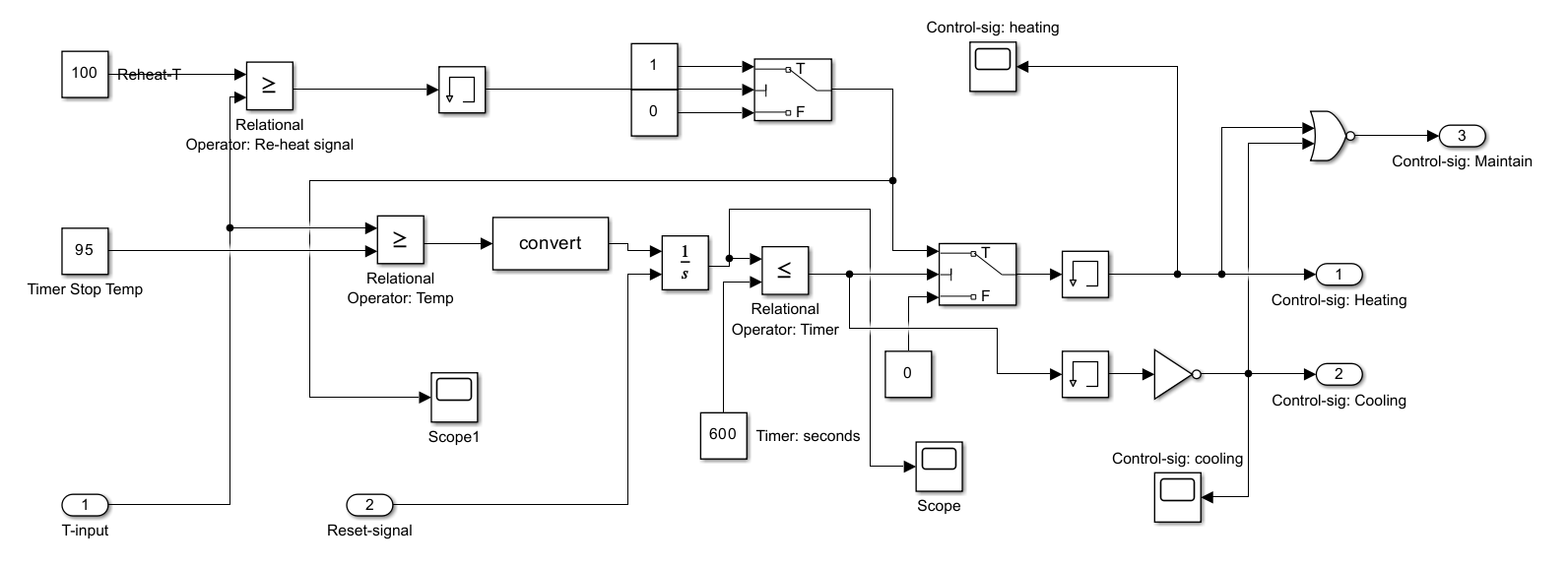}
    \caption{\textbf{MATLAB Simscape model: DAC regeneration control subsystem system by receiving optimization signals.}}
    \label{tab:figS9}
\end{figure}

\begin{figure}[p!]
    \centering
    \includegraphics[width=1\linewidth]{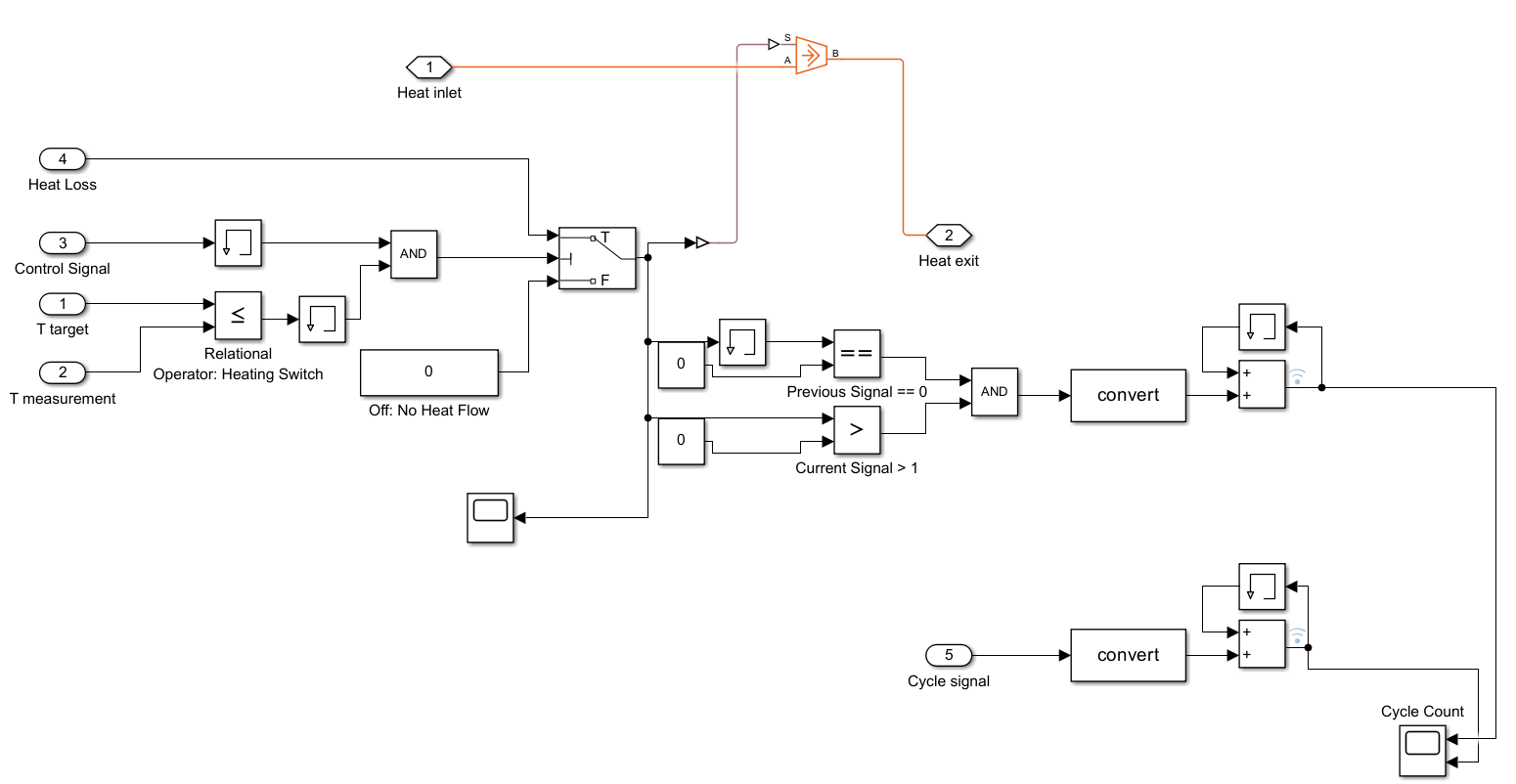}
    \caption{\textbf{MATLAB Simscape model: DAC temperature maintain control subsystem.}}
    \label{tab:figS10}
\end{figure}

\begin{figure}[p!]
    \centering
    \includegraphics[width=1\linewidth]{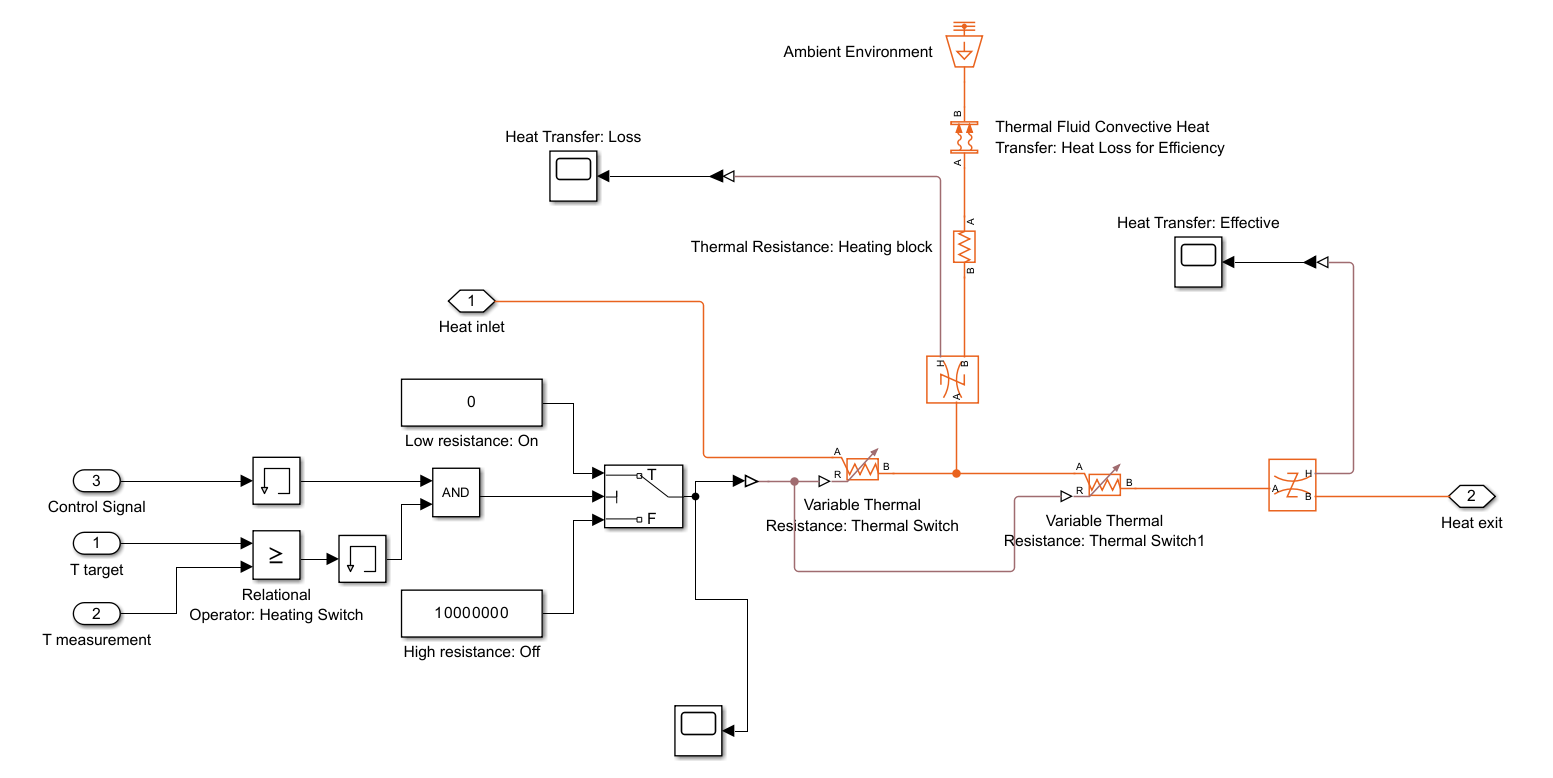}
    \caption{\textbf{MATLAB Simscape model: thermal switch subsystem to control heat transfer using variable thermal resistor.}}
    \label{tab:figS11}
\end{figure}

\begin{figure}[p!]
    \centering
    \includegraphics[width=1\linewidth]{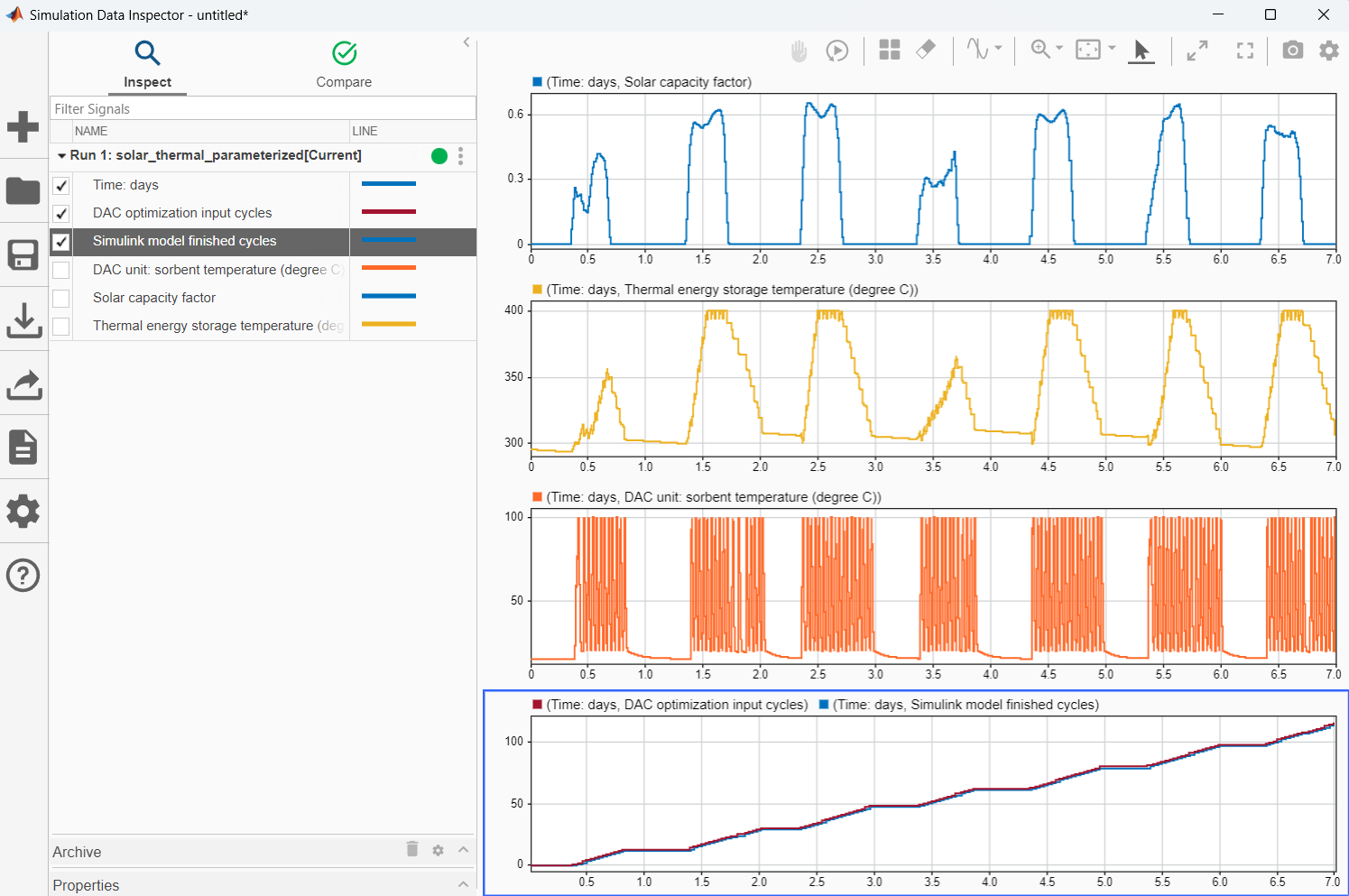}
    \caption{\textbf{MATLAB Simscape model: automatic data inspector with 7-day simulation sample results.}}
    \label{tab:figS12}
\end{figure}

\clearpage
\bibliographystyle{plain}
\bibliography{DAC_main}
